\documentclass[preprint,11pt,nofootinbib]{article}
\usepackage{jheppub}
\usepackage{overpic}
\usepackage[english]{babel}
\usepackage{graphicx}
\usepackage{epsfig}
\usepackage{amssymb}
\usepackage{amsmath}
\usepackage{ulem}
\usepackage[T1]{fontenc} % if needed
\usepackage[utf8]{inputenc}
\usepackage{color}

\usepackage[titletoc]{appendix}

\begin{document}

\title{Pion quasiparticles and QCD phase transitions at finite temperature and isospin density from holography}

\author[a]{Xuanmin Cao,}
\emailAdd{caoxm@jnu.edu.cn}
\author[a]{Hui Liu,}%\note{Corresponding author.}
\emailAdd{tliuhui@jnu.edu.cn}
%\email[Corresponding author: ]{tliuhui@jnu.edu.cn}
\author[a,1]{Danning Li}\note{Corresponding author.}
\emailAdd{lidanning@jnu.edu.cn}
%\email[Co-corresponding author: ]{lidanning@jnu.edu.cn}
\affiliation[a]{
 Department of Physics and Siyuan Laboratory, Jinan University, Guangzhou 510632, China}

\abstract{
Spectra of pions, which are known as the pseudo-Goldstone bosons of spontaneous chiral symmetry breaking, as well as their relationship with chiral phase transition and pion superfluidity phase transition, have been investigated in the framework of soft-wall AdS/QCD. In chiral limit, it is proved both numerically and analytically that pions are massless Goldstone bosons even at finite temperature, which was usually considered as an assumption in soft-wall models. Above $T_c$, at which chiral condensate $\langle \bar{q}q\rangle$ vanishes,  the spectra of pions and scalar mesons merge together, showing the evidence of the restored chiral symmetry in hadronic spectrum level. Extending to finite quark mass, pion masses increase with quark mass. Further, it is more interesting to observe that the pole masses of pions decrease with temperature below $T_c$, which agrees with the analysis in Phys.Rev.Lett.88(2002)202302. Meanwhile, symmetry restoration above $T_c$ could be seen in the spectra of scalar and pseudo-scalar mesons. With finite temperature and isospin chemical potential $\mu_I$, it is shown that the masses of charged pions would split.  The mass of positive charged pion $\pi^+$ decreases  almost linearly to zero when $\mu_I$ grows to  $\mu_{I}^c$, where pion condensation starts to form. This reveals the Goldstone nature of $\pi^+$ after pion superfluidity transition, which are closely related to the experimental observation.
}

\maketitle

\newcommand{\limit}[3]
{\ensuremath{\lim_{#1 \rightarrow #2} #3}}

\section{Introduction}
Relativistic heavy ion collisions(RHIC) provide an important approach to probe the possible new state of nuclear matter in the laboratory \cite{Adams:2005dq}. The fire ball, created by collisions of high energy nuclei, is far away from equilibrium and it exists for only few $\text{fm}/c$, which makes it difficult to be detected directly.  Instead, the distributions and correlations of hadrons emitted after freeze-out and hadronization, are the direct observables to probe the hot/dense nuclear matter. To understand the experimental data, the in-medium properties of particles are of critical importance. For example, the variation of meson mass at finite temperature might significantly change the final distribution of hadrons\cite{Cleymans:2006xj,Andronic:2008gu}. Therefore, it is quite essential to get full understanding of particles in medium.

Among the different probes, pion is an important and special one both theoretically and phenomenologically. It is the lightest meson, which shows its nature of pseudo-Goldstone bosons(PGB). As a concequence, it is closely related with chiral phase transition, which happens between chiral symmetry broken and restored phases. Thus, it is considered as a possible probe of the transition\cite{Dumitru:1994vc}. Furthermore, the detecting of a significant fraction of coherent charged pions\cite{Abelev:2013pqa} at LHC energy, suggests a possible superfluid phase, consisting of condensed charged pions\cite{Begun:2013nga,Begun:2015ifa}. Besides, the ratio of multiplicity of charged pions $\pi^-$ to $\pi^+$ provides the possibility for extracting symmetry energy, which plays important roles in neutron stars\cite{Sako:2014usa}.  Thus, to study the properties of pions in medium have attracted lots of attentions.

Due to the pesudo-Goldstone nature, the relevant physics of pion spectrum in medium is naturally nonperturbative. Actually, it is usually studied within non-pertubative frameworks, like lattice simulations\cite{Ishii:2016dln,Brandt:2014qqa,Brandt:2015sxa}, Dyson-Schwinger Equations(DSE)\cite{Fischer:2018sdj,Gao:2020hwo}, functional renormalization group(FRG)\cite{Tripolt:2013jra,Wang:2017vis}, Nambu-Jona-Lasinio models(NJL)\cite{Ebert:1992jx,Xia:2013caa,Xia:2014bla,Chao:2018ejd,Liu:2018zag,Xu:2020yag}. Generally, due to the effective interacting (attractive or repulsive) with the medium, pions might have different masses and widths at different temperatures and densities, and they are termed `quasipions' in Ref.\cite{Shuryak:1990ie}. At relatively high temperature, above chiral transition temperature $T_c$, most of the studies obtain growing masses of pion quasiparticles with the increasing temperature\cite{Fischer:2018sdj,Gao:2020hwo,Tripolt:2013jra,Xia:2013caa,Xia:2014bla}. Such a behavior might originate partly from pions' Goldstone nature. In chiral limit, $m_\pi$ remains zero at temperature below $T_c$, and it could only increase with $T$ above $T_c$(though the excitation modes might change). Then, extending to physical quark mass, it is quite reasonable to understand the increasing of them. However, below $T_c$, there are debates on the temperature behavior of $m_\pi$ with physical quark mass in the literature. Son and  Stephanov give a general analysis in Refs.\cite{Son:2001ff,Son:2002ci}, and  decreasing of $m_\pi$ below $T_c$ is proposed. An estimation of about $30\%$ reduction of $m_\pi$ from its vacuum value to its values around $T_c$ is suggested. Qualitatively, this result is supported by lattice simulations in Refs.\cite{Brandt:2014qqa,Brandt:2015sxa} and model study in NJL model with gluon condensate in Ref.\cite{Ebert:1992jx}. Nevertheless, the studies from other research groups using different methods, including NJL model\cite{Xia:2013caa,Xia:2014bla}, FRG\cite{Tripolt:2013jra}, lattice simulations\cite{Ishii:2016dln}, DSEs\cite{Gao:2020hwo}, obtain a contrast result, showing an increasing of $m_\pi$ below $T_c$. A conclusion on this issue is still hard to be drawn, and more information from other approaches might be useful and necessary.

Fortunately, in recent decades, the discovery of anti-desitter/conformal field theory correspondence (AdS/CFT)\cite{Maldacena:1997re,Gubser:1998bc,Witten:1998qj} offers another possibility to solve the strong coupling problems of Quantum Chromodynamics(QCD). The shear viscosity to entropy density ratio $\eta/s$ is worked out as small as ${1}/{(4\pi)}\approx0.08$\cite{Kovtun:2004de} , which agrees with the values used in fitting elliptic flow $v_2$ of RHIC data\cite{Teaney:2000cw,Huovinen:2001cy,Hirano:2005xf,Romatschke:2007mq}. In addition, the application of holographic methods in QCD is shown to be powerful in many other models, like the top-down brane systems\cite{Karch:2002sh,Babington:2003vm,Kruczenski:2003uq,Sakai:2004cn,Sakai:2005yt} and the bottom-up hard wall model\cite{Erlich:2005qh}, soft-wall model\cite{Karch:2006pv},  Light-front holographic QCD\cite{deTeramond:2005su}, Einstein-Maxwell-Dilaton systems\cite{Gubser:2008ny,Gubser:2008yx,DeWolfe:2010he,Gursoy:2007cb,Gursoy:2007er} and so on(for a review, please refer to Refs.\cite{Aharony:1999ti,Erdmenger:2007cm,deTeramond:2012rt,Adams:2012th,Brodsky:2014yha}).

Among those models, the hard-wall model and soft-wall model provide a good start point to describe hadronic spectrum, as well as chiral phase transition.  The spectrum of pions in the vacuum extracted in their extended models agree very well with the experimental data\cite{Gherghetta-Kapusta-Kelley,Gherghetta-Kapusta-Kelley-2,YLWu,YLWu-1,Li:2012ay,Li:2013oda,Colangelo:2008us,Ballon-Bayona:2020qpq,FolcoCapossoli:2019imm,Contreras:2018hbi}. Chiral symmetry breaking is well characterized both by a non-zero chiral condensate and by the mass splits of chiral partners, $(\rho, a_1)$ and $(f_0, \pi)$\cite{Gherghetta-Kapusta-Kelley,Gherghetta-Kapusta-Kelley-2,Li:2012ay,Li:2013oda}. Extended to finite temperature, it is shown that chiral condensate could be determined dynamically by soft-wall model itself, and chiral phase transition could be well depicted\cite{Colangelo:2011sr,Dudal:2015wfn,Chelabi:2015cwn,Chelabi:2015gpc,Fang:2015ytf,Li:2016gfn,Li:2016smq,Bartz:2016ufc,Fang:2016nfj,Bartz:2017jku,Fang:2018vkp}. For pion quasiparticles at finite temperature, a few investigations has been made in hard-wall and soft-wall extending models\cite{Ghoroku:2005kg,Cui:2013zha,Cui:2014oba}. Those studies show that pion mass would decrease with temperature at relatively low temperature. But as pointed out in Ref.\cite{Ghoroku:2005kg}, a constant chiral condensate is input and the restoration of chiral symmetry is neglected at high temperature. Thus, the relationship between spectrum of pion quasiparticle and chiral phase transition in those models are still unknown. As well, the high temperature(above chiral transition temperature $T_c$) behavior has not been investigate, and it is unclear whether chiral symmetry could be restored in hadronic spectrum. Therefore, it is still interesting to investigate pion spectrum at finite temperature in hard-wall and soft-wall models. From theoretical aspects, one should check whether $m_\pi$ remains zero below $T_c$ in chiral limit, to guarantee the goldstone theorem\cite{Nambu:1961tp,Nambu:1961fr}. Phenomenologically, moving to the realistic case with physical quark mass, the temperature dependent behavior of $m_\pi$ under the effect of a dynamically determined chiral condensate is still interesting to be studied. Finally, it is interesting to check whether chiral symmetry could be restored in hadronic spectrum. It is meaningful to check whether scalar meson $\sigma$ and pion could become degenerate above $T_c$ or not. In this work, since chiral condensate could be determined dynamically in soft-wall models, we will consider all these issues in the soft-wall framework.

Besides the temperature effects, the density effects are of great interest as well. To seek the possible critical end point(CEP) in the $T-\mu_B$(temperature-baryon number chemical potential) plane is the primary goal of the beam energy scan(BES) project\cite{Aggarwal:2010cw,Odyniec:2013aaa,Luo:2017faz}. The recent experimental data also suggest the increasing effect of isospin density $n_I$ at LHC energy\cite{Abelev:2013pqa,Begun:2013nga,Begun:2015ifa}. At large isospin chemical potential $\mu_I$, a transition from the normal phase to a pion superfluidity phase, consisting of condensed charged pions,  might play important role at this energy. Thus, there are growing interests on the density dependent behavior of pion mass. In bottom-up holographic framework, the pure isospin density effect have been studied in hard wall model\cite{Albrecht:2010eg,Lee:2013oya,Nishihara:2014nva,Nishihara:2014nsa,Mamedov:2015sha}. A transition from normal phase to pion condensed phase is shown to occur at $\mu_I=m_\pi, T=0$. Also, mass splitting of mesons are shown in hard wall model. Since those studies focus on finite isospin density only and the mutual effect from temperature is unclear, we extend those studies to finite temperature in soft-wall model and get the $T-\mu_I$ phase diagram\cite{Lv:2018wfq,Cao:2020ske}. Here, we will also continue our studies and consider the mutual effect of isospin densities and temperature on pion quasiparticles.

The paper is organized as follows. In Sec.\ref{sec-chiral-pion}, after a brief review on chiral phase transition in soft wall model, we will extract mass spectra of pion quasiparticles and scalar mesons from spectral functions at finite temperature. Both numerically and analytically, in chiral limit, we prove the Goldstone nature of pions at low temperature. An extension of the Gell-Mann-Renner(GOR) relation at finite temperature and with physical quark mass will be given. In Sec.\ref{sec-pion-superfluidity}, we will study the mutual effect of temperature and isospin density on pion spectrum. In Sec.\ref{sum}, a conclusion will be given.

\section{Quasipions and chiral phase transition at finite temperature in soft-wall model}
\label{sec-chiral-pion}
As mentioned above, both the hard-wall model and soft-wall model provide a good start point to deal with hadronic spectrum and chiral phase transition. Incorporating the global symmetry of QCD, it could be naturally extended to cases with multiple-flavors\cite{Li:2016smq,Bartz:2017jku,Fang:2018vkp}, finite temperature\cite{Colangelo:2011sr,Dudal:2015wfn,Chelabi:2015cwn,Chelabi:2015gpc,Fang:2015ytf,Li:2016gfn,Li:2016smq,Bartz:2016ufc,Fang:2016nfj,Bartz:2017jku,Fang:2018vkp}, finite baryon number density $\mu_B$\cite{Colangelo:2011sr,Bartz:2017jku}, finite isospin number density $\mu_I$\cite{Lv:2018wfq,Cao:2020ske},different space-time dimensions\cite{Rodrigues:2018chh,Rodrigues:2018pep} and so on. Since the soft-wall model could be imposed on the radial excitations and chiral condensate could be self-consistently determined by the equation of motion, the soft wall model provides a better framework to study the relationship of the spectrum and phase transition. Here, we will briefly review the soft-wall model first.

The model starts from an action with $SU_L(N_f)\times SU_R(N_f)$ gauge symmetry, which reads\footnote{If one considers the baryon number density, a $U(1)$ related to the $U(1)_B$ symmetry could be added.}
\begin{equation}\label{action}
S=\int d^5 x \sqrt{g} e^{-\Phi(z)}{\rm{Tr}}\left\{|D_MX|^2-V_X(|X|)-\frac{1}{4g_5^2}(F_L^2+F_R^2)\right\}.
\end{equation}
Here, $g$ is the determinant of the background metric $g_{MN}$; $\Phi(z)$ is the dilaton field, which depends only on the fifth dimension  $z$; $X$ is the matrix-valued scalar field, which is dual to the operator $\bar{q}^\alpha q^\beta$, with $\alpha,\beta$ the index in flavor space; $V_X=m_5^2|X|^2+\lambda |X|^4$ is the potential of $X$, with $m_5^2$ the 5D mass of $X$ and $\lambda$ a free parameter of the potential; $g_5$ is the gauge coupling, which can be determined to be $g_5=2\pi$ by comparing the large momentum expansion of vector current correlator ($J_\mu^a=\bar q \gamma_\mu t^a q$, $\mu=(0,1,2,3)$) to perturbative calculation~\cite{Erlich:2005qh}.  In this work, we will foucus on $N_f=2$, with the lightest two flavors. Thus $t^a (a=1,2,3)$ is taken as the generators of $SU(2)$. The covariant derivative $D_M$, $M=(\mu,5)$, and gauge field strengths $F_{MN}^{L/R}$ are defined as
\begin{subequations}
\begin{eqnarray}
    D_M &=&\partial_M X-i L_MX+iX R_M,\\
    F_{MN}^{L}&=&\partial_ML_N-\partial_N L_M-i[L_M,L_N],\\
    F_{MN}^{R}&=&\partial_MR_N-\partial_N R_M-i[R_M,R_N],
\end{eqnarray}
\end{subequations}
where the gauge potentials $L_M=L_{M}^a t^a$, $R_M=R_{M}^at^a$ are dual to left- and right-handed current in the 4D field theory at the boundary, i.e. $L_{\mu}^a\leftrightarrow \bar{q}_L\gamma_{\mu}t^aq_L$ and $R_{\mu}^a\leftrightarrow \bar{q}_R\gamma_{\mu}t^aq_R$~\cite{Erlich:2005qh}. For later convenience, one can redefine the chiral gauge fields into the vector gauge field and the axial-vector gauge field,
\begin{subequations}\label{gaugef}
    \begin{eqnarray}
        V_M=\frac{L_M+R_M}{2},\\
        A_M=\frac{L_M-R_M}{2},
    \end{eqnarray}
\end{subequations}
where the vector $V_M^a$ and the axial field $A_{M}^a$ corresponds to the vector and axial vector current, $J_{V\mu}^a$ and $J_{A\mu}^a$, respectively. After a transformation, the covariant derivative and the transformed gauge field strength are deformed as
\begin{subequations}
    \begin{eqnarray}
        D_M X &=&\partial_M X-i[V_M,X]-i[{A_M,X}],\\
        F^V_{MN}&=&\frac{1}{2}(F^L_{MN}+F^R_{MN}),\\
        F^A_{MN}&=&\frac{1}{2}(F^L_{MN}-F^R_{MN}).
    \end{eqnarray}
\end{subequations}

In this section, we will focus on the temperature effect on scalar and pseudo-scalar mesons. Thus, all the gauge field will be set to zero. In QCD vacuum, only the diagonal components of the operator  $\bar{q}^\alpha q^\beta$ have non-vanishing  expectation value. Accordingly, in the dual gravity side, $X$ would be taken as
\begin{equation}\label{chi-vac}
    X=\chi t^0,
\end{equation}
where we have $t^0=I_2/2$, with $I_2$  a $2\times 2$ identity matrix. Up to now, if one couples the soft-wall action Eq.(\ref{action}) with certain gravity systems, like the Einstein-Maxwell-Dilaton systems\cite{Gubser:2008ny,Gubser:2008yx,DeWolfe:2010he,Gursoy:2007cb,Gursoy:2007er}, one can solve the background gravity and extract the information for 4D field theory. However, the full numerical process is difficult. Here, we follow previous studies\cite{Colangelo:2011sr,Dudal:2015wfn,Chelabi:2015cwn,Chelabi:2015gpc,Fang:2015ytf,Li:2016gfn,Li:2016smq,Bartz:2016ufc,Fang:2016nfj,Bartz:2017jku,Fang:2018vkp} and take Eq.(\ref{action}) as a probe. In this way, the profile of the dilaton field, the background geometric would be considered as an input generated from certain Einstein-Maxwell-Dilaton system\footnote{Generally, one might solve the background field from a fixed dilaton potential like in Refs.\cite{Gubser:2008ny,Gubser:2008yx,DeWolfe:2010he,Gursoy:2007cb,Gursoy:2007er,Finazzo:2014cna,Rougemont:2017tlu,Zollner:2018uep} or use the potential reconstruction approach to construct such kind of background like in Refs.\cite{Li:2011hp,Cai:2012xh,Chen:2019rez,He:2020fdi,Ballon-Bayona:2020xls,Mamani:2020pks}.}. Since the AdS-Schwarzchild black hole solution has been widely tested, we will follow those studies and take the geometric as
\begin{eqnarray}\label{metric-T-1}
ds^2&=&e^{2A(z)}\left(f(z)dt^2-dx^i dx_i-\frac{1}{f(z)}dz^2\right),
\end{eqnarray}
with\footnote{In the current work, the topology of the boundary is flat and the AdS radius $L$ will be scale out everywhere. Thus, we have set $L=1$.}
\begin{eqnarray}\label{metric-T-2}
A(z)&=&-\ln(z),\\
f(z)&=&1-\frac{z^4}{z_h^4}.
\end{eqnarray}
The temperature is encoded by the following equation
\begin{equation}
T=\left|\frac{f'(z_h)}{4\pi}\right|=\frac{1}{\pi z_h}.
\end{equation}
In the original soft-wall model, the dilaton field are taken as
\begin{eqnarray}\label{dilaton}
\Phi(z)=\mu_g^2 z^2,
\end{eqnarray}
with $\mu_g$ a free parameter, which will be determined from meson spectrum.

It is shown that there is no spontaneous chiral symmetry breaking under such a dilaton and constant 5D mass $m_5^2$. Either modifying the dilaton field\cite{Chelabi:2015cwn,Chelabi:2015gpc} or modifying the 5D mass\cite{Fang:2016nfj}\footnote{Actually, one can consider it as a modification on the interaction between the scalar $\chi$ and dilaton $\phi$.} are necessary. Since the scenario in \cite{Fang:2016nfj} also gives good description on meson spectra, here we will follow this scenario and take the 5D mass as the following function
\begin{equation}\label{m5}
    m_5^2(z)=-3-\mu_c z^2,
\end{equation}
with $\mu_c$ another free parameter. According to the study in Ref.\cite{Fang:2016nfj},
\begin{eqnarray}\label{parameters}
\mu_g=440\rm{MeV}, \mu_c=1450\rm{MeV}, \lambda=80,
\end{eqnarray}
gives the best fitting of the meson spectra. Thus, we will take this group of parameters in the following calculation.

Inserting Eqs.(\ref{chi-vac}-\ref{dilaton}) into Eq.(\ref{action}), one could easily derive the equation of motion for $\chi$ as
\begin{eqnarray}\label{EOMchi}
        \chi ''+\left(3 A'+\frac{f'}{f}-\Phi '\right)\chi ' +\frac{e^{2 A}}{f} \left[\left(3+\mu_c^2 z^2\right)-\frac{\lambda \chi ^2}{2}\right]\chi  =0.
\end{eqnarray}
The above equation has two boundaries, the ultraviolet(UV)$z=0$ and the infrared(IR)$z=z_h$. The asymptotic behaviors at the two boundaries could be obtained as
    \begin{subequations}\label{aexpand-T}
        \begin{eqnarray}
            \chi(z\rightarrow 0 )&=& m_q \zeta z+\frac{m_q \zeta}{4}  \left(-2 {\mu_c}^2+4 {\mu_g}^2+m_q^2\zeta ^2 \lambda  \right)z^3 \ln(z)+\frac{\sigma}{\zeta }z^3+\mathcal{O}(z^4),\\
            \chi(z\rightarrow z_h)&=& c_0 +\frac{c_0 \left(2 \mu _c^2 z_h^2-c_0^2\lambda +6\right)}{8 z_h-4 \gamma  \mu ^2 z_h^3}\left(z-z_h\right)+\mathcal{O}[(z-z_h)^2],
        \end{eqnarray}
    \end{subequations}
with $m_q$ and $\sigma$ the two independent integral constants at UV, dual to quark mass and chiral condensate respectively. $\zeta$ is a normalization constant and it is worked out as $\zeta=\sqrt{N_c}/2\pi$ according to Ref.\cite{Cherman:2008eh}. Here, we will take $N_c=3$ for realistic QCD. $c_0$ is the integral constant generating a regular solution at the horizon $z_h$, while the other integral constant at IR is dropped out since it leads to divergent solution at $z_h$.  It is interesting to observe that, with a fix quark mass $m_q$ one can solve $\sigma,c_0$ simultaneously from Eq.\eqref{EOMchi}. At both sides, with certain values of $\sigma$ or $c_0$, one can solve the equation numerically and get solutions $\chi_\sigma$ or $\chi_{c_0}$, regular at UV and IR repectively. Requiring the smooth connection of  $\chi_\sigma$ and $\chi_{c_0}$, one gets two conditions $\chi_\sigma=\chi_{c_0}$ and $\chi^\prime_\sigma=\chi^\prime_{c_0}$. Then, $\sigma$ and $c_0$ could be solved and a full solution regular at both sides is obtained. Imposing such a `shooting' method, one can obtain the temperature dependent chiral condensate and get the information of chiral phase transition.

\subsection{Chiral phase transition in soft-wall model}
\label{sec-chiralPT-T}
Taking the value of parameters in Eq.(\ref{parameters}), one can obtain the temperature and quark mass dependent behaviors of chiral condensate. Because chiral phase transition reflects the breaking of chiral symmetry $SU(2)\times SU(2)$, which is exact symmetry in 4D theory only in chiral limit, with vanishing quark masses. A finite value of quark mass would always generate slight breaking of the symmetry. As a theoretical check, one should check whether symmetry breaking appears in this limit first.
\begin{figure}[h]
    \centering
    \begin{overpic}[scale=0.36]{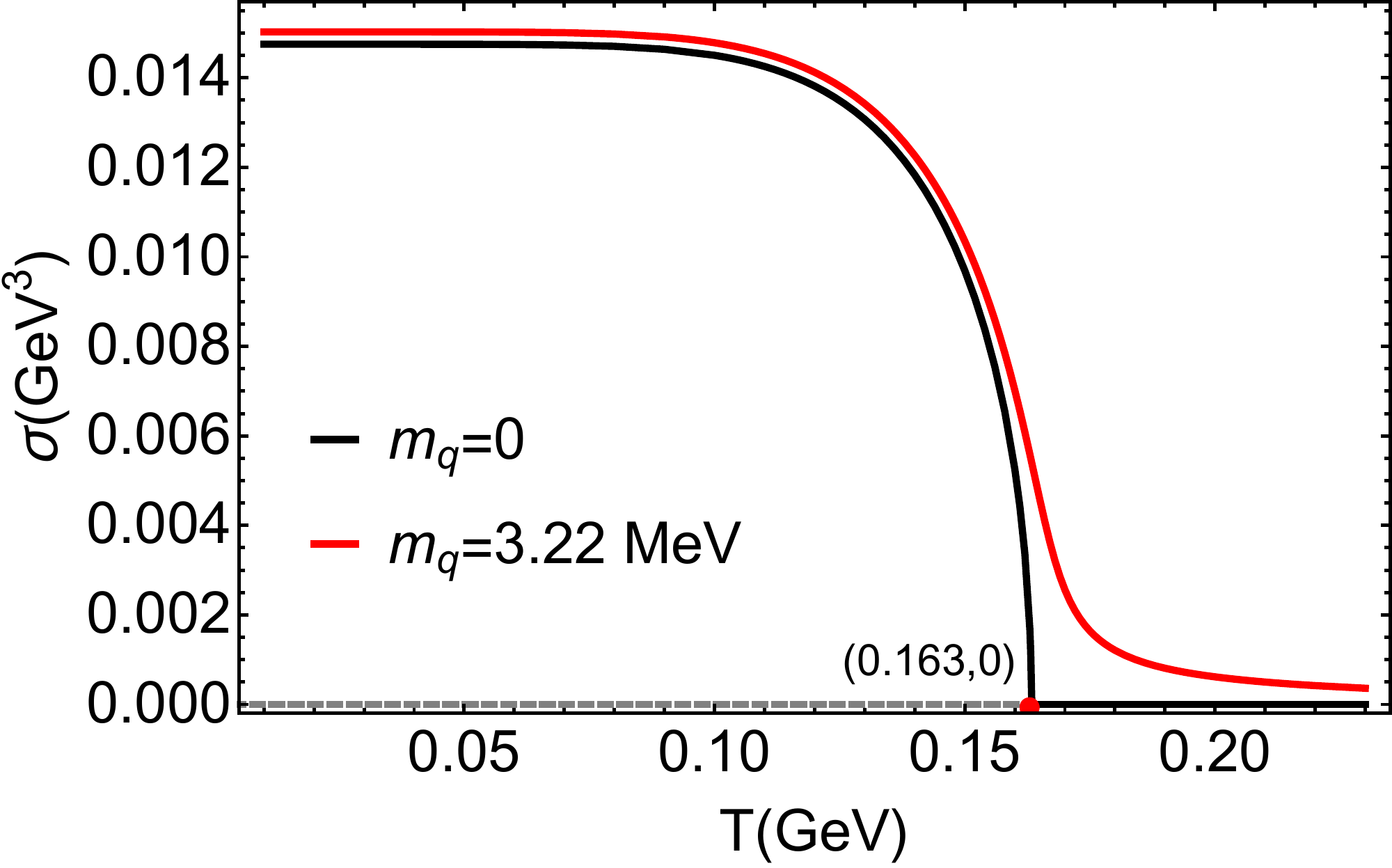}
        \put(90,56){\bf{(a)}}
    \end{overpic}
    \hfill
    \begin{overpic}[scale=0.36]{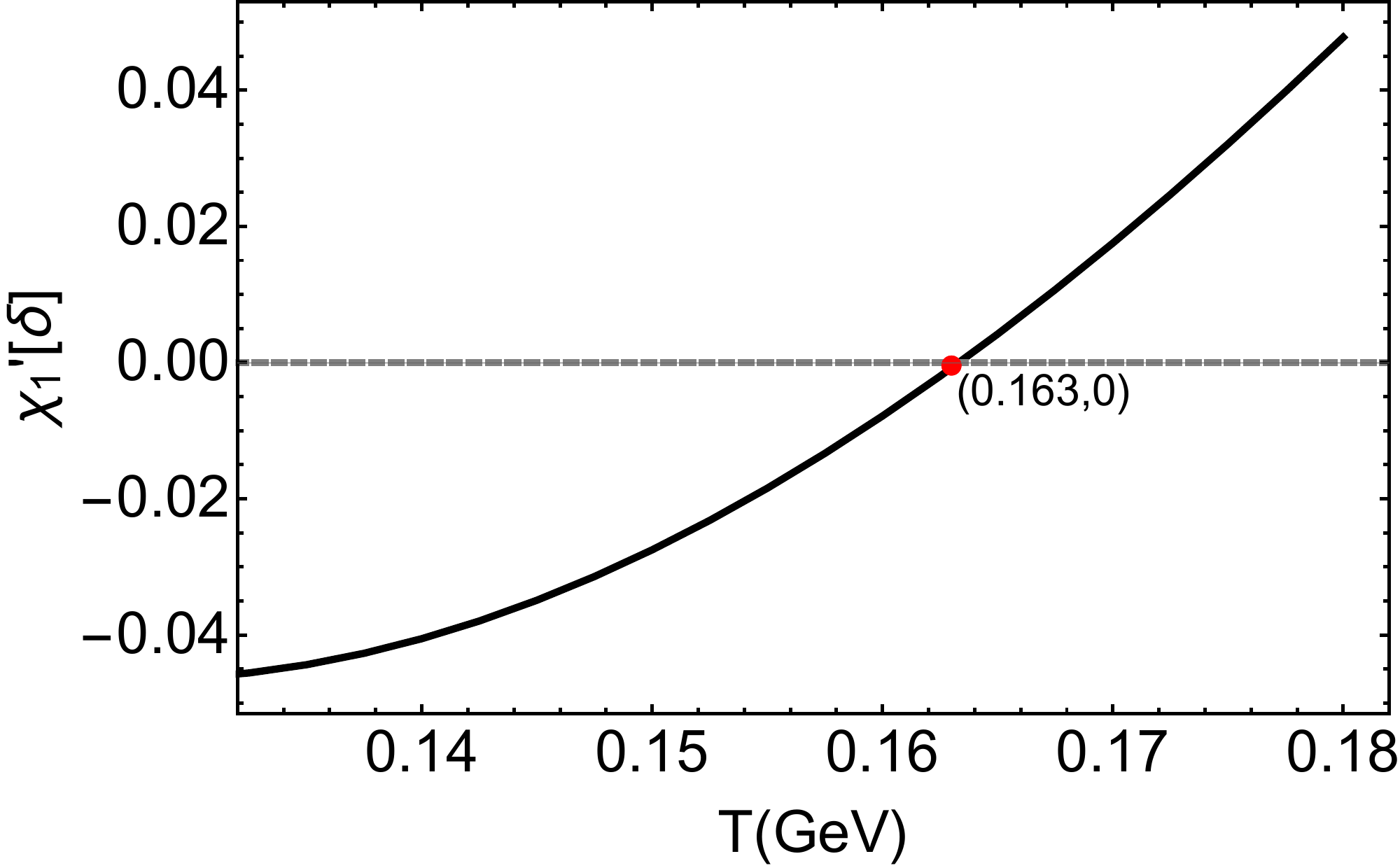}
        \put(20,56){\bf{(b)}}
    \end{overpic}
    \hfill
    \caption{\label{abc}(a) The temperature dependent behaviors of chiral condensation. The black line stands for chiral limit, $m_q=0$, where the critical temperature is labeled with a red solid dot, $T_c=0.163\rm{GeV}$. The red line stands for $m_q=3.22\rm{MeV}$ with pesudo-critical temperature $T_{cp}=0.164\rm{GeV}$\cite{Fang:2016nfj}. (b) The temperature dependent behavior of $\chi_1'(\delta)$ from Eq.(\ref{linear-chi}), where $\delta$ is a vanishing small constant and set to $10^{-8}\rm{GeV}^{-1}$ in our numerical calculations. The red dot, where $\chi_1'(\delta)=0$, locates at $(T=0.163\rm{GeV},0)$. }
\end{figure}
Therefore, we take $m_q=0$ in Eq.(\ref{aexpand-T}) and extract $\sigma$ at different temperatures. The result is shown with the black line in Fig.\ref{abc}(a). From the figure, we could see that $\sigma\approx (0.246 \rm{GeV})^3=0.0148\rm{GeV}^3\neq0$ at temperature near zero. Thus, spontaneous chiral symmetry breaking is realized at low temperature. We could also see that $\sigma$ decrease with $T$ and it vanishes at $T_c=0.163 \rm{GeV}$. Above $T_c$, $\sigma$ becomes zero and chiral symmetry is restored. Actually, from the study in Ref.\cite{Chen:2018msc}, the existence of $T_c$ is governed by the following linearized equation
\begin{eqnarray}\label{linear-chi}
        \chi_1 ''+\left(3 A'+\frac{f'}{f}-\Phi '\right)\chi_1 ' +\frac{e^{2 A}}{f} \left(3+\mu_c^2 z^2\right)\chi_1  =0,
\end{eqnarray}
which is the expansion at $T_c$ around the leading solution $\chi_0\equiv0$. The existence of $T_c$ requires the existence of solution to the above equation, with the boundary condition
\begin{eqnarray}\label{chi1-bdy}
       \chi_1^\prime(z=0)=0,\chi_1^\prime(z=z_h)=\text{finite}.
\end{eqnarray}
This conclusion is proved in Ref.\cite{Chen:2018msc}. Here, as a numerical check, we take $\chi_1(z_h)=1$ and solve Eq.(\ref{linear-chi}) numerically. Then we plot $\chi^\prime(z=0)$ as a function of $T$ in Fig.\ref{abc}. It is easy to see that $T_c$ in Fig.\ref{abc} is exactly locating at the temperature where $\chi_1^\prime(z=0)=0$ in Fig.\ref{abc}. So we get the conclusion that at $T_c$ Eq.(\ref{linear-chi}) has solution satisfying $\chi_1^\prime(z=0)=0$ and $\chi_1^\prime(z=z_h)=\text{finite}$.

After the theoretical check in chiral limit, it is also important to move to case with physical quark mass. According to Ref.\cite{Fang:2016nfj}, $m_q=3.22\rm{MeV}$ gives the best fitting of experimental data. Thus, we take this value of quark mass, and solve Eq.(\ref{EOMchi}) to obtain the temperature dependent $\sigma(T)$. The result is shown with red line in Fig.\ref{abc}(a). From this figure, we could see that the second order phase transition turns to be crossover type. Due to the finite quark mass, the exact chiral symmetry in 4D Lagrangian level is explicit broken, though slightly. Thus, the phase transition is weakened. Furthermore, one could extract the pseudo-critical temperature $T_{cp}=0.164\rm{GeV}$ where $|\sigma^\prime(T)|$ reaches its maximum.

\subsection{Spectral function for $S$ and $\pi$ at finite temperature }
\label{sec-spectral-T}
In the above section, the phase transition could be well described by the order parameter $\sigma$. As mentioned in previous section, it is also expected to be seen from the hadron spectrum. In chiral limit, as the goldstone of the symmetry breaking below $T_c$, one might expect that pion mass $m_\pi$ remains zero at any temperature below $T_c$. Also, since scalar meson, $f_0$ or $\sigma$ meson, is the chiral partner of pion, one might ask whether they would become degenerate at temperature above $T_c$. Therefore, we will try to study the temperature behavior of pion and scalar meson. The particles are excitation of the vacuum in 4D theory. In the dual description, they are perturbations around the background field in gravity side. Thus, to describe pion and scalar meson, we should consider the following perturbations
\begin{equation}\label{chi-pert}
    X=(\chi+S)t^0 e^{2 i \pi^a t^a},
\end{equation}
with $\pi^a$ and $S$ the pesudo-scalar and scalar perturbations respectively. In addition, one can check that the pseudo-scalar perturbation will couple with the longitudinal part($\varphi^i$) of axial vector
\begin{subequations}\label{pseudoscalaraction}
    \begin{eqnarray}
        &a_\mu^i = a_\mu^{T,i}+\partial_\mu \varphi^i,\\
        &\partial^\mu a_\mu^{T,i} = 0.
    \end{eqnarray}
\end{subequations}
Thus $\varphi^i$ should be considered as well.

Inserting the above perturbation into the action Eq.(\ref{action}) and keeping the quadratic terms, one gets the effective part for scalar perturbation as
\begin{equation}\label{scalar-T}
    S_S=\frac{1}{2}\int d x^5 \sqrt{g}e^{{-\Phi}}\bigg [ g^{\mu\nu}\partial_{\mu}S\partial_{\nu} S+g^{zz}(\partial_z S)^2-m_5^2 S^2-\frac{3\lambda}{2}\chi^2(S)^2\bigg ],
\end{equation}
and the pseudo-scalar part as
\begin{eqnarray}\label{pi-T}
    S_{PS}&=&-\frac{1}{4 {g_5}^2}\int d^5x \sqrt{g}e^{-\Phi}\sum _{i=1}^3 \bigg\{{g_{\mu \nu }} {g_{zz}} {\partial_z }{\partial_\mu \varphi^i}{\partial_z }{\partial_\nu \varphi^i} +{g_5}^2 {g_{\mu \nu }} \chi ^2 \partial_\mu\varphi^i \partial_\nu\varphi^i+\nonumber\\
    & & {g_5}^2 \chi ^2 \left({g_{\mu \nu}} \partial_\mu\pi^i \partial_\nu\pi^i+{g_{zz}} {(\partial_z\pi^i)}^2 -2 {g_{\mu \nu}} {g_5}^2\chi ^2
    \partial_\mu \varphi^i \partial_\nu \pi^i\right)\bigg\}.
\end{eqnarray}

Generally, $S$ and $\pi^i, \varphi^i$ are functions of all the coordinates. Here, since we would focus on the pole mass of the quasi-particle at finite temperature, we assume that all the perturbations only depends on $t$ and $z$. Thus, one can transform the perturbations to the frequency space by fourier transformation
\begin{equation}
    S(t,z)=\frac{1}{2\pi}\int d \omega\  e^{-i \omega t} {S}(\omega,z),
\end{equation}
and
\begin{eqnarray}
 \pi^i(t,z)=\frac{1}{2\pi}\int d\omega \ e^{-i\omega t} \pi^i(\omega, z),\\
\varphi^i(t,z)=\frac{1}{2\pi}\int d\omega \ e^{-i\omega t} \varphi^i(\omega, z).
\end{eqnarray}
Under these conditions, one can get the equation of motion for the scalar quasi-particle as
\begin{equation}\label{EOMscalar-T}
    S''+\left(3 A'+\frac{f'}{f}-\Phi'\right)S' +\left(\frac{\omega ^2}{f^2}-\frac{2 m_5^2+3 \lambda f   \chi ^2}{2 f}A'^2 \right)S=0,
\end{equation}
and the pseudo-scalar quasi-particle as\footnote{We note that the coupling to $\varphi$ is neglected in Refs.\cite{Cui:2013zha,Cui:2014oba}, which is different from our scenario.}
\begin{subequations}\label{EOMpi-T}
    \begin{eqnarray}
        \varphi^{''} + (A'- \Phi ')\varphi ^{'}-\frac{e^{2 A} g_5^2 \chi ^2}{f}\left(\varphi -\pi \right)&=&0,\\
        \pi ^{''}+\left(3 A'+\frac{f'}{f}-\Phi '+\frac{2 \chi '}{\chi }\right)\pi^{'}-\frac{\left(\varphi-\pi \right) \omega ^2}{f^2}& =& 0.\label{EOMpi-Tb}
    \end{eqnarray}
\end{subequations}
Here, we have neglected the isospin index $i$ in $\pi^i, \varphi^i$, due to the isospin symmetry at finite temperature.

At finite temperature, due to the interaction with the hot medium, the spectra of particles are broaden and a good description is the spectral function. The quasi-excitation appears as the peak of the spectral function.

The spectral function could be extracted from the Retarded-Green function $G^R(\omega)$,
\begin{equation}\label{rhoGrelation}
    \rho(\omega)=-\frac{1}{\pi}{\rm{Im}} G^R(\omega).
\end{equation}
The holographic correspondence states that the 4D operator $\mathcal{O}(x)$ and 5D field $\phi(x,z)$ are connected through the relation between 4D generating function with a external source $\phi_0(x)$ and  the classical action $S_{5D}$ in the AdS space, as
\begin{equation}
 \left\langle e^{i \int d^4 \phi_0(x)\mathcal{O}(x)}\right\rangle_{CFT}=\left .e^{iS_{5D}[\phi^{cl}]}\right |_{\phi^{cl}(x,z=0)=\phi_0},
\end{equation}
where $\phi^{cl}$ is the classical sotluions of $S_{5D}[\phi^{cl}]$ with its boundary value equaling the external source $\phi_0(x)$ \cite{Gubser:1998bc,Witten:1998qj,Kovtun:2004de}. Therefore, the Green's functions can be obtained by differentiating the 5D effective action with respect to the external sources. Here, for scalar mode, the on-shell action becomes
\begin{eqnarray}\label{onshellactionofscalar}
    S_S^{on}&=&\left .-\frac{1}{2}\int d \omega f(z) S(-\omega, z) e^{3 A(z)-\Phi (z)} S'(\omega, z)\right |_{z=\epsilon}^{z=z_h},
\end{eqnarray}
with $\epsilon$ a UV cutoff. For pseudo-scalar mode, the on-shell action has the following form
\begin{eqnarray}\label{onshellactionofpi}
    S_{\pi}^{on}= - \frac{1}{4 g_5^2}\int d\omega\  e^{A-\Phi }\left [e^{2 A} g_5^2 f  \chi ^2\pi(-\omega,z)  {\pi^{'}}(\omega, z)-\omega ^2\varphi(-\omega, z)  \varphi^{'}(\omega, z) \right ] \bigg |_{z=\epsilon}^{z=z_h}.\nonumber\\
\end{eqnarray}

Up to now, the main task is to solve the solution of $S,\pi, \varphi$ from Eqs.(\ref{EOMscalar-T},\ref{EOMpi-T}). For scalar mode, from Eqs.(\ref{EOMscalar-T}), one can get the the UV asymptotic solution of $S$ as
\begin{equation}\label{uv-s0}
    S(z\rightarrow 0)=s_1 z+s_3 z^3-\frac{1}{4} s_1 z^3\bigg[2 \text{$\mu $c}^2-4 \text{$\mu $g}^2-3 \zeta ^2 \lambda  m_q^2+2 \omega ^2\bigg ]\log (z)+\mathcal{O}(z^3).
\end{equation}
Here, $s_1$ and $s_3$ are the two integral constants of the second order ordinary derivative equation(ODE). From the holographic principle, we assume that $s_1$ corresponds to the external source while $s_3$ corresponds to the scalar operator $\bar{q}q$. Inserting this expansion into the on-shell action Eq.(\ref{onshellactionofscalar}), one gets the Retarted Green function and spectral function of scalar mode as
\begin{eqnarray}\label{scalarspectral}
    G^R_S(\omega)&=&\left .\frac{\delta^2 S_{S}^{on}}{\delta s_1\delta s_1}\right|_{z=\epsilon},\\
    \rho_S(\omega)&=&-\frac{1}{\pi}{\rm{Im}}G^R_S(\omega)=\frac{2}{\pi}\rm{Im}\left [\frac{s_3}{s_1}\right ].
    \left(\mu_c^2-2\mu_g^2-\omega ^2\right)
\end{eqnarray}
Here, since we focus on the imaginary part of the Green function, we have erased the real part inside the `$\text{Im}$' function. Right now, both $s_1$ and $s_3$ are free integral constants. The Retarded Green function property of Eq.(\ref{scalarspectral}) is related to the IR boundary at the horizon $z_h$. In fact, the $\omega^2\over{f^2}$ leads to the in-falling and out-going boundary conditions at the horizon. According to the study in Ref.\cite{Son:2002sd}, to get the Retarded Green function, one has to impose the in-falling boundary condition. Then, we have
\begin{equation}\label{bdyh-s-T}
S(z\rightarrow z_h)\sim (z_h-z)^{-i \omega/4\pi T}.
\end{equation}
Taking this boundary condition and solving the equation of motion Eq.(\ref{EOMscalar-T}), one can obtain $s_1$, $s_3$, and the spectral function.

Similarly, for pseudo-scalar mode, from Eqs.(\ref{EOMpi-T}), one can obtain the asymptotic expansion at UV boundary as
\begin{subequations}\label{UVmu0}
    \begin{eqnarray}
        \varphi(z\rightarrow0)&=& c_f-\frac{1}{2} \zeta ^2 g_5^2 m_q^2 \pi_0 z^2 \log (z)+\varphi _2 z^2+\mathcal{O}(z^3),\\
        \pi(z\rightarrow 0) &=& \pi_0+c_f-\frac{1}{2} \pi_0 \omega ^2 z^2 \log (z)+\pi _2 z^2+\mathcal{O}(z^3).
    \end{eqnarray}
\end{subequations}
where $c_f$, $\varphi_2$ and $\pi_0$, $\pi_2$ are the four integral constants of the two second ODEs. We note that Eqs.(\ref{EOMpi-T}) is invariant under the transform $\pi\rightarrow \pi+c_f, \varphi\rightarrow \varphi+c_f$. Thus, it is easy to understand $c_f$ corresponds to a redundant degree of freedom and we will set it to be zero. Then only the three integral constants $\pi_0, \varphi_2, \pi_2$ are relevant. Substituting the asymptotic solutions Eqs.~\eqref{UVmu0} into Eq.~\eqref{onshellactionofpi}, one gets
\begin{eqnarray}\label{onshellactionofpiz0}
    S_{\pi}^{on}=\int d\omega \left [\frac{1}{4} \pi _0^* \zeta ^2 m_q^2 \left(\pi _0 \omega ^2-4 \pi _2\right)+\frac{1}{2}  \zeta ^2 m_q^2 \omega ^2 |\pi _0|^2\log (z)\right ]\bigg|_{z=\epsilon}.
\end{eqnarray}
According to the holographic dictionary, the coupled 5D fields of $\varphi$ and $\pi$ can be decomposed in terms of bulk-to-boundary propagators~\cite{Abidin:2008hn,Abidin:2019xwu}, as
\begin{subequations}
    \begin{eqnarray}
        \varphi(\omega, z)&=&\frac{iJ(\omega)}{\omega} \widetilde{\varphi}(\omega, z),\\
        \pi(\omega, z)&=& \frac{i J(\omega)}{\omega}\widetilde{\pi}(\omega, z),
    \end{eqnarray}
\end{subequations}
where $J(\omega)$ is the external source of the longitidinal component of the axail current operator, $J^0_{A\mu}$, or the value of $\varphi$ field on the boundary. Comparing to ~\eqref{UVmu0}, one can identify the boundary value $\pi_0$ as the source $J(\omega)$. Then, the Retarded Green's function could be extracted from the second derivative of the action with respect to the source. From~\eqref{onshellactionofpiz0}, it reads
\begin{eqnarray}\label{green-pi-T}
    G^R_{\pi}(\omega)&=&-\frac{\delta^2 S_{\pi}^{on}}{\delta \pi_0^*\delta \pi_0}\nonumber\\
    &=&\left.-\frac{1}{4\pi _0}  \zeta ^2 m_q^2 \left(\pi _0 \omega ^2-4 \pi _2\right)-\frac{1}{2} \  \zeta ^2 m_q^2 \omega ^2 \log (z)\right |_{z=\epsilon}.
\end{eqnarray}
Taking the relation in Eq.~\eqref{rhoGrelation}, one can get the spectral function of pion as,
\begin{equation}\label{spectral-pi-T}
    \rho_{\pi}(\omega)=-\frac{1}{\pi}{\rm{Im}} G^R_{\pi}(\omega)=\frac{m_q^2\zeta ^2}{\pi}\rm{Im}\left [\frac{\pi_2}{\pi_0}\right ].
\end{equation}
Like in the scalar mode, to get the Retarded Green function, one has to impose the in-falling boundary condition at $z_h$. On this side, the asymptotic expansion reads
\begin{subequations}\label{bdyh-pi-T}
    \begin{eqnarray}
    \varphi(z\rightarrow z_h) & =& \left(z_h-z\right)^{\frac{i \omega  z_h}{2   \mu ^2 z_h^2-4}}\bigg\{ \frac{2 i c_0^2 g_5^2 \pi _{\text{h0}}  \left(  \mu ^2 z_h^2-2\right)\left(z-z_h\right)}{\omega  z_h^2 \left(2  \mu ^2 z_h^2+i \omega  z_h-4\right)}+\mathcal{O}[(z-z_h)^2]\bigg\}+c_{h0},\nonumber\\
    \\
    \pi(z\rightarrow z_h) &=& \left(z_h-z\right)^{\frac{i \omega  z_h}{2   \mu ^2 z_h^2-4}}\bigg\{\pi_{h0}+\mathcal{O}(z-z_h)\bigg\}+c_{h0}.
    \end{eqnarray}
\end{subequations}
Here, $\pi_{h0},c_{h0}$ are the two integral constants describing the in-falling mode, while another two describing the out-going mode are omitted. Parameter $c_0$ is the integral constant as shown in Eq.~\eqref{aexpand-T}. Since the ODEs are linear, one could take $\pi_{h0}=1$ without shifting the spectra. The other integral constant $c_{h0}$, as well as the integral constants at UV, $\pi_0, \varphi_2, \pi_2$ ,  could be determined by requiring $c_f=0$ applying the `shooting' algorithm. Then one obtains the spectral function for pseudo-scalar mode. With all these preparation, we will extract and discuss the spectral functions and the spectra of the scalar and pseudo-scalar quasiparticles in the next three subsections.

\subsection{Chiral limit: Goldstone bosons and symmetry restoration of chiral partners }

\begin{figure}[h]
    \centering
    \begin{overpic}[scale=0.41]{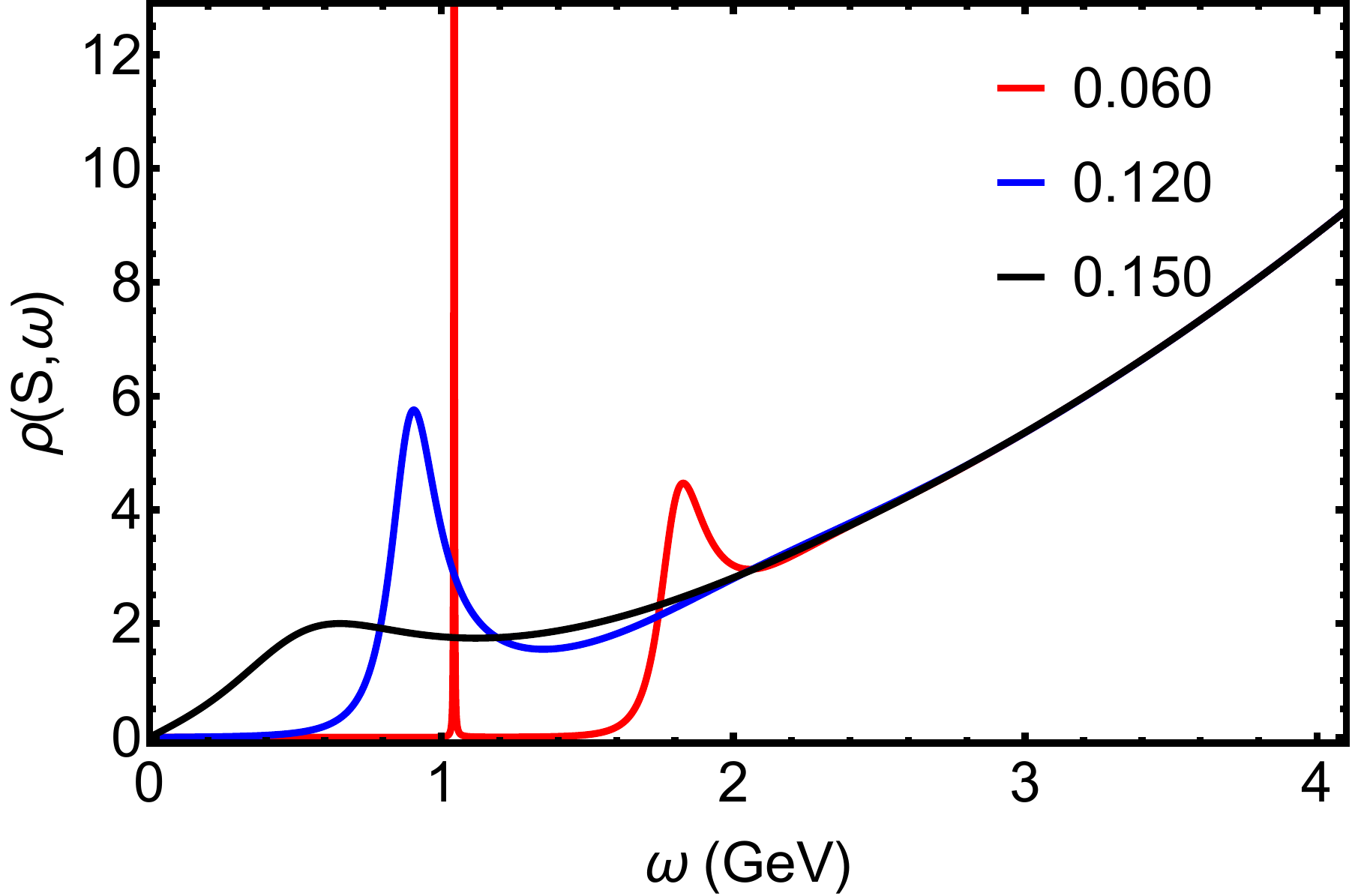}
        \put(90,60){\bf{(a)}}
    \end{overpic}
    \hfill
    \begin{overpic}[scale=0.403]{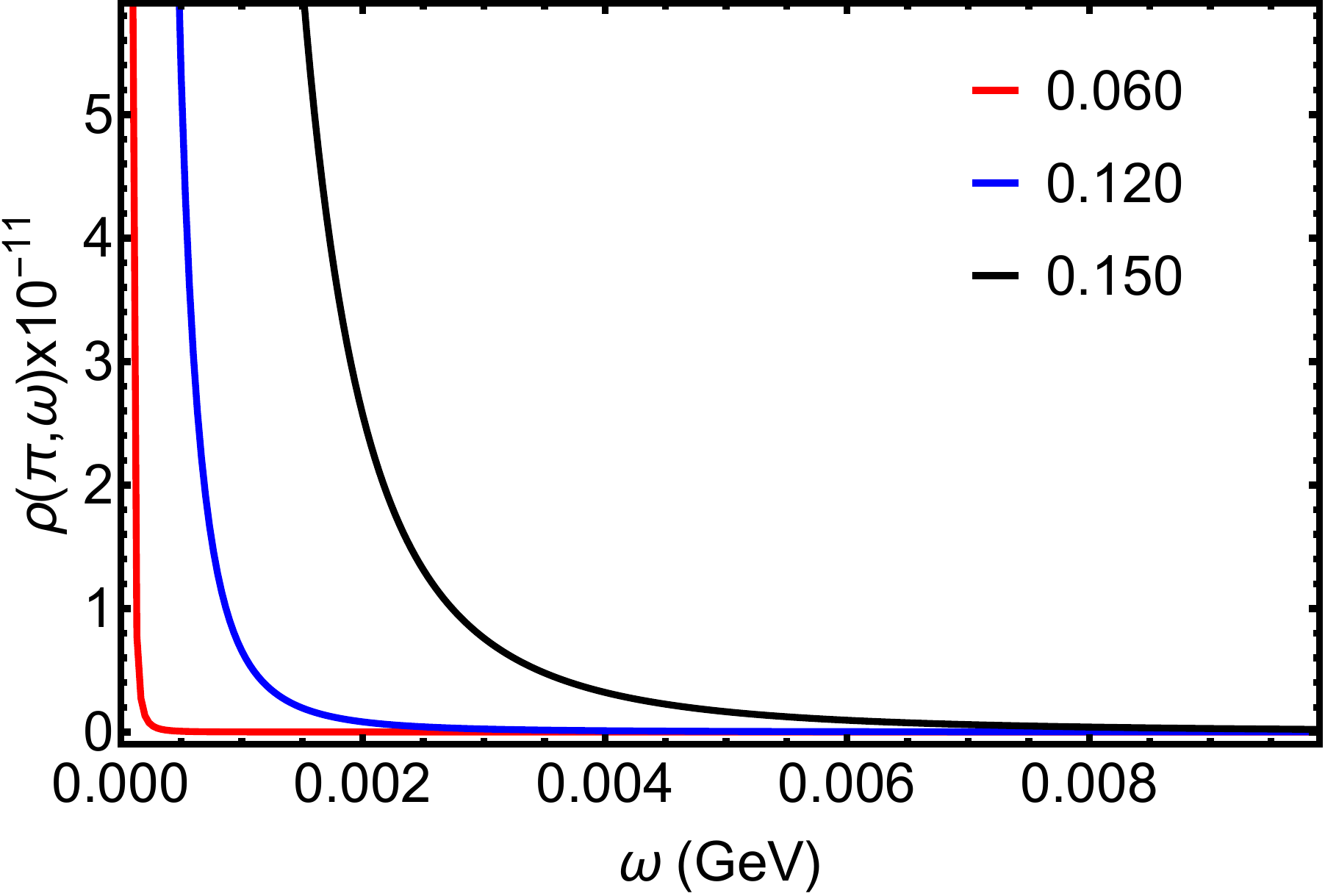}
        \put(90,60){\bf{(b)}}
    \end{overpic}
    \hfill
    \begin{overpic}[scale=0.4]{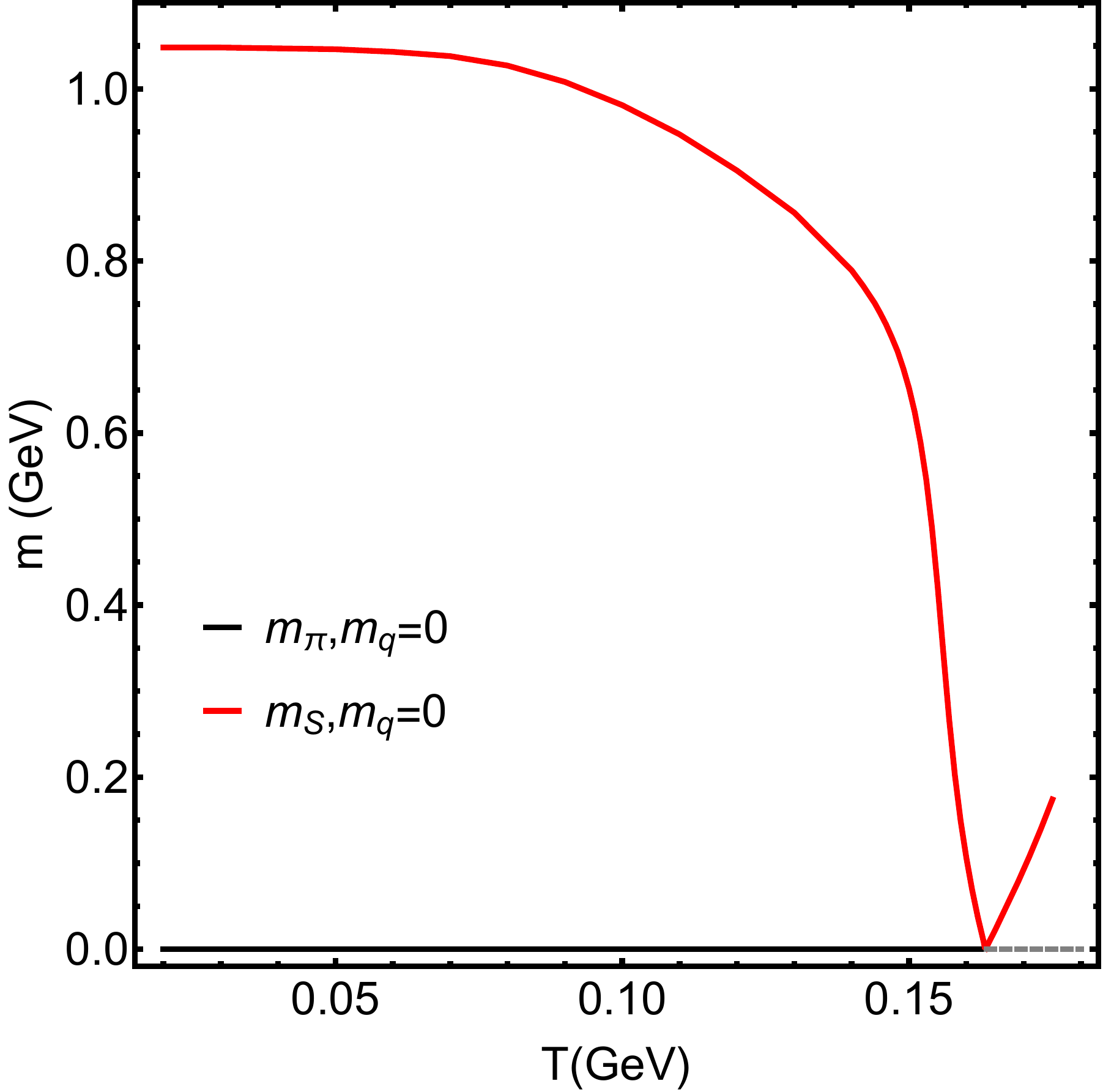}
        \put(90,90){\bf{(c)}}
    \end{overpic}
    \hfill
    \caption{\label{spectral-chiral-limit} Spectral functions in frequency space of (a) scalar meson, $\rho(S,\omega)$, and (b) pion, $\rho(\pi, \omega)$, in chiral limit at $T=0.06$, $0.12$ and $0.15$GeV. (c) The temperature dependence of lowest $m_S$ and $m_{\pi}$ in chiral limit.}
\end{figure}

As mentioned above, theoretically, it is important to check whether the pseudo-scalar mode at finite temperature is the Goldstone mode or not. If it is not, there would be theoretical inconsistency in soft-wall model. To check that, the quark mass $m_q$ is set to zero, i.e. in chiral limit, to guarantee the exact chiral symmetry in 4D Lagrangian. Then we solve the solution of $\chi$ applying the process describing in Sec.\ref{sec-chiralPT-T}. Substituting $\chi$ into the process describing in Sec.\ref{sec-spectral-T}, one could obtain the spectral functions for both scalar mode and pseudo-scalar mode. We take $T=0.060, 0.120$ and $0.150 \rm{GeV}$ as an example and plot the spectral functions in Fig.\ref{spectral-chiral-limit}(a) and (b). We could see that there are peaks in the spectral functions. Those peaks are related to the thermal excitations. Thus, we would extract the masses from the location of these peaks. From Fig.\ref{spectral-chiral-limit}(b), we could see that for pseudo-scalar mode, below $T_c$, there is always a divergence locating at $\omega=0$, which reveals $m_\pi=0$. The temperature dependent behavior of $m_\pi$ and $m_S$ are shown in Fig.\ref{spectral-chiral-limit}(c). From the figure, we can see that $m_S$ is about $1.05\rm{GeV}$ near $T=0$. It decreases at low temperature, and reaches zero at $T=0.163 \rm{GeV}$, which is exactly the chiral transition temperature $T_c$ in chiral limit.  Moreover, it is very interesting to observe that,  above $T_c$, $m_\pi$ and $m_S$ should emerge to the same line and increase together very fast. Numerically, in soft-wall model, the Goldstone nature of pion are checked and chiral symmetry restoration has been realized in the hadronic level.

Actually, this conclusion can also be proved analytically. Firstly, it is quite easy to understand why $m_S=0$ at $T_c$ in chiral limit. We have proved in Sec.\ref{sec-chiralPT-T} that, at and above $T_c$, $\chi\equiv0$ in chiral limit. Substituting this equation into Eq.\eqref{EOMscalar-T}, and setting $\omega=0$, it is easy to see that Eq.(\ref{EOMscalar-T}) becomes exactly the same as the $T_c$ criteria Eq.(\ref{linear-chi}). In fact, the solution of Eq.(\ref{linear-chi}) under the boundary condition Eq.(\ref{chi1-bdy}) is just the wave function of massless scalar mode at $T_c$. Physically, in chiral limit, when $m_S=0$, instability appears. Actually, this is the physical meaning of the $T_c$ criteria Eq.(\ref{linear-chi}).

Then, since $\chi\equiv0$ above $T_c$, from Eq.(\ref{chi-pert}), the coupling between $\pi$ and $\varphi$ disappears. Since the modular of $X$ vanishes, the expansion of $X$ could no longer be Eq.(\ref{chi-pert}). Instead, the scalar expansion and the pseudo-scalar expansion should have the same form. As a result, the masses of the two modes are naturally the same. In this sense, the red piece above $T_c$ in Fig.\ref{spectral-chiral-limit}(c) is for $m_\pi$ as well, though it is obtained from the scalar sector actually.

Finally, we will try to get an analytical understanding on masses of pions below $T_c$. As we have known, below $T_c$, the pseudo-scalar mode satisfies Eqs.(\ref{EOMpi-T}). Our main goal is to check the existence of the massless mode with $\omega=0$. Under this condition, we see that Eq.~\eqref{EOMpi-Tb} becomes
\begin{eqnarray}\label{EOMpi-T-1}
    \pi^{''}+\left(3 A'+\frac{f'}{f}-\Phi '+\frac{2 \chi '}{\chi }\right)\pi^{'}=0,
\end{eqnarray}
which could be directly solved as
\begin{eqnarray}
\pi(z)=p_1+p_2 \int_0^{z} \frac{e^{-3A+\Phi}}{f\chi^2} dz^\prime,
\end{eqnarray}
with $p_1, p_2$ the two integral constants of the second derivative ODE. However, the $p_2$ branch is not physically acceptable at both UV and IR boundary.

At UV, in chiral limit, the leading expansions are $\chi\sim z^3, f\sim 1, e^{-3A}=z^3, e^{\Phi}\sim1$. Thus the integral kernel is divergent as $z^{-3}$ at $z=0$. Only with physical quark mass, this part could contribute to the wave function.

At the horizon $z_h$, the leading expansions are $\chi\sim \text{cosntant}, f\sim (z-z_h), e^{-3A}=\text{cosntant}, e^{\Phi}\sim \text{cosntant}$, and it is also divergent.  Therefore, the physical acceptable solution for $\pi$ is $\pi\equiv\text{cosntant}$. Since in calculating spectral functions we will normalize $\pi_0=1$, here the constant should be chosen as $1$.  Thus, the existence of the massless mode is equivalent to the existence of solution to the following equation, subject to $\varphi(0)=0$,
\begin{eqnarray}
        \varphi^{''} + (A'- \Phi ')\varphi ^{'}-\frac{e^{2 A} g_5^2 \chi ^2}{f}(\varphi-1)=0,
\end{eqnarray}
which comes from Eq.~\eqref{EOMpi-Tb} by replacing $\pi$ with $1$. Redefining $\tilde{\varphi}=\varphi+1$, one reaches a linear ODE
\begin{eqnarray}
        \tilde{\varphi}^{''} + (A'- \Phi ')\tilde{\varphi}^{'}-\frac{e^{2 A} g_5^2 \chi ^2}{f}\tilde{\varphi}=0.
\end{eqnarray}
The asymptotic expansion of the above second order ODE can be extracted easily as
\begin{eqnarray}
        \tilde{\varphi}=p_0+p_2z^2+\mathcal{O}(z^4)
\end{eqnarray}
at UV and
\begin{eqnarray}
        \tilde{\varphi}=p_{h0}(1+\frac{g_5^2\chi(z_h)^2}{4z_h}(z-z_h)\ln(z_h-z))+p_{h1}(z-z_h)+\mathcal{O}((z-z_h)^2)
\end{eqnarray}
at IR, with $p_0,p_2, p_{h0},p_{h1}$ the corresponding integral constants at both sides. The $p_{h0}$ branch should be dropped out, since it leads to a divergent $\varphi^\prime$ at $z_h$. Thus, only $p_{h1}$ can be nonzero. Considering the linearity of the equation, one can take $p_{h1}=1$ and solve $\tilde{\varphi}$ from the equation. After one get the solution, one can normalize $p_0$ to $1$, again from the linearity of the equation. Then, actually one has obtained the massless wave function $\varphi=\tilde{\varphi}-1,\pi\equiv1$ subject to the boundary condition $\varphi(0)=0$. Therefore, we have prove that the massless mode always exists below $T_c$, and it is the Goldstone mode or the massless pion at finite temperature. It should be pointed out that the above proof could not be extend to finite quark mass. The reason is that with finite quark mass, there would be terms $\frac{p_0 g_5^2m^2\zeta^2}{2}z^2\ln(z)$ between $p_0$ and $p_2z^2$, which leads to a divergent on-shell action. So it is not dual to the physical mode. As a result, with finite quark mass, the massless mode does not exist for most of the cases in soft-wall model\footnote{Of course, if one choose the background field properly, with finite quark masses, massless mode might appear at certain value of  temperature. But it is still hard to guarantee the existence of Goldstone mode at any temperature below $T_c$. We have checked that in the current model, there will not be such weird result.}. Also, it can not be extended to temperature above $T_c$. When $\chi\equiv0$ above $T_c$, the equations of motion for pseudo-scalar and scalar modes should be the same, and the two kinds of excitations would become degenerate.

In a short summary,  by numerical calculation and analytical analysis, we prove that in chiral limit, pion is always massless and it is the Goldstone mode of the spontaneous symmetry breaking below $T_c$. The analytical proof is quite general for most of the models in soft-wall AdS/QCD framework. Moreover, the expected degeneration in spectral of chiral partners is observed.

\subsection{Physical quark mass: the pole mass of quasipions}

In the above section, we have checked the theoretical consistence of soft-wall model in chiral limit. To be more realistic and to get information for understanding the current experimental data, we would consider the situation with physical quark mass. According to the study at zero temperature\cite{Fang:2016nfj}, $m_q=3.22\rm{MeV}$ gives the best fitting of experimental data for meson spectral, and we will take this value as the physical quark mass in the current model. As mentioned above, with finite quark mass, below $T_c$, pion might gain certain mass due to the explicit chiral symmetry breaking.  We will present the numerical results first as in chiral limit.

\begin{figure}[htbp]
    \centering 
    \begin{overpic}[width=.49\textwidth]{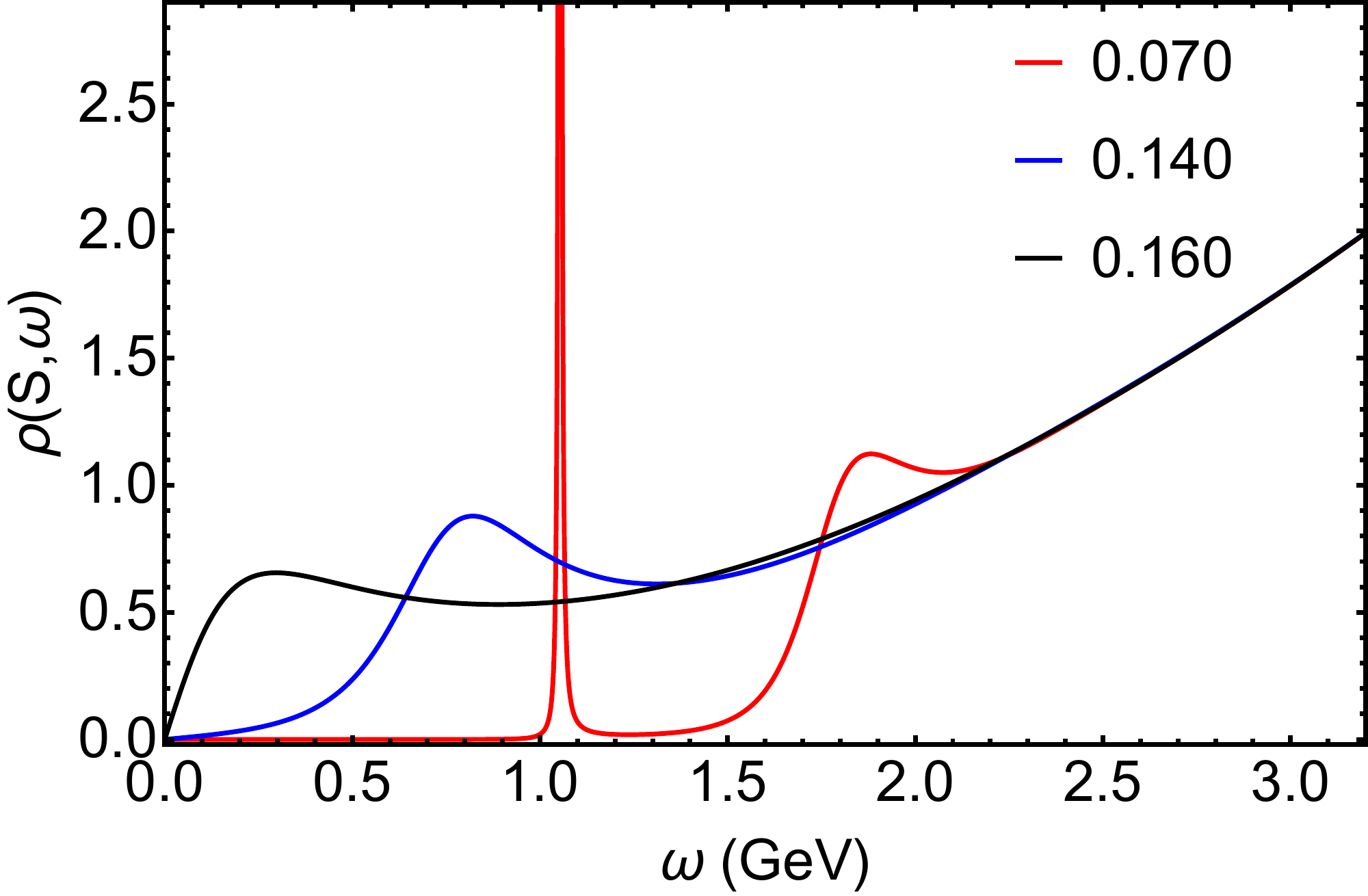}
        \put(90,60){\bf{(a)}}
    \end{overpic}
    \hfill
    \begin{overpic}[width=.475\textwidth]{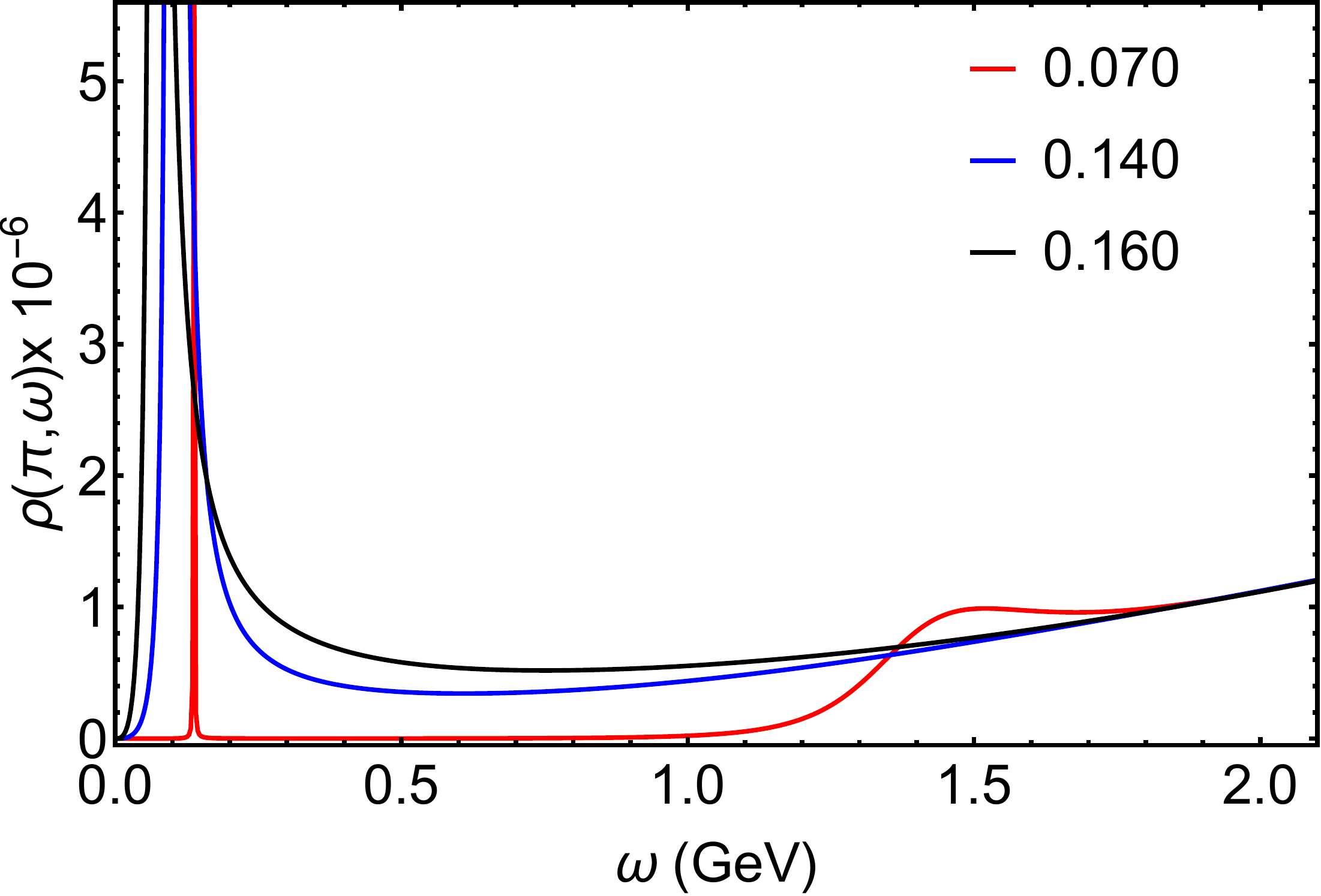}
        \put(90,62){\bf{(b)}}
    \end{overpic}
    \hfill
    \begin{overpic}[width=.49\textwidth]{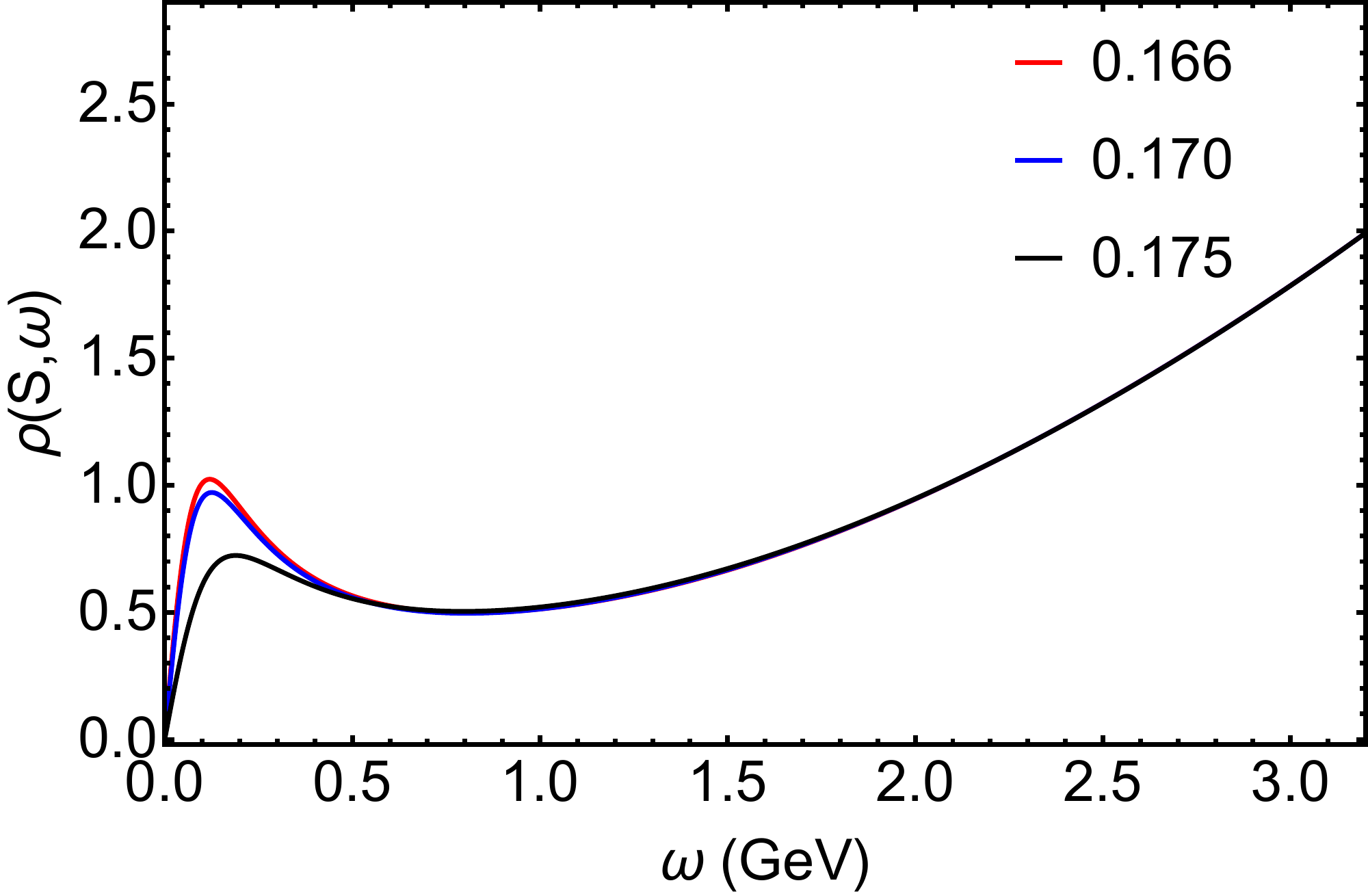}
        \put(90,60){\bf{(c)}}
    \end{overpic}
    \hfill
    \begin{overpic}[width=.475\textwidth]{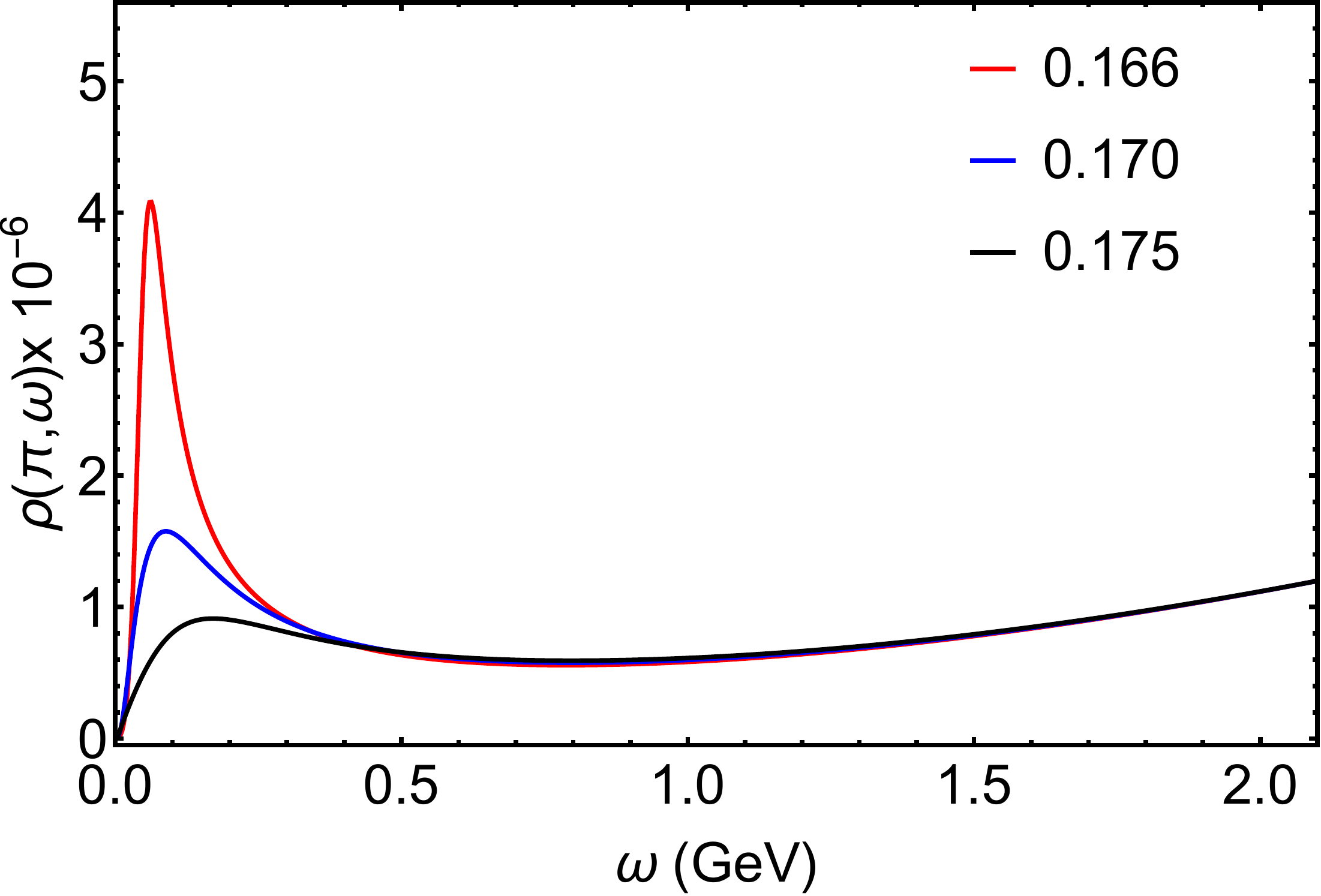}
        \put(90,62){\bf{(d)}}
    \end{overpic}
    \caption{\label{pispectralfigmu0varyT} Spectral functions of (a) scalar mode $\rho(S,\omega)$ at $T=0.070$, $0.140$ and $0.160\rm{GeV}$, (b) pion $\rho(\pi,\omega)$  at $T=0.070$, $0.140$ and $0.160\rm{GeV}$, (c) scalar mode $\rho(S,\omega)$ at $T=0.166$, $0.170$ and $0.175\rm{GeV}$, (d) pion $\rho(\pi,\omega)$ at $T=0.166$, $0.170$ and $0.175\rm{GeV}$. }
\end{figure}

Taking $m_q=3.22\rm{MeV}$ and the values of parameters in Eq.~(\ref{parameters}), one can solve $\chi$ from Eq.~\eqref{EOMchi}. Then, imposing the IR boundary condition Eq.~\eqref{bdyh-s-T} and Eq.~\eqref{bdyh-pi-T}, and solving Eqs.~\eqref{EOMscalar-T} and~\eqref{EOMpi-T}, after normalizing $s_1=1,\pi_0=1$, one can obtain $s_3, \varphi_2, \pi_2$ , which are defined in the UV expansion Eqs.~\eqref{uv-s0} and \eqref{UVmu0}. Inserting those results in the expression of spectral function in Eqs.~\eqref{scalarspectral}, \eqref{green-pi-T} and \eqref{spectral-pi-T}, one could obtain the finite temperature spectral functions for scalar meson and pion.

From the numerical calculation, we find that the behavior of spectral function at temperature below the pseudo-transition temperature $T_{cp}=0.164\rm{GeV}$(in Fig.\ref{abc}(a)) is different from that at temperature above $T_{cp}$. To show this difference, we take $T=0.070,0.140,0.160\rm{GeV}$, and $0.166, 0.170, 0.175\rm{GeV}$, and extract the spectral functions for scalar meson and pion in Fig.~\ref{pispectralfigmu0varyT}. The left two panels are results of scalar spectral functions, while the right two for pseudo-scalar spectral functions. The upper two panels give the low temperature behavior. At very low $T$, e.g. $T=0.07\rm{GeV}$(the red lines in the upper two panels), for scalar and pseudo-scalar spectral functions, there are sharp peaks at around $\omega=1.05\rm{GeV}$ and $\omega=0.137\rm{Gev}$ respectively, very close to the vacuum values of scalar and pseudo-scalar mesons. To the right of the first peaks, at around $\omega=1.88\rm{GeV}$ for scalar function and $\omega=1.52\rm{GeV}$ for pesudo-scalar function, wide peaks, corresponding to the radial excitations at zero temperature, appear in the two spectral functions. Then when temperature increases to $T=0.14\rm{GeV}$, we could see that the centres and the heights of the left peaks decrease, while widths of the peaks are broadened(though still quite sharp). It shows the decrease of masses of scalar meson and pion at low temperature.  Moreover, the right peaks at $T=0.14\rm{GeV}$ could not be identified at this temperature, due to the fast decreasing of the heights and the rapid increasing of the widths. This might correspond to the melting of the higher excitations at finite temperature. Actually, this is quite reasonable.  It is easier to destroy the weaker binding of the higher excitations, so they melt before the ground states. When temperature increases to $T=0.16\rm{GeV}$, still slightly lower than $T_{cp}$, we find that the masses continue to decrease, while the peaks becomes very wide. Thus, we could see the decreasing of meson mass at temperature below $T_{cp}$, which is similar with the decreasing of masses in chiral limit. In some sense, the chiral limit behavior governs the small quark masses behavior also. The results for chiral condensate and for meson mass with small quark masses just differ slightly from that in chiral limit.

Then, we increase the temperature further. From the lower two panels in Fig.\ref{pispectralfigmu0varyT}, we could see that at temperature above $T_{cp}$(but not far away from $T_{cp}$), the left peaks are still alive. However, the temperature behaviors are totally different. From $T=0.166\rm{GeV}$ to $T=0.175\rm{GeV}$, we could see that both the masses of scalar and pseudo-scalar mesons increases, together with the broadening of the widths. Such a behavior is consistent with the 4D effective model studies\cite{Fischer:2018sdj,Gao:2020hwo,Tripolt:2013jra,Xia:2013caa,Xia:2014bla}. Physically, this is probably related to the transition from bound states to resonances, though the exact details from holographic framework is still unclear and need further study in the future.

\begin{figure}[htbp]
    \centering 
    \includegraphics[width=.48\textwidth]{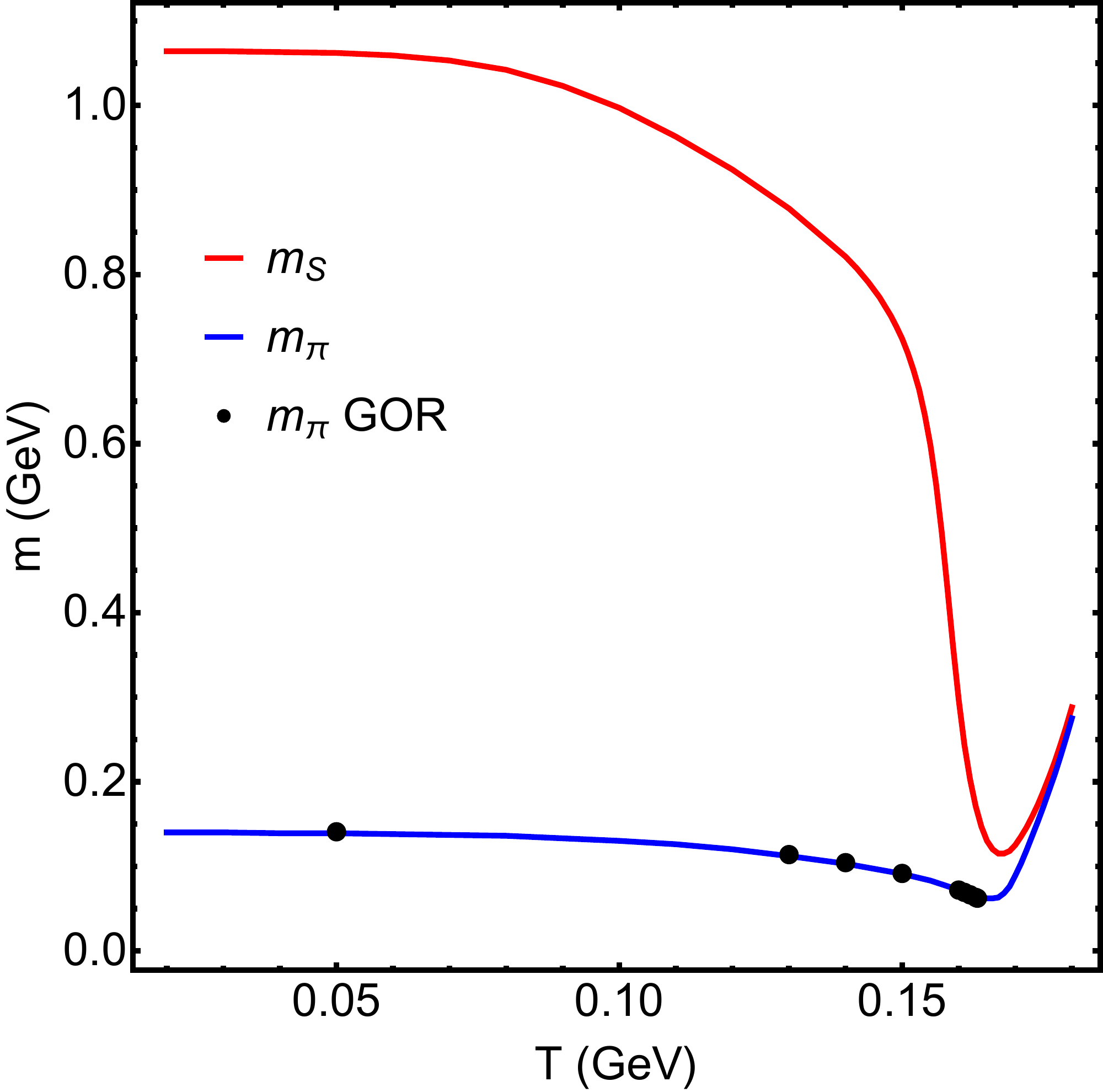}
    \hfill
    \caption{\label{massfigmu0varyT} Masses of scalar meson and pion  at finite temperature. The red and blue lines are extracted from the peaks of spectral functions, representing  results of scalar meson and pion respectively. The black dots are extracted through Eq.(\ref{GOR-T}).}
\end{figure}

To be clearer, we extract the temperature dependence of masses of scalar meson($m_S$) and pion ($m_{\pi}$) in Fig.\ref{massfigmu0varyT}. In the figure, the red line and the blue line present $m_S(T)$ and $m_\pi(T)$ respectively. From the figure, $m_S$ decrease rapidly below $T_{cp}$, from its vacuum value $m_S=1.06\rm{GeV}$ to $m_S=0.115\rm{GeV}$ around $T_{cp}$, almost $90\%$ reduction.  As for pion, $m_\pi$ decreases below $T_{cp}$ also, from its vacuum value about $m_\pi=0.140\rm{GeV}$ to $m_\pi=0.062\rm{GeV}$ around $T_{cp}$, almost $60\%$ reduction. It is also quite interesting to see that qualitatively the results for pion is consistent with Son and Stephanov's prediction\cite{Son:2001ff,Son:2002ci}, and the lattice simulations in Refs.\cite{Brandt:2014qqa,Brandt:2015sxa}. Quantitatively, the holographic model gives almost double the reduction, with a rate of $60\%$, while the reduction rates from Refs.\cite{Son:2001ff,Son:2002ci} and Refs.\cite{Brandt:2014qqa,Brandt:2015sxa} are around $30\%$ and $20\%$ respectively. Since the reduction of pion mass would enhance the low momentum distribution of pion, the holographic prediction would contribute more fraction in the relevant physics. Of course, the exact value of the enhancement depend on the evolution model of the fireball, and it is out of the scope of this work. Also, the holographic prediction is also consistent with the NJL model prediction with gluon condensation~\cite{Ebert:1992jx}, and it is contrast to that in NJL model without gluon condensate. In some sense, the gluon dynamics has been considered correctly, though in an implicit way. Finally, one could easily find that above $T_{cp}$, $m_\pi$ increases as well as scalar meson, which is consistent with the 4D studies\cite{Fischer:2018sdj,Gao:2020hwo,Tripolt:2013jra,Xia:2013caa,Xia:2014bla}. Moreover, the degenerate of scalar meson and pion is observed in the mass spectrum above $T_{cp}$, which reveals the restoration of the breaking symmetry in hadronic spectrum level.

\subsection{GOR relation at finite temperature}
\label{sec-GOR-T}
In the above section, we have extracted the masses from the spectral functions. Here, we will follow Ref.\cite{Erlich:2005qh} and try to derive a different way to calculate the mass below $T_c$. Going back to  Eq.(\ref{EOMpi-T}), one can prove that it is equivalent to
\begin{subequations}\label{GOR-start}
    \begin{eqnarray}
        \varphi^{''}&& + (A'- \Phi ')\varphi ^{'}-\frac{e^{2 A} g_5^2 \chi ^2}{f}\left(\varphi -\pi \right)=0,\label{GOR-starta}\\
        \pi ^{'}&&-\frac{\omega^2 e^{-2A} \varphi^\prime}{g_5^2 f\chi^2} = 0.\label{GOR-startb}
    \end{eqnarray}
\end{subequations}
The main observation is that with $f(z_h)=0$ in Eq.~\eqref{GOR-startb}, the boundary condition at IR could not be the normal real regular condition.  The wave like solution, i.e. the in-falling and out-going boundary conditions, appear and one has to calculate the spectral functions.  Thus, we will make an naive assumption that the peak location for pion below $T_c$ is not related to $f(z)$ in Eq.~\eqref{GOR-startb}, which only generates the widths at temperature below $T_c$. Under such kind of assumption, one replace $f(z)$ in Eq.\eqref{GOR-startb} with $f(z)\equiv 1$ as zero temperature, and reaches
\begin{subequations}\label{GOR-startm}
    \begin{eqnarray}
        \varphi^{''}&& + (A'- \Phi ')\varphi ^{'}-\frac{e^{2 A} g_5^2 \chi ^2}{f}\left(\varphi -\pi \right)=0,\\
        \pi ^{'}&&-\frac{m_\pi^2 e^{-2A} \varphi^\prime}{g_5^2\chi^2} = 0,
    \end{eqnarray}
\end{subequations}
where we have replaced $\omega^2$ with $m_\pi^2$ explicitly. Though actually only $\omega^2$ corresponding to the normalizable mode can be considered as pion mass.

Then the process is similar to the derivation of GOR relation at zero temperature case in Ref.\cite{Erlich:2005qh}. Considering in chiral limit $m_q\rightarrow0$, $m_\pi\rightarrow0$, one can construct the solution with very small quark mass from the Goldstone mode in chiral limit. Given the leading solution $\tilde{\varphi}, \pi\equiv0$, one could try to construct the next order for $\pi$ as
\begin{eqnarray}
\delta\pi(z)=\int_0^z du \frac{m_\pi^2u^3}{\chi^2} \frac{\tilde{\varphi}^\prime}{g_5^2 u}.
\end{eqnarray}
It is easy to check, if $m_q$ equals zero exactly, the above integration is divergent at UV. However, if there is any finite $m_q$, the divergent would be move. Thus, one can parameterize the divergence in $m_q$. Considering a small $m_q$ in $\chi$, the divergence in chiral limit tells us that the contribution of the integration is mainly from the UV. Actually, with a very small $m_q$, the contribution is from small $z$. Thus, one can get
\begin{eqnarray}
\delta\pi(z)\sim-m_\pi^2f_{\pi,T}^2\int_0^z du \frac{u^3}{\chi^2},
\end{eqnarray}
where the temperature dependent pion decay constant $f_{\pi,T}$ is defined as $f_{\pi,T}^2=-\frac{\tilde{\varphi}^\prime}{g_5^2 z}|_{z\rightarrow0}$.
In small $m_q$ limit, the integration could be obtained as ${1}/{(2m_q\sigma})$.  Thus, we have
\begin{eqnarray}
\delta\pi(z)\sim-\frac{m_\pi^2f_{\pi,T}^2}{2m_q\sigma}.
\end{eqnarray}
Considering in exact chiral limit, this solution should be the massless Goldstone mode with $\pi\equiv1$, we have the GOR relation at finite temperature,
\begin{eqnarray}\label{GOR-T}
m_\pi^2f_{\pi,T}^2=2m_q\sigma.
\end{eqnarray}
Then for a small $m_q$, one can extract $f_{\pi,T}, \sigma$ from the solution $\tilde{\varphi}$ and $\chi$. Then use this relation to get $m_\pi$ at finite temperature.

Since the above derivation depends on the assumption at the beginning of this section, we make a numerical check of this relation. The black dots in Fig.\ref{massfigmu0varyT} are obtained by the finite $T$ GOR relations. From  Fig.\ref{massfigmu0varyT}, we could see that it agrees very well with the blue line from spectral functions. This might be considered as a numerical check of our assumption. It might provide a simpler way to extract the pole mass at  temperature below $T$. Of course, when $T_c$ is above $T_c$, the Goldstone mode in chiral limit disappears and the GOR relation derived here could not be used. From the discussion in this section, we could see that the coupling with $\varphi$ is quite important to realize the Goldstone nature of pion. The scenario neglecting such kind of coupling in Refs.\cite{Cui:2013zha,Cui:2014oba} might  not be a good approximation at low temperature, especially in chiral limit.

\section{Pion quasiparticles at finite isospin density}
\label{sec-pion-superfluidity}
In the above section, we have presented a careful analysis on scalar type quasiparticles at finite temperature only. However, in heavy ion collisions, the nuclear matter density might be alao very important at certain colliding energy. The recent experimental project, BES\cite{Aggarwal:2010cw,Odyniec:2013aaa,Luo:2017faz}, is designed mainly to probe the baryon number density effect, especially to seek the CEP in $T-\mu_B$ plane. Besides baryon number, the isospin number is another conserved charge in QCD and its density might change the property of medium also. For example, finite isospin density might be generated by the different number of proton and neutron in the initial nuclei. This might lead to the imbalance between charged pions in the final distribution\cite{Li:1997px}. Also, another phase, the pion superfluid phase, consisting of condensed charged pions,  might have been produced in the experiment\cite{Abelev:2013pqa,Begun:2013nga,Begun:2015ifa}. Thus, to investigate the isospin density $n_I$(or the isospin chemical potential $\mu_I$) effect has attracted growing attractions.

At finite $T$ and $\mu_B$, one of the main interests is the chiral phase transition, which is related to the breaking symmetry from $SU(2)_V\times SU(2)_A$(or equivalently $SU(2)_L\times SU(2)_R$) to $SU(2)_V$. As shown in previous sections, under this transition, the neutral pion $\pi^0$ and the two charged pions $\pi^+, \pi^-$ would play the role of Goldstone bosons in chiral limit and pseudo-Goldston bosons with finite quark mass. The study\cite{Son:2000xc} from Son and Stephanov suggests that with sufficient large $\mu_I$, nuclear matter might transit from the normal phase to the pion superfluid phase. It is supported both from lattice simulations\cite{Brandt:2017oyy} and model studies\cite{Xia:2013caa,Xia:2014bla}. Under the transition, the $SU(2)_V$ symmetry would be broken to $U_I(1)$, which is the Abelian subgroup of $SU(2)_V$. Since the isospin numbers $I_3$ of $\pi^+, \pi^0, \pi^-$ are different,  the three pions might split at finite $\mu_I$. One of the charged pions become the Goldstone boson of this symmetry breaking process. Since one of the main goals of this work is to investigate the relation ship of hadron spectrum and phase transitions, in this section we will focus on the isospin density effect, which provides a possibility to probe another kind of phase transition.

\begin{figure}[h]
    \centering
    \includegraphics[scale=0.45]{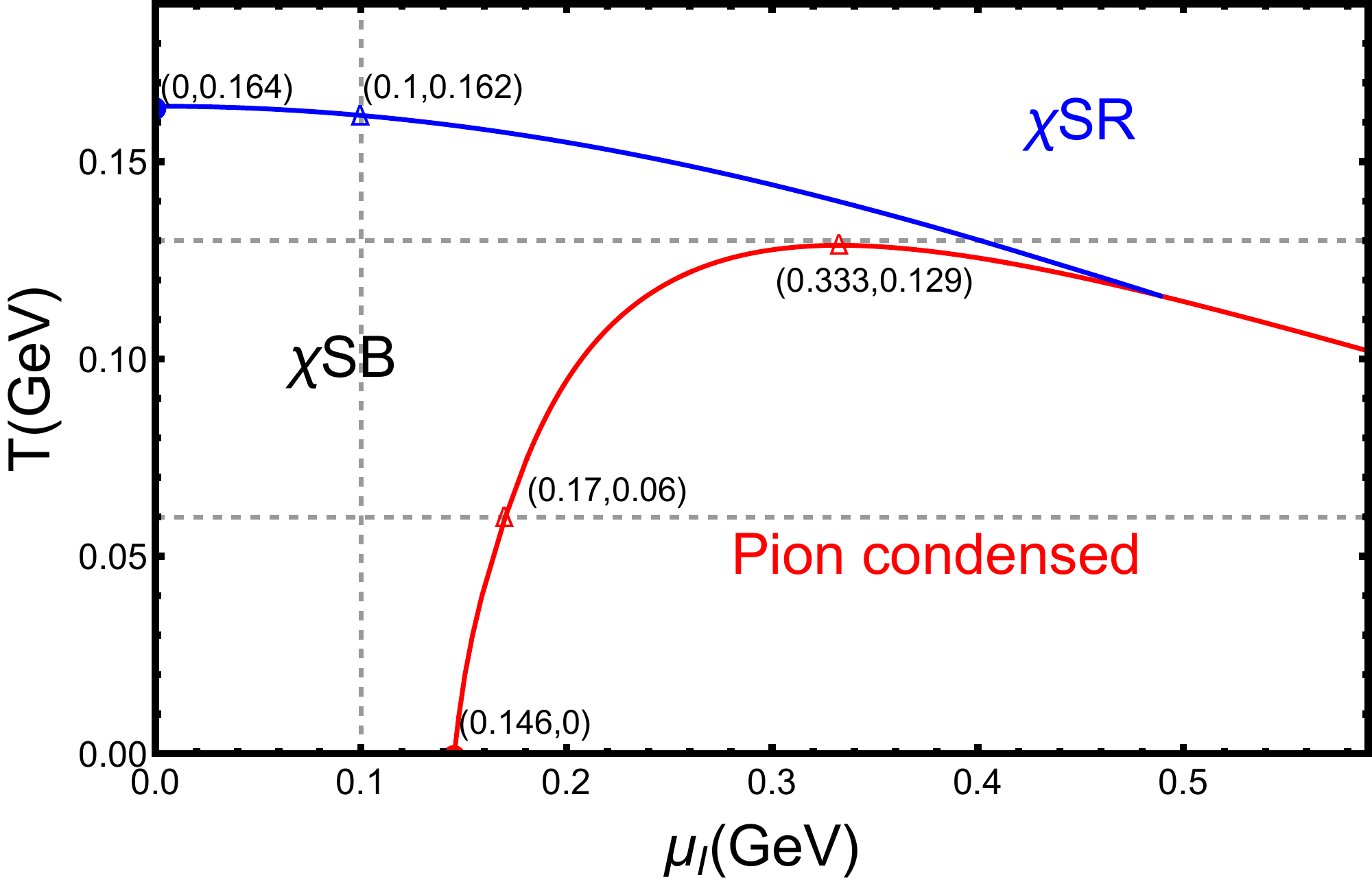}
    \hfill
    \caption{\label{phasediagram} The QCD phase diagram at finite temperature and finite isospin chemical potential ~\cite{Cao:2020ske}. The blue line is the phase boundary between normal chiral symmetry broken($\chi$SB) phase and chiral symmetry restored($\chi$SR) phase. The pion condensed phase is bounded by the Red line and the $\mu_I$-axis. The maximal temperature of pion condensation is $T_{c,top}=129 \rm{MeV}$, labeled at the peak of the red line. Masses of pions along the three grey lines will be studied in Sec.\ref{sec-mass-muI}.  }
\end{figure}

In holographic framework, the isospin density effect has been studied in hard-wall model in Refs.\cite{Albrecht:2010eg,Lee:2013oya,Nishihara:2014nva,Nishihara:2014nsa,Mamedov:2015sha}. It has been shown that above $\mu_I$ charged pions tend to form the Bose-Einstein condensation.  In our previous work\cite{Cao:2020ske}, we extend this study to both finite temperature and isospin density, and we get the phase diagram in $T-\mu_I$ plane, as shown in Fig.\ref{phasediagram}. In the figure, a $\Lambda$ type of phase boundary(the red solid line) is extracted. Below the phase boundary, i.e. at relatively large $\mu_I$ and low $T$, the pion condensed phase would form. Outside this region, the nuclear matter is in the normal phase. The chiral transition line is also presented in blue line.

\subsection{Spectral functions of pseudo-scalar mode}
According to the holographic recipe, the conserved current is dual to the gauge field in Eq.~\eqref{gaugef}. The isospin current $\bar{q}\gamma_\mu t^3 q$ is dual to $V_\mu^3$. At finite isospin density, one has to consider nonzero $V_\mu^3$. Generally, the solution should be solved from certain kind of gravity system coupled with the soft wall AdS/QCD model action. But it is difficult to solve the action with the full back-reaction. Thus, for simplicity, in the sense of probe limit, we take the AdS-Reissner-Nordström(AdS-RN) metric solution of the action, which couples the $F^2_L+F^2_R$ terms in Eq.~\eqref{action} with the 5D Einstein-Hilbert action. Thus, the metric in Eqs.~\eqref{metric-T-1} and \eqref{metric-T-2} would be replaced with
\begin{subequations}
    \begin{eqnarray}
    ds^2&=&e^{2A(z)}\left(f(z)dt^2-dx^i dx_i-\frac{1}{f(z)}dz^2\right),\\
    f(z)&=&1-(1+ \mu_I^2z_h^2)\frac{z^4}{z_h^4}+ \mu_I^2\frac{z^6}{z_h^4},
    \end{eqnarray}
\end{subequations}
together with a nonzero $V_0^3$ of the following form
\begin{eqnarray}
    V_0^3(z)=\mu_I\left(1-\frac{z^2}{z_h^2}\right).
\end{eqnarray}
Here, $\mu_I$ is the isospin density, and $z_h$ is the horizon where $f(z)=0$. In this case, the temperature is defined by the surface gravity as
\begin{equation}
T=\left|\frac{f'(z_h)}{4\pi}\right|=\frac{2-\mu_I^2z_h^2
}{2\pi z_h}.
\end{equation}
Since we will focus on the pion superfluid transition, in which the pseudo-scalar mode is relevant, we will consider only pions in this section. At finite temperature, the three pseudo-scalar modes $\pi^1, \pi^2, \pi^3$ in Eq.~\eqref{chi-pert} is symmetric under rotation in isospin space. However, with finite isospin chemical potential $\mu_I$, such a symmetry is broken and we expect the splits of these three modes. This could be read easily from the effective action
\begin{eqnarray}\label{actionpi-mu}
    S_{PS}&=&-\frac{1}{4 {g_5}^2}\int d^5x \sqrt{-g}e^{-\Phi}\bigg[\sum _{i=1}^3 \bigg\{{g_{\mu \nu }} {g_{zz}} {\partial_z }{\partial_\mu \varphi^i}{\partial_z }{\partial_\nu \varphi^i} +{g_5}^2 {g_{\mu \nu }} \chi ^2 \partial_\mu\varphi^i \partial_\nu\varphi^i+\nonumber\\
    & & {g_5}^2 \chi ^2 \left({g_{\mu \nu}} \partial_\mu\pi^i \partial_\nu\pi^i+{g_{zz}} {(\partial_z\pi^i)}^2 \right)\bigg\}+{g_5}^2\chi ^2\bigg\{{g_{00}} \left((V_{0}^3)^2 \left((\pi^1)^2+(\pi^2)^2\right)\right)\nonumber\\
    & & +2 g_{00}V_0^3 \left(\pi^1 \partial_0\pi^2-\pi^2 \partial_0\pi^1\right)-2 {g_{00}} V_0^3 \left(\pi^1\partial_0\varphi^2-\pi^2\partial_0\varphi^1\right)\nonumber\\
    & & \left.\left.-2 {g_{\mu \nu}} \sum _{i=1}^3
    \partial_\mu \varphi^i \partial_\nu \pi^i\right\}\right].
\end{eqnarray}
For later convenience, we redefine $\pi^1, \pi^2$ as
\begin{eqnarray}\label{pidefinepm}
    \pi^{\pm}=\frac{1}{\sqrt{2}}(\pi^1\mp \pi^2), \ \ \ \ \ \ \ \  \ \ \ \ \ \ \ \  \varphi^{\pm}=\frac{1}{\sqrt{2}}(\varphi^1\mp \varphi^2),
\end{eqnarray}
where $\pi^{\pm},\varphi^{\pm}$ represent the degree of freedom for charged pions. We denote the mode $\pi^3,\varphi^3$ as $\pi^0,\varphi^0$ to specify the charge difference.

Like the case at finite temperature, we only consider time $t$ and $z$ dependence of the three modes, and transform all of them to the frequency space as
\begin{subequations}
\begin{eqnarray}
    \pi^{\pm}(t,z)&=&\frac{1}{2\pi}\int d\omega_{\pm} \ e^{-i\omega_{\pm} t} \pi^{\pm}(\omega_{\pm}, z),\ \    \pi^0(t,z)=\frac{1}{2\pi}\int d\omega_0 \ e^{-i\omega_0 t} \pi^0(\omega_0, z),\\
    \varphi^{\pm}(t,z)&=&\frac{1}{2\pi}\int d\omega_{\pm} \ e^{-i\omega_{\pm} t} \varphi^{\pm}(\omega_{\pm}, z),\  \      \varphi^0(t,z)=\frac{1}{2\pi}\int d\omega_0 \ e^{-i\omega_0 t} \varphi^0(\omega_0, z).
\end{eqnarray}
\end{subequations}
Then one can obtain the equation of motion for the three modes as
\begin{subequations}\label{EOMpi0-mu}
    \begin{eqnarray}
        \varphi^{0''} + (A'- \Phi ')\varphi ^{0'}-\frac{e^{2 A} g_5^2 \chi ^2}{f}\left(\varphi ^0-\pi ^0\right)&=&0,\\
        \pi ^{0''}+\left(3 A'+\frac{f'}{f}-\Phi '+\frac{2 \chi '}{\chi }\right)\pi^{0'}-\frac{\left(\varphi ^0-\pi ^0\right) \omega ^2}{f^2}& =& 0,
    \end{eqnarray}
\end{subequations}
 and
 \begin{subequations}\label{EOMpi+pi-mu}
    \begin{eqnarray}
        \varphi ^{\pm ''}+(A'-\Phi')\varphi ^{\pm '} -\frac{e^{2 A} g_5^2 \chi ^2}{\omega_{\pm} f}\left[\omega_{\pm} \varphi ^{\pm }-(\omega_{\pm} \pm V_0^3)\pi ^{\pm }\right ]=0,\\
        \pi ^{\pm ''}+\left(3 A'+\frac{f'}{f}-\Phi '+\frac{2 \chi '}{\chi }\right)\pi ^{\pm '}-\frac{(\omega_{\pm} \pm V_0^3)\left[\omega_{\pm} \varphi ^{\pm }-(\omega_{\pm} \pm V_0^3)\pi ^{\pm }\right]}{f^2}=0.
    \end{eqnarray}
\end{subequations}

Also, it is not difficult to derive the on-shell action for the neutral pion as
\begin{eqnarray}\label{onshellactionofpi0-mu}
    S_{\pi 0}^{on}= - \frac{1}{4 g_5^2}\int d\omega\  e^{A-\Phi }\left [e^{2 A} g_5^2 f  \chi ^2\pi^0(-\omega,z)  {\pi^{0'}}(\omega, z)-\omega ^2\varphi^0(-\omega, z)  \varphi^{0'}(\omega, z) \right ] \bigg |_{z=\epsilon}^{z=z_h},\nonumber\\
\end{eqnarray}
The charged pions' on-shell actions are
\begin{eqnarray}\label{onshellactionofpipm-mu}
    S_{\pi\pm}^{on}=-\frac{1}{4 g_5^2}\int d \omega\  e^{A-\Phi }\left[e^{2 A} g_5^2 f \chi^2  \pi ^{\pm}(-\omega, z) \pi ^{\mp'}(\omega, z)-\omega ^2\varphi ^{\pm}(-\omega, z)\varphi ^{\mp'}(\omega, z)\right]\bigg|_{z=\epsilon}^{z=z_h}.\nonumber\\
\end{eqnarray}

To get the on-shell action, one has to solve the equation of motions. Before that, the UV and IR boundary condition should be specified. At UV boundary, the asymptotic expansion could be derived as
\begin{subequations}\label{UVmu1}
    \begin{eqnarray}
        \varphi^0(z\rightarrow0)&=&c_f+\varphi _2 z^2-\frac{1}{2} \zeta ^2 g_5^2 m^2 \pi _0 z^2 \log (z)+\mathcal{O}(z^3),\\
        \pi^0(z\rightarrow 0) &=&\pi_0+c_f+\pi _2 z^2-\frac{1}{2} \pi _0 \omega ^2 z^2 \log (z)+\mathcal{O}(z^3),
    \end{eqnarray}
\end{subequations}
and
\begin{subequations}
    \begin{eqnarray}
        \varphi ^{\pm }(\omega, z\to 0)&=&\varphi _0^{\pm }+\varphi _2^{\pm }z^2+\frac{g_5^2m^2\zeta ^2\left[\omega \varphi _0^{\pm }-(\omega \pm \mu_I) \pi _0^{\pm }\right]z^2\log (z)}{2 \omega }+\mathcal{O}(z^3),\\
        \pi^{\pm}(\omega, z\to 0)&=&\pi _0^{\pm }+\pi _2^{\pm }z^2+\frac{1}{2}(\omega \pm \mu) \left[\varphi _0^{\pm }\omega -(\omega \pm \mu_I) \pi _0^{\pm }\right]z^2\log (z)+\mathcal{O}(z^3),\nonumber\\
    \end{eqnarray}
\end{subequations}
for $\pi^0,\varphi^0$ and $\pi^{\pm},\varphi^{\pm}$ respectively, where $\pi_0$, $c_f$, $\pi_2$, $\varphi_2$, $\varphi_0^{\pm}$, $\varphi_2^{\pm}$, $\pi_0^{\pm}$, $\pi_2^{\pm}$ are integral constants.

To get the Retarded Green functions, the in-falling condition would be imposed at IR. Thus, the expansions at IR are
\begin{subequations}
    \begin{eqnarray}
    \varphi^0(z\rightarrow z_h) & =& \left(z_h-z\right)^{\frac{i \omega  z_h}{2\mu_I ^2 z_h^2-4}}\bigg\{ \frac{2 i c_0^2 g_5^2 \pi _{\text{h0}}  \left(\mu_I ^2 z_h^2-2\right)\left(z-z_h\right)}{\omega  z_h^2 \left(2   \mu_I ^2 z_h^2+i \omega  z_h-4\right)}+\mathcal{O}[(z-z_h)^2]\bigg\}+c_{h0},\nonumber\\
    \\
    \pi^0(z\rightarrow z_h) &=& \left(z_h-z\right)^{\frac{i \omega  z_h}{2  \mu_I ^2 z_h^2-4}}\bigg\{\pi_{h0}+\mathcal{O}(z-z_h)\bigg\}+c_{h0},
    \end{eqnarray}
\end{subequations}
and
\begin{subequations}
    \begin{eqnarray}
        \varphi^{\pm}(z\rightarrow z_h)&=& \left(z_h-z\right)^{\frac{i \omega  z_h}{2  \mu_I ^2 z_h^2-4}}\bigg\{\frac{2ic_0^2g_5^2(z-z_h)(\mu_I ^2 z_h^2-2)\pi _{\text{h0}}^{\pm }}{\omega  z_h^2 \left(2  \mu_I ^2 z_h^2+i \omega  z_h-4\right)}+\mathcal{O}[(z-z_h)^2]\bigg\}+c_{h0}^{\pm},\nonumber\\
        \\
        \pi^{\pm}(z\rightarrow z_h)&=&\left(z_h-z\right)^{\frac{i  z_h}{2 \mu_I ^2 z_h^2-4}}\big\{\pi _{\text{h0}}^{\pm }+\mathcal{O}(z-z_h)\big\}+c_{h0}^{\pm},
    \end{eqnarray}
\end{subequations}
where $\pi_{h0}$, $c_{h0}$, $\pi_{h0}^{\pm}$, and $c_{h0}^{\pm}$ are integral constants. Under this conditions, one can solve all the integral constants, and get the spectral function through the following expressions\footnote{The parameter $c_f$ and $\varphi_0^{\pm}$ are set to zero. Certain real terms inside the imaginary function are neglected.}

\begin{eqnarray}
    \rho(\pi^0,\omega)&=&-\frac{1}{\pi}{\rm{Im}}G_{\pi^0}^R(\omega)\nonumber\\
    &=&{\rm{Im}}\bigg[ \frac{1}{8\pi\pi _0}  \zeta ^2 m_q^2 \left(\pi _0 \omega ^2-4 \pi _2\right)\bigg ].
\end{eqnarray}
and
\begin{eqnarray}
    \rho(\pi^{\pm}, \omega)&=&-\frac{1}{\pi} {\rm{Im}}G^R_{\pi^{\pm}}(\omega)\nonumber\\
   &=&{\rm{Im}}\left\{\frac{1}{8\pi\pi _0^{\pm }}m_q^2\zeta ^2\left [(\omega \pm \mu)^2 \pi _0^{\pm }-4\pi _2^{\pm }\right ]\right\}.
\end{eqnarray}

\subsection{Mass spectrum of pion quasiparticles and the Goldstone boson}
\label{sec-mass-muI}

The pion superfluid transition is connected with the breaking of $SU(2)_V$ symmetry, which is an exact symmetry in both chiral limit and cases with finite quark masses. So, in this section, we only consider the realistic case with physical quark mass. Taking $m_q=3.22\rm{MeV}$, and solving the equations of motion, one can obtain the spectral functions.

\begin{figure}[thbp]
    \centering 
    \begin{overpic}[scale=0.35]{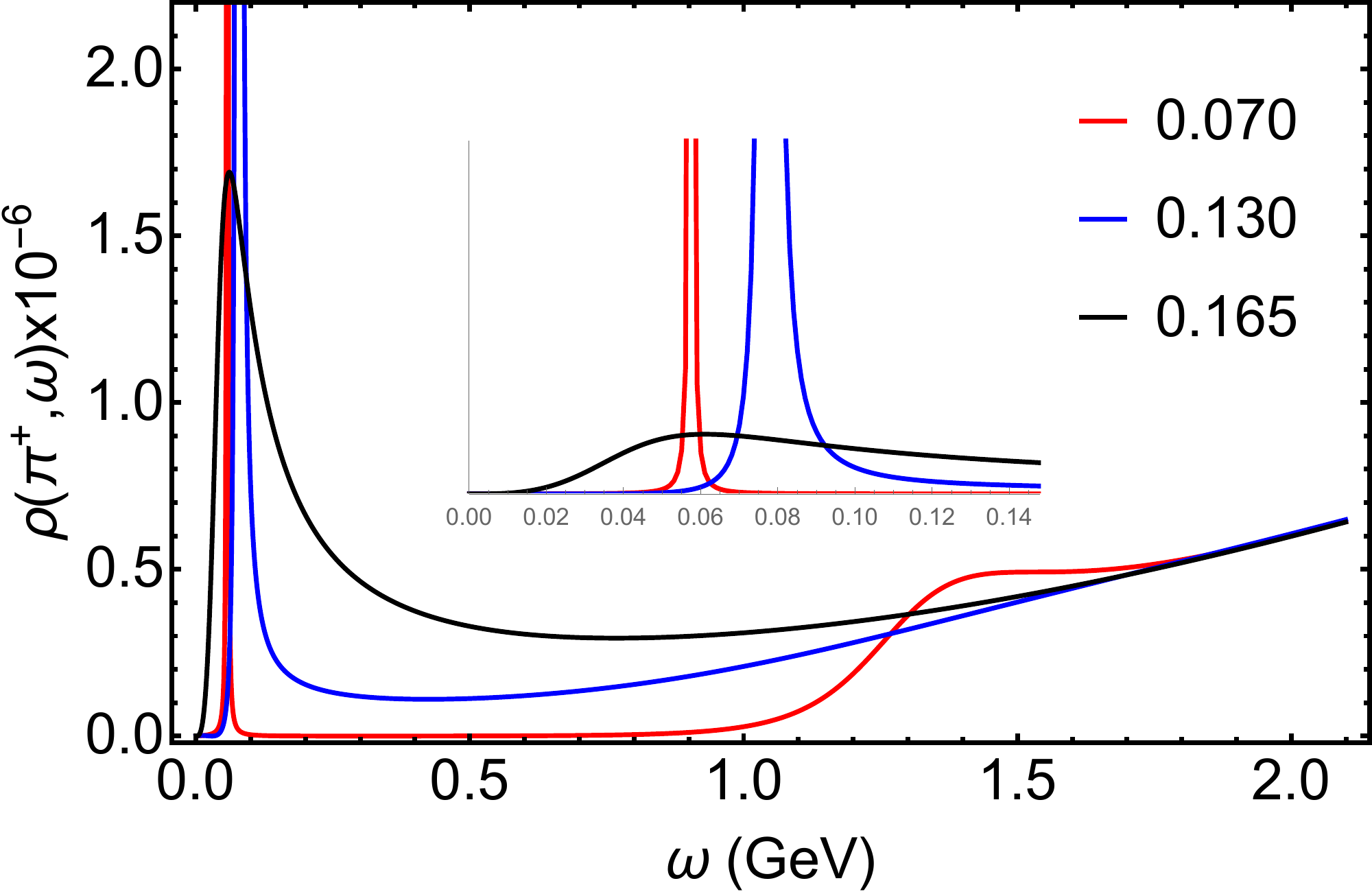}
        \put(90,60){\bf{(a)}}
    \end{overpic}
    \hfill
    \begin{overpic}[scale=0.35]{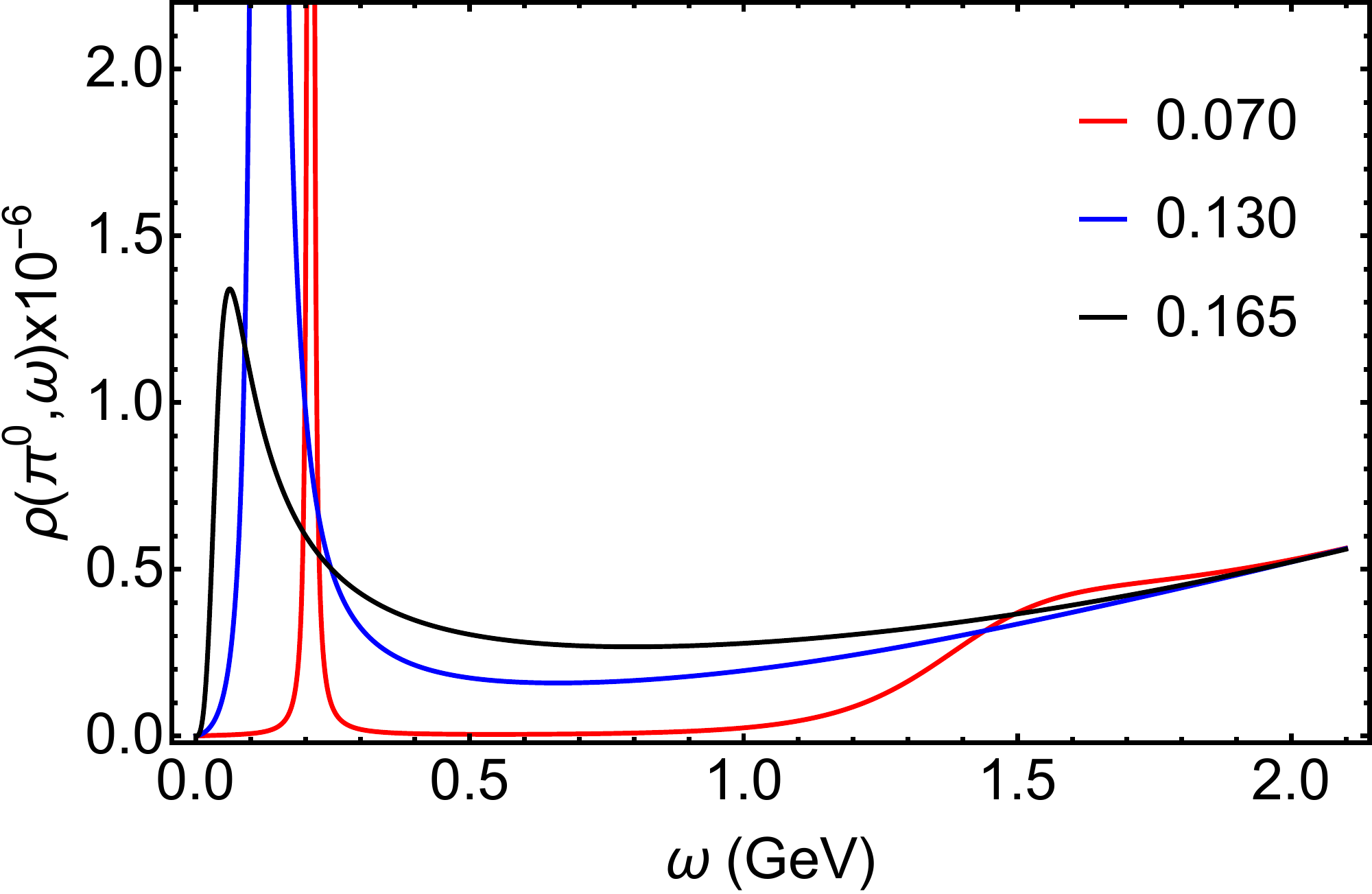}
        \put(90,60){\bf{(b)}}
    \end{overpic}
    \hfill
    %\includegraphics[width=.45\textwidth]{mu01spectra4.pdf}
    %\hfill
    %\includegraphics[width=.48\textwidth]{figmu01pi-.pdf}
    \begin{overpic}[scale=0.35]{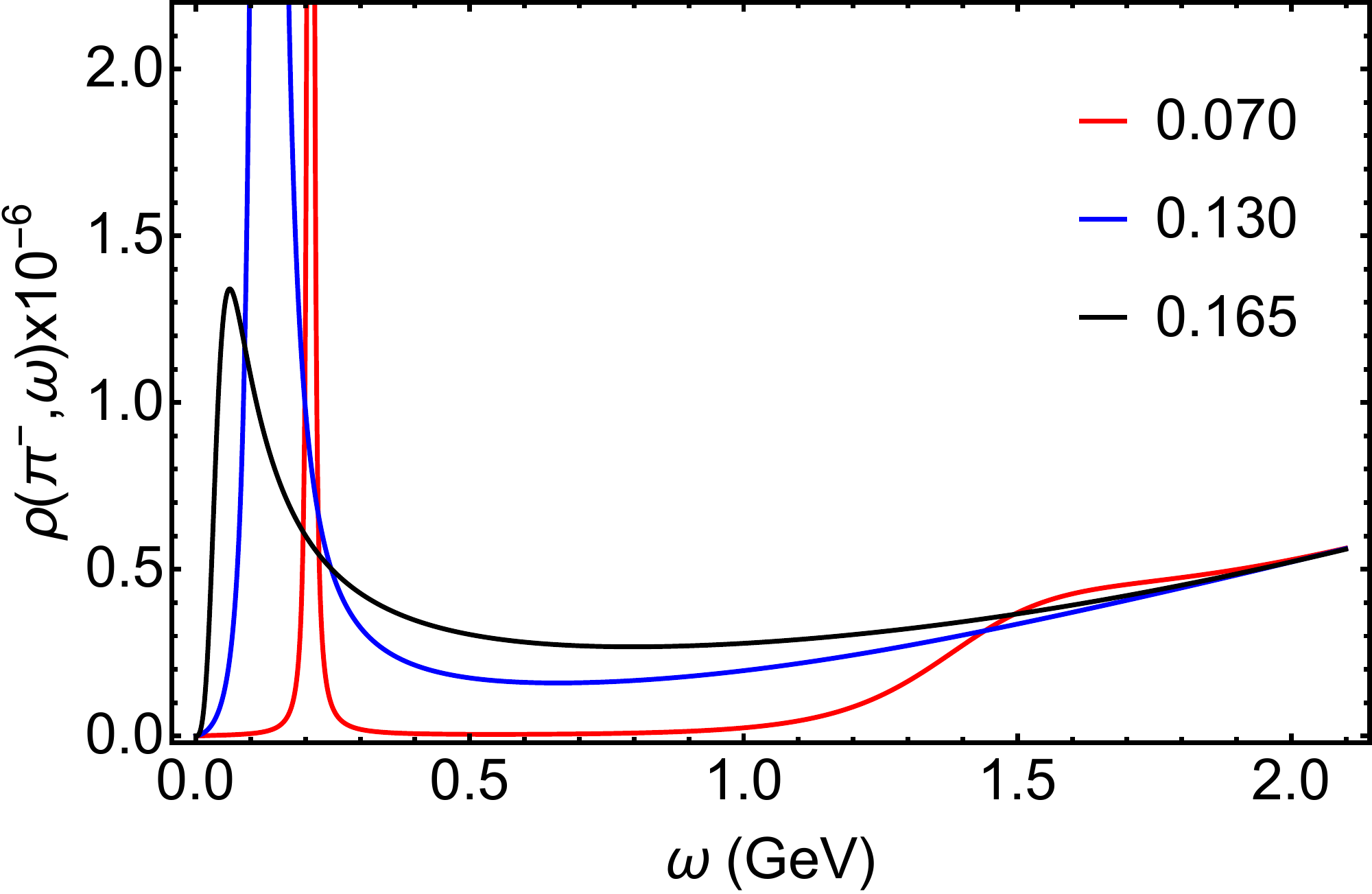}
        \put(90,60){\bf{(c)}}
    \end{overpic}
    \hfill
    \begin{overpic}[scale=0.35]{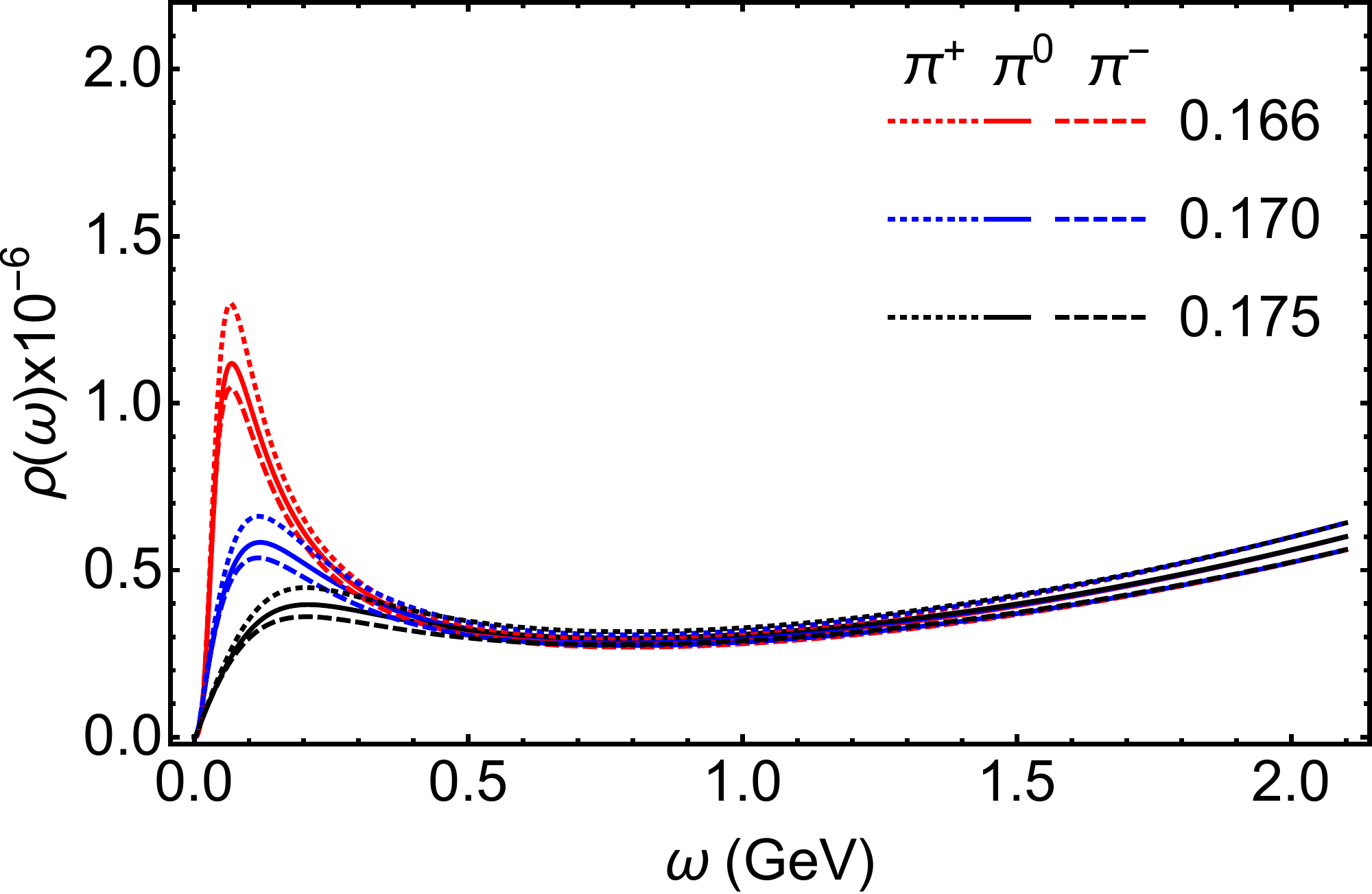}
        \put(90,60){\bf{(d)}}
    \end{overpic}
    \hfill
    \caption{\label{pispectralfigmu01varyT}  Spectral functions of (a) $\rho(\pi^+, \omega)$(an enlarged view for frequency at the interval of $(0,0.15)$ is shown), (b) $\rho(\pi^0, \omega)$, and (c) $\rho(\pi^-, \omega)$, at $\mu_I=0.1$GeV. The red, blue and black solid lines represent results below $T_{cp}$, at $T=0.07, 0.13$ and $0.165\rm{GeV}$, respectively. (d) Spectra functions at $\mu_I=0.1$GeV and $T$ above $T_{cp}$, with the dotted line for $\pi^+$, solid line for $\pi^0$ , dashed line for $\pi^-$. The red, blue and black lines represent results of $T=0.166, 0.170$ and $0.175\rm{GeV}$, respectively. }
\end{figure}

Firstly, we fix $\mu_I=0.1\rm{GeV}$ and investigate the temperature dependent behavior of the three modes, i.e. along the vertical gray line in Fig.\ref{phasediagram}. The spectral functions for $\pi^+,\pi^0$, and $\pi^-$  are shown in Fig.\ref{pispectralfigmu01varyT}(a), (b), and (c) respectively, taking temperatures as $T=0.07, 0.13$, and $0.165 \rm{GeV}$. From the location of the peaks, we could see that at low temperature and finite isospin chemical potential, the masses of $\pi^+,\ \pi^0$, and $\pi^-$ split. For example, at $T=0.07\rm{GeV}$, $m_{\pi^+}\approx 0.057\rm{GeV}, m_{\pi^0}\approx 0.136\rm{GeV}$, smaller than their vacuum values, while $m_{\pi^-}\approx 0.210\rm{GeV}$, larger than its vacuum value. Moreover, from the location of the peaks, one can see that, with the increasing of temperature, the masses of $\pi^-,\pi^0$  decrease monotonically, while $m_{\pi^+}$ increases at low temperature. This result might be reasonable. Considering the pion condensed phase appear at $T_{c,\pi}=0, \mu_I\approx 0.146\rm{GeV}$, possible coherent fraction of $\pi^+$ might form at $\mu_I=0.1\rm{GeV}$, and it reduces the energy to excite the $\pi^+$. The increasing temperature tends to destroy the coherence of particles, thus it would cause the contrast effect on $m_{\pi^+}$.

\begin{figure}[htbp]
    \centering 
    \includegraphics[width=.62\textwidth]{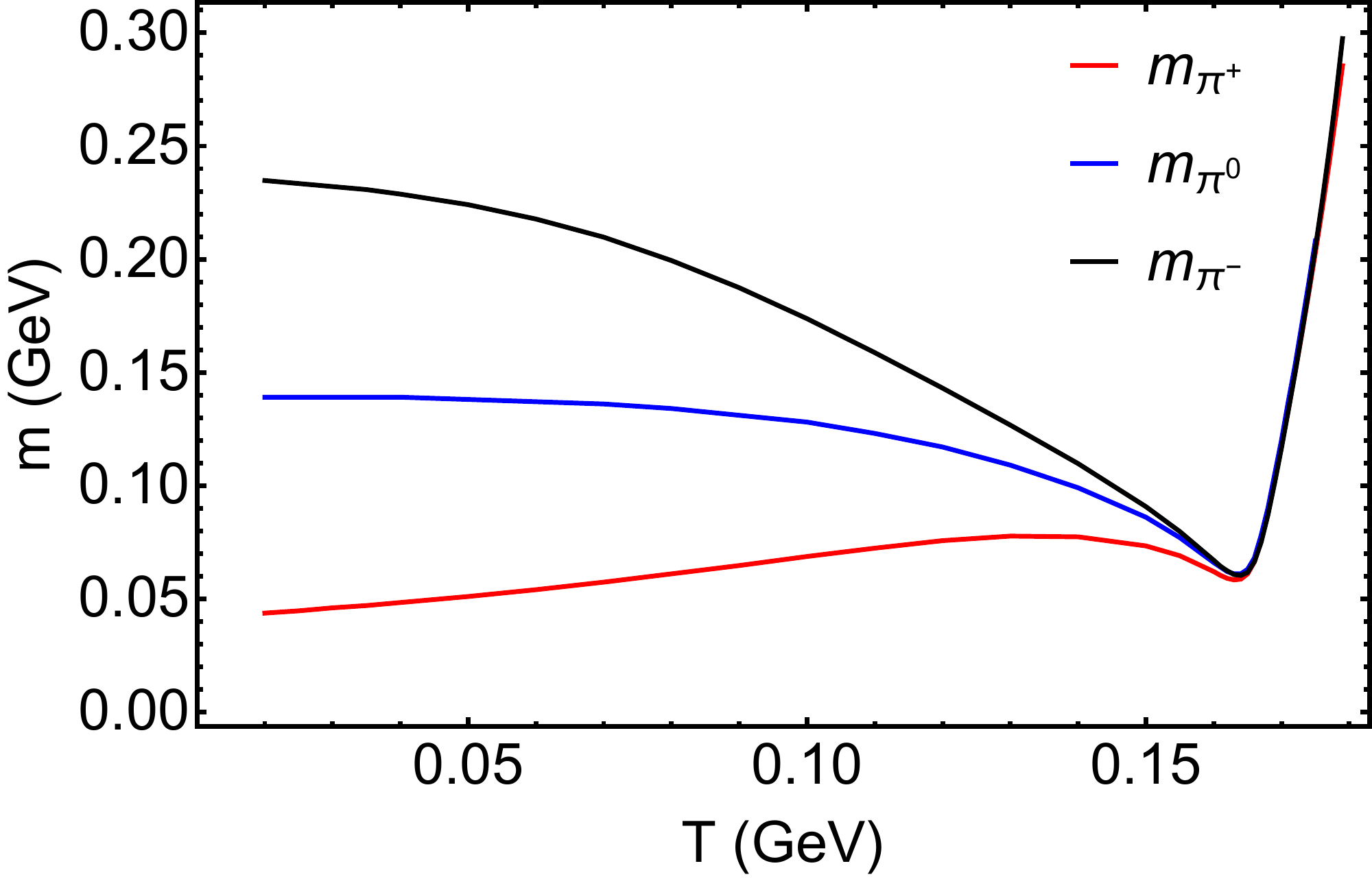}
    \hfill
    \caption{\label{massfigmu01varyT}  The temperature dependent behaviors of pions masses at $\mu_I=0.1\rm{GeV}$. The black line respects for $m_{\pi^-}$, the blue line respects for $m_{\pi^0}$ and the red line respects for $m_{\pi^+}$. }
\end{figure}

For temperatures above the blue line in Fig.\ref{phasediagram}, we take $T=0.166,0.170, 0.175\rm{GeV}$ as examples and show the results in Fig.\ref{pispectralfigmu01varyT}(d). From the figure, we could see that though the heights of the peaks are different for the three modes, the location of the peaks are almost the same. It shows that the high temperature modes are still governed by chiral phase transition.

To be clearer, we extracted the temperature dependent masses and plot them in Fig.\ref{massfigmu01varyT}. The decreasing of $m_{\pi^-}, m_{\pi^0}$ could be clearly seen from the black and blue lines respectively.  The result of $m_{\pi^0}$ is almost the same as that at $\mu_I=0$, since $I_3$ number of $\pi^0$ is zero. For $m_{\pi^-}$, enhanced by $\mu_I$, near $T=0$, it increases to about $0.235\rm{GeV}$. With the increasing of temperature, it decreases to $0.061\rm{GeV}$ near $T_{cp}$. For $m_{\pi^+}$, a little bump appears near the chiral crossover point. At a first glance, this is quite strange. Actually, it could also be reasonable. Here, the increasing of temperature has two effects. One is destroying the coherent fraction of $\pi^+$, while the other one is decreasing chiral condensate. The former effect would lead to the increase of $m_{\pi^+}$, while the latter one leads to its decreasing, as discussed in Sec.\ref{sec-chiral-pion}. Therefore, the bump is the result of the competition of these two effects. Then, it is quite obvious that the three pions become degenerate at temperature above $T_{cp}$, which is mainly controlled by chiral phase transition.

\begin{figure}[htbp]
    \centering 
        \begin{overpic}[scale=0.35]{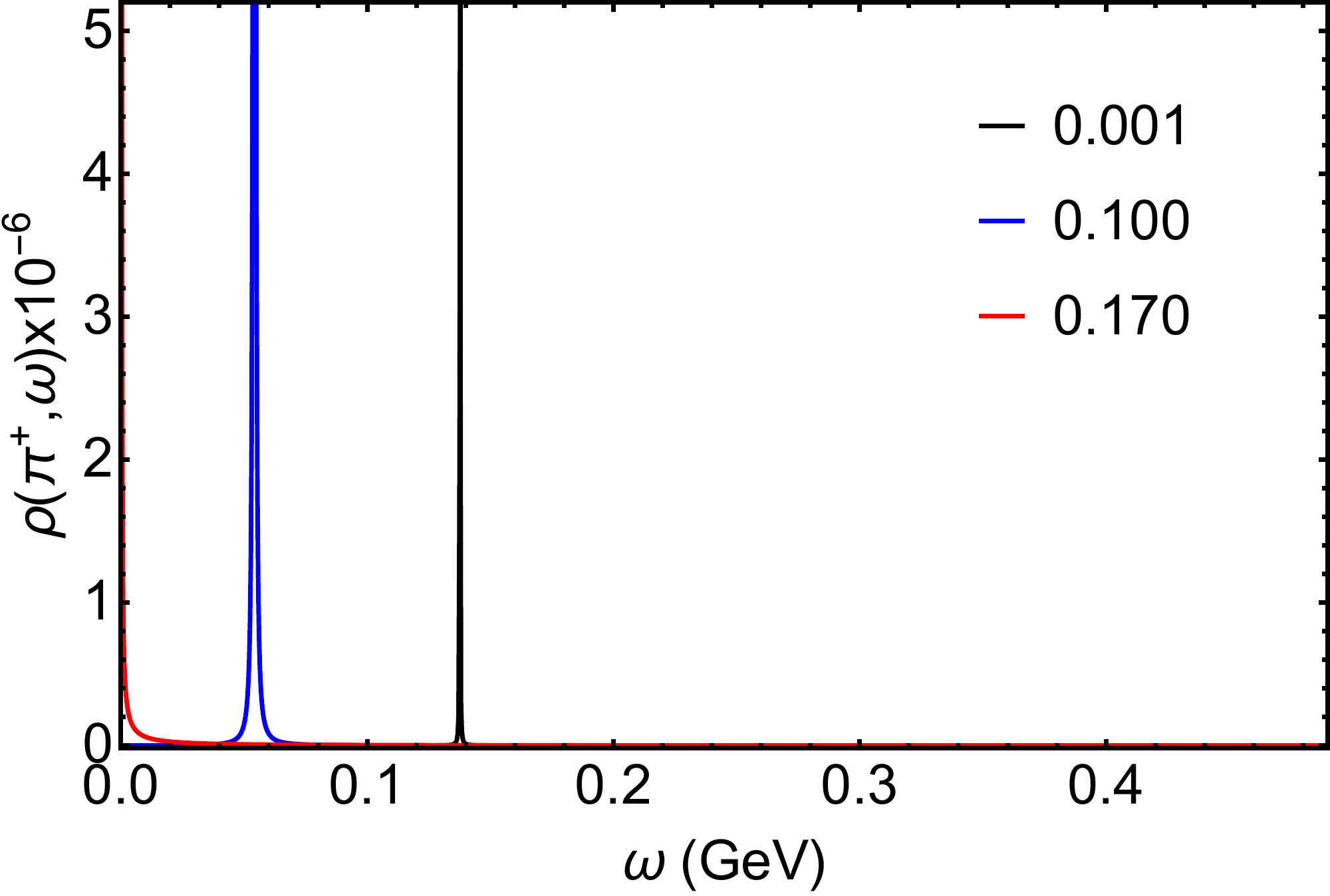}
        \put(90,62){\bf{(a)}}
        \end{overpic}
        \hfill
        \begin{overpic}[scale=0.35]{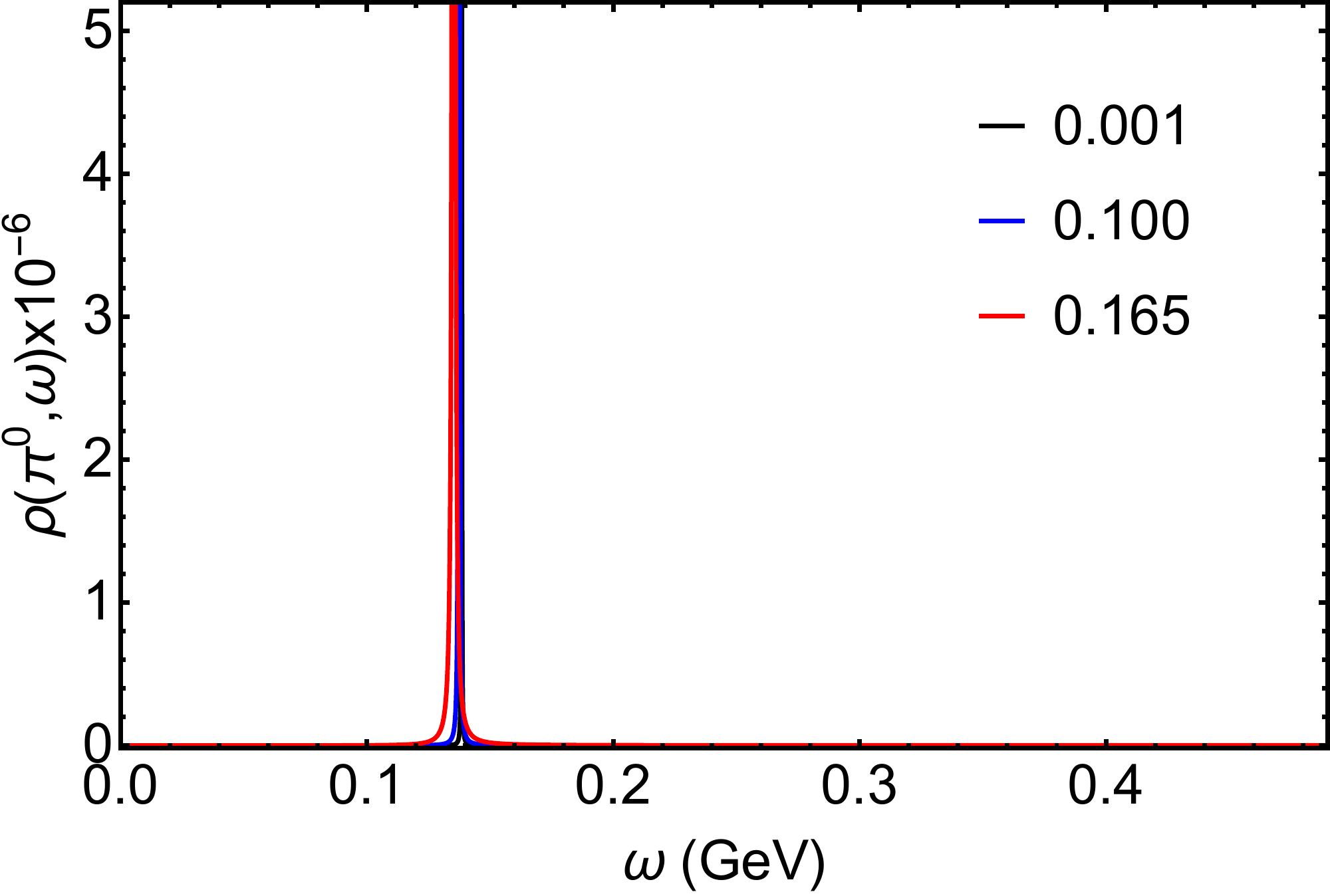}
        \put(90,62){\bf{(b)}}
        \end{overpic}
        \hfill
        \begin{overpic}[scale=0.35]{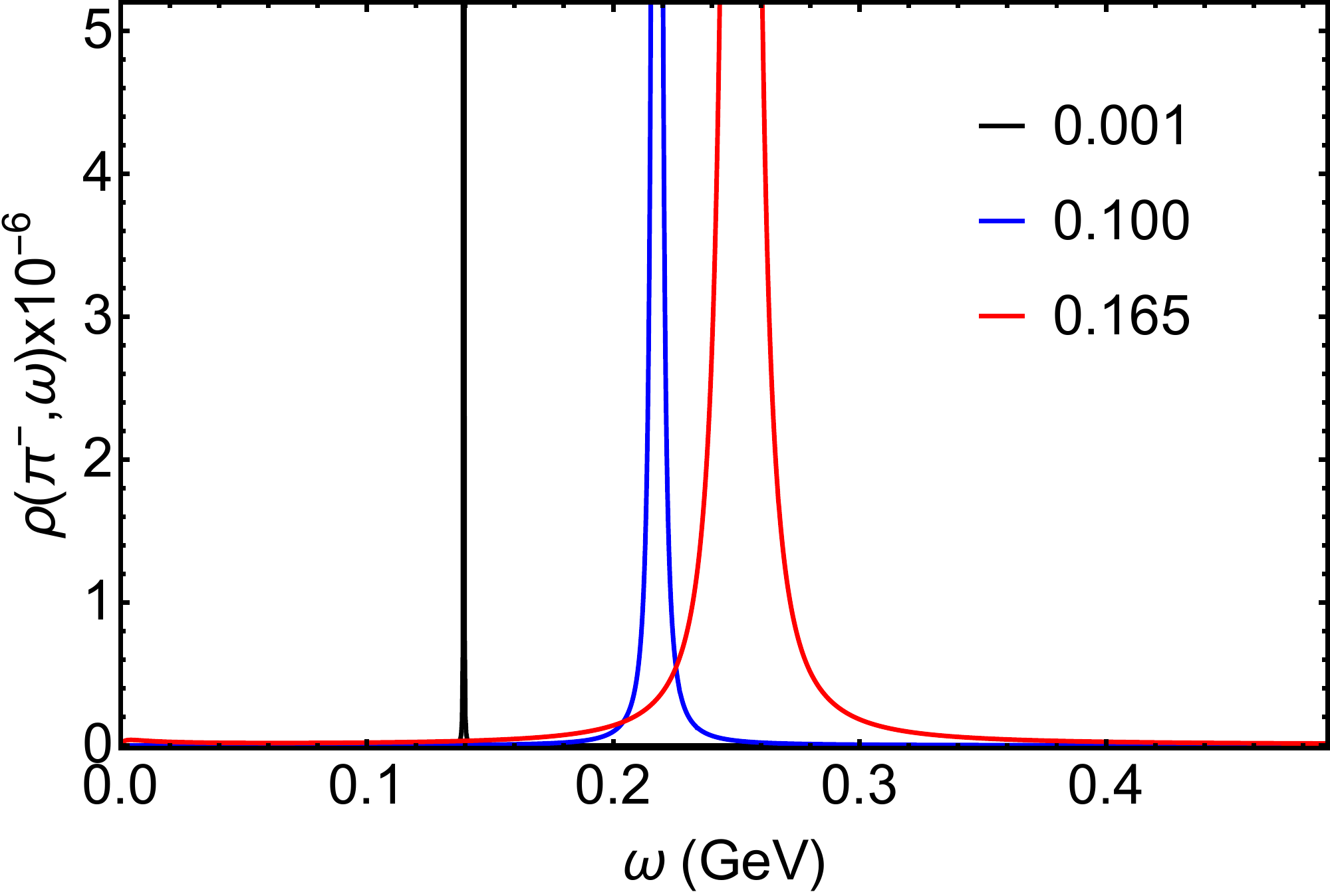}
        \put(90,62){\bf{(c)}}
        \end{overpic}
    \caption{\label{figT60} Spectral functions for (a) $\rho(\pi^+, \omega)$, (b) $\rho(\pi^0, \omega)$, and (c) $\rho(\pi^-, \omega)$, at $T=0.06\rm{GeV}$. The black and blue lines represent results at $\mu_I=0.001$, and $0.100$, respectively. The red line represents results at $\mu_I=0.170\rm{GeV}$ in (a) and $\mu_I=0.165\rm{GeV}$ in (b) and (c).}
\end{figure}

Then, we will turn to the $\mu_I$ dependence of the three modes. We fix $T=0.06\rm{GeV}$ and vary $\mu_I$, %$ take \mu_I=0.001, 0.1, 0.165$
, i.e. along the lower gray horizontal line in Fig.\ref{phasediagram}. The spectral functions of $\pi^+, \pi^0$, and $\pi^-$ are presented in Fig.\ref{figT60}(a), (b), and (c) respectively. From Fig.\ref{figT60}(b), again, we find the dependence of $m_{\pi^0}$ on $\mu_I$ is rather weak at low temperature. The location of the peaks just move slightly towards $\omega=0$. In Fig.\ref{figT60}(a),(c), the $\mu_I$ dependence of $m_{\pi^+}$ is contrast to that of $m_{\pi^-}$. The former one decreases with $\mu_I$ while the latter one increases. In addition, the most interesting thing is the appearance of massless $\pi^+$ at $\mu_I\approx 0.170 \rm{GeV}$. Another observation is that the widths of the peaks increase with $\mu_I$ at such a low temperature.

\begin{figure}[htbp]
    \centering
    \includegraphics[width=.48\textwidth]{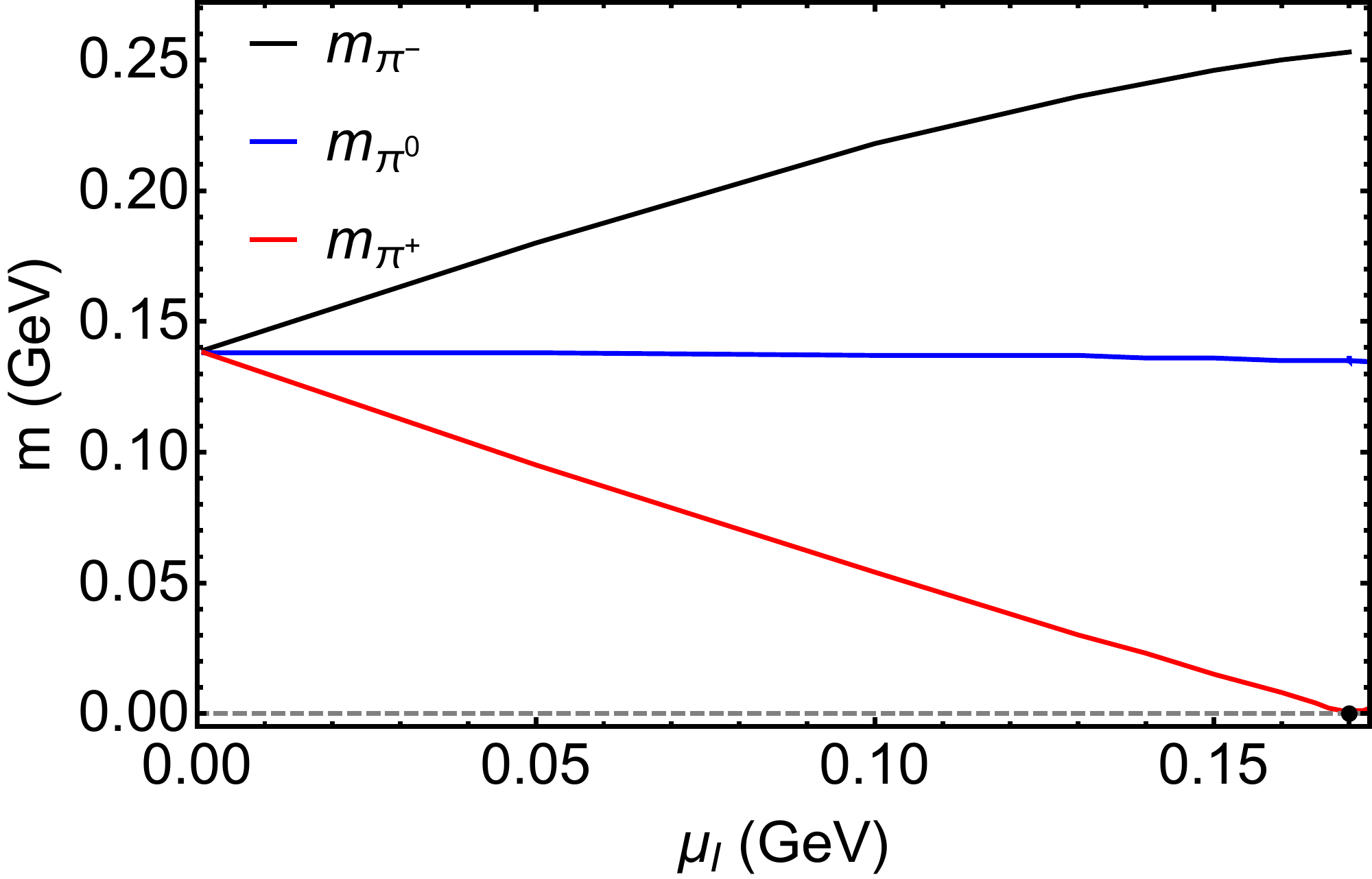}
    \caption{\label{figT60mass} $\mu_I$ dependence of pions' masses at $T=0.06\rm{GeV}$. The black, blue, red lines represent results for $m_{\pi^-}$, $m_{\pi^0}$ and $m_{\pi^+}$ respectively. $m_{\pi^+}$ vanishes at $0.170\rm{GeV}$  as shown by the black dot.}

\end{figure}

To be clearer, we also display the $\mu_I$ dependence of the quasiparticles' masses in Fig.\ref{figT60mass}. Qualitatively, the picture of the mass splitting is in agreement with previous study in hard-wall model\cite{Lee:2013oya,Nishihara:2014nsa,Mamedov:2015sha}. At $T=0.06\rm{GeV}, \mu_I=0$, the masses of the three pions are degenerate at $m\approx 0.138\rm{GeV}$. When $\mu_I$ increases to $0.170\rm{GeV}$, the $\pi^+$ becomes massless. This is related to certain instability. In fact, $T=0.06\rm{GeV}, \mu_I=0.17\rm{GeV}$ locates exactly at the phase boundary between pion condensed phase and normal phase. Above $\mu_I=0.17\rm{GeV}$, the $SU(2)_V$ symmetry will be broken to its subgroup $U_{I}(1)$, and $\pi^+$ becomes the massless Goldstone boson of the symmetry breaking. Interestingly, we observe the realization of the Goldstone theorem in holographic approach at both finite temperature and isospin density. As for $\pi^-$, its mass increase from $0.138\rm{GeV}$ to around $0.253\rm{GeV}$. The increasing mass split of $\pi^+$ and $\pi^-$ might affect the distribution of the final charged pions, if the fireball passes such an  intermediate state.

\begin{figure}[htbp]
        \centering % \begin{center}/\end{center} takes some additional vertical space
        %%\includegraphics[width=.48\textwidth]{massT60.pdf}
        %\hfill
        \includegraphics[width=.48\textwidth]{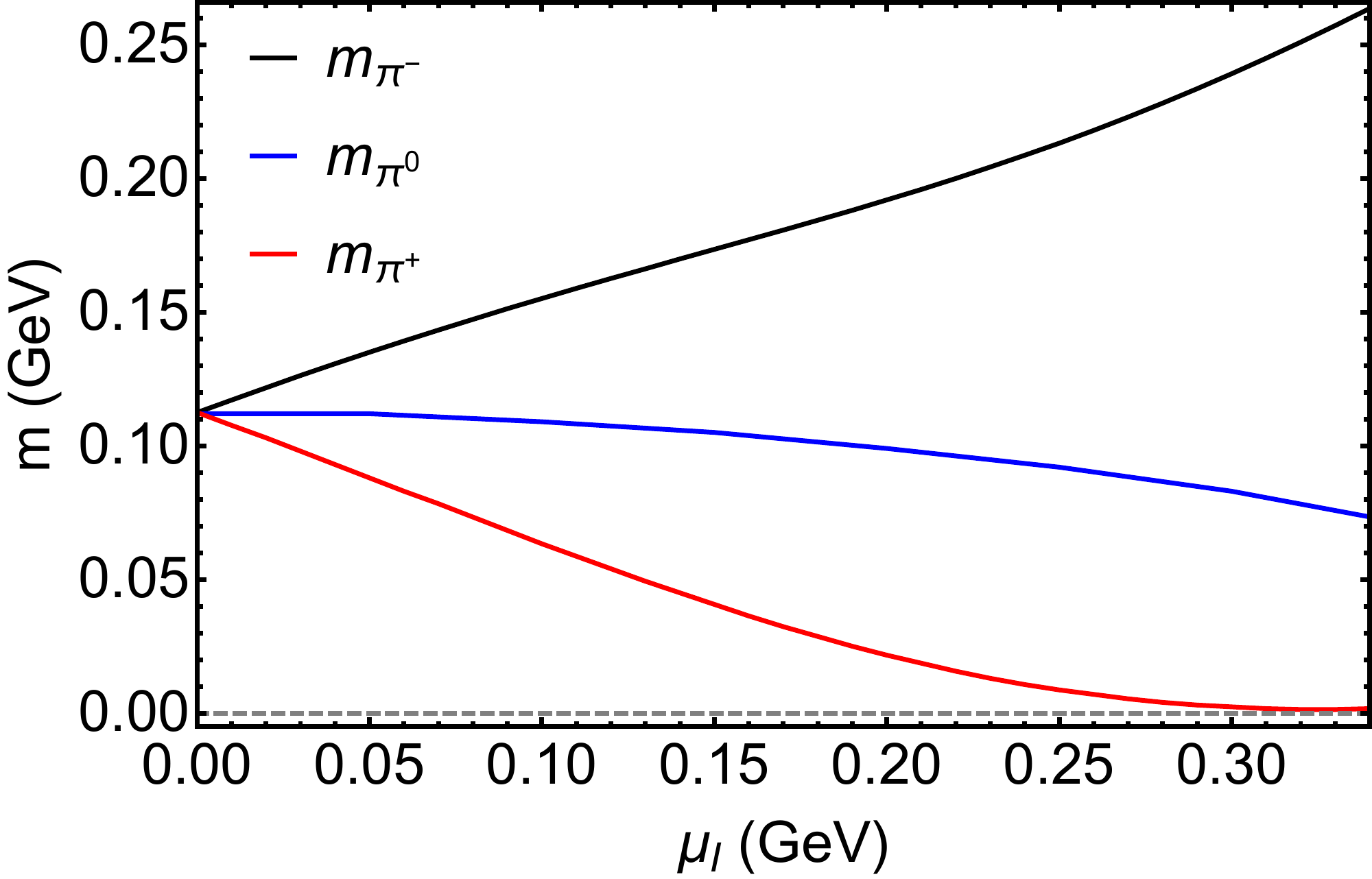}
        \hfill
        %\includegraphics[width=.45\textwidth,origin=c,angle=180]{figmu0.eps},trim=0 380 0 200,clip
        % "\includegraphics" is very powerful; the graphicx package is already loaded
        \caption{\label{figT130mass} $\mu_I$ dependence of pions' masses at $T=0.13\rm{GeV}$. The black, blue, red lines represent results for $m_{\pi^-}$, $m_{\pi^0}$ and  $m_{\pi^+}$ respectively. }
\end{figure}

Finally, as shown in Fig.\ref{phasediagram}, when the temperature is higher than $0.129\rm{GeV}$, the condensed phase would be destroyed, even with very large $\mu_I$. Thus, we will also investigate the behavior of the mass spectral at high temperature region. We take $T=0.13\rm{GeV}$ as an example. From our calculation, with larger widths, the peaks in spectral functions are similar with the ones at $T=0.06\rm{GeV}$. So we give the $\mu_I$ dependence of pion masses only. It is shown in Fig.\ref{figT130mass}. Qualitatively, the results for $\pi^0, \pi^-$ are similar to those at $T=0.06\rm{GeV}$. As for $\pi^+$, the massless mode disappears even at very large $\mu_I$, which is consistent with the absence of pion superfluid at high temperature.

\section{Conclusion}
\label{sum}

In this work, extracted from the spectral functions, masses of light (pseudo-)scalar mesons at finite temperature $T$ and isospin chemical potential $\mu_I$, as well as their relationship with chiral phase transition and pion superfluid transition, have been investigate in a two-flavor soft-wall AdS/QCD model.

At finite temperature and in chiral limit, charged and neutral pions are proved numerically and analytically to be massless Goldstone modes, for any temperature below the critical temperature $T_c$ of chiral phase transition. The mass of scalar meson decreases from its vacuum value about $1.05\rm{GeV}$ to zero at $T_c=0.163\rm{GeV}$. Above $T_c$, the masses of scalar and pseudo-scalar mesons become degenerate and start to increase with temperature. This could be considered as the realization of chiral symmetry restoration in hadronic level, which is consistent with the signal from chiral condensate $\langle\bar{q}q\rangle$. Qualitatively, this picture is in good agreement with the Goldstone theorem\cite{Nambu:1961tp,Nambu:1961fr} and expectation from theoretical analysis. In this sense, the theoretical consistency of soft-wall model has been checked.

Then, to be more realistic, we turn to cases with physical quark mass, when the second order chiral transition turns to chiral crossover. Qualitatively, the mass dependence of scalar meson is similar with that in chiral limit. But instead of zero, its mass reaches a finite minimum,  around the pseudo-critical temperature $T_{cp}= 0.164\rm{GeV}$ of chiral crossover. The main difference comes from pions. Below $T_{cp}$, their masses decrease from the vacuum value $0.140\rm{GeV}$ to $0.062\rm{GeV}$ at $T_{cp}$. Qualitatively, this result agrees well with the general analysis by Son and Stephanov in Refs.\cite{Son:2001ff,Son:2002ci}, as well as lattice simulations in Refs.\cite{Brandt:2014qqa,Brandt:2015sxa}. Quantitatively, the reduction rate from our holographic study is about $60\%$, larger than $30\%$ from Refs.\cite{Son:2001ff,Son:2002ci} and $20\%$ from Refs.\cite{Brandt:2014qqa,Brandt:2015sxa}. This behavior might lead to a larger contribution from the mass decreasing of pions to the enhancement of low momentum distribution of pion in heavy ion collisions. For temperature above $T_{cp}$, the degeneration of scalar and pseudo-scalar modes is observed again, and both the spectral increase with temperature, which is coincident with studies from 4D models\cite{Fischer:2018sdj,Gao:2020hwo,Tripolt:2013jra,Xia:2013caa,Xia:2014bla}. Phenomenologically, it might be a possible candidate to find the signal of phase transition in experiments.

Considering the growing interests on the effect of isospin density, we extend the above study to situations with finite isospin chemical potential $\mu_I$. In previous soft-wall model study Ref.\cite{Cao:2020ske}, extending the finite temperature above to finite $\mu_I$, a phase consisting of condensed charged pions is observed at large $\mu_I$ when temperature is below $T=129\rm{MeV}$. In this work, we extend this study and extract the $T$ and $\mu_I$ dependence of pion mass by calculating the spectral functions.

With a fixed $\mu_I$, e.g. $\mu_I=0.1\rm{GeV}$,  there would be mass splits of $\pi^+, \pi^0, \pi^-$ at low temperature. Masses of $\pi^+$ and $\pi^-$ are shifted down and up respectively, while $m_{\pi^0}$ is kept. This result is consistent with previous holographic study at $T=0, \mu_I\neq0$\cite{Lee:2013oya,Nishihara:2014nsa,Mamedov:2015sha}. Phenomenologically, the mass splits of $\pi^\pm$ would affect the distribution of $\pi^\pm$ and contribute to the ratio of multiplicity of charged pions $\pi^-$ to $\pi^+$ detected in experiments. Below $T_{cp}$, with the increasing of temperature, $m_{\pi^-}, m_{\pi^0}$ decrease. $m_{\pi^+}$ increases to a maximum value first, and then decrease with temperature. Above $T_{cp}$, the three modes become degenerate and they increase with temperature, as shown in Fig.\ref{massfigmu01varyT}.

With a fixed low temperature, e.g. $T=0.06\rm{GeV}$, $m_{\pi^+}$($m_{\pi^-}$) decreases(increases) with $\mu_I$, while $m_{\pi^0}$ depends weakly on $\mu_I$. The mass splits of pions increase rapidly with $\mu_I$. When $m_{\pi^+}=0$ at $T=0.06\rm{GeV}$ and $ \mu_I=0.17\rm{GeV}$, an instability occurs and the pion condensed phase appears. There is a spontaneous symmetry breaking from $SU(2)_V$ to $U_{I}(1)$, and $\pi^+$ is the Goldstone mode. With a fixed higher temperature, e.g. $T=0.13\rm{GeV}$, qualitatively, the $\mu_I$ dependence of pions is similar to $T=0.06\rm{GeV}$ with relatively small $\mu_I$. However, one could no longer find the massless Goldstone mode  at large $\mu_I$. This is consistent with our previous study that no condensed phase exists above $T=0.129\rm{GeV}$.

The current study check the theoretical consistency of soft-wall holographic framework. The qualitative behavior of chiral phase transition and pion superfluidity transition could be well realized both from the order parameters and hadronic spectral. Furthermore, the extracted $T, \mu_I$ dependence of light scalar and pseudo-scalar mesons might have interesting consequences in heavy ion collisions. From the LHC data, which indicate that the chemical freeze-out temperature $T_{ch}$ is about $0.156$ GeV~\cite{Stachel:2013zma}, at which the hadron abundances are fixed. From the current results, the decrease of the pion mass around this temperature might contribute to the overpopulation of pions at low momenta. Also, the effect of pion splits at finite $\mu_I$ might contribute to the charge imbalance in the final particle spectral.  Though the highest pion condensed temperature $T_{top}=0.129$ GeV is a bit lower than $T_{ch}$, it is still higher than the thermal freeze-out temperature $T_{th}$, which is estimated at $0.1-0.12$ GeV~\cite{Pratt:1999ku, Melo:2015wpa,Prorok:2015vxa}.  Thus, it is possible for the formation of pion condensation before thermal freeze-out. Therefore, in the final detection, the low energy pion and the coherent fraction would be enhanced. The quantitative relation between our results and the experimental data requires further study, and we leave it to the future.

\vspace{12pt}
\noindent {\bf Acknowledgments:}
We would like to thank the useful discussion with Xinyang Wang and Lang Yu. H.L. is supported by the National Natural Science Foundation of China under Grant No. 11405074. D.L. is supported by the National Natural Science Foundation of China under Grant No.11805084, the PhD Start-up Fund of Natural Science Foundation of Guangdong Province under Grant No. 2018030310457 and Guangdong Pearl River Talents Plan under Grant No. 2017GC010480.

%\clearpage


\begin{thebibliography}{ABC}	
%%%%%%%%%%%%%%%%%%%%%%%%%%%%%%%%%%%%%%%%%%%%%%%%%%%	

\bibitem{Adams:2005dq}
  J.~Adams {\it et al.} [STAR Collaboration],
  ``Experimental and theoretical challenges in the search for the quark gluon plasma: The STAR Collaboration's critical assessment of the evidence from RHIC collisions,''
  Nucl.\ Phys.\ A {\bf 757} (2005) 102
  %doi:10.1016/j.nuclphysa.2005.03.085
  [nucl-ex/0501009].
  %%CITATION = doi:10.1016/j.nuclphysa.2005.03.085;%%

\bibitem{Cleymans:2006xj}
  J.~Cleymans, I.~Kraus, H.~Oeschler, K.~Redlich and S.~Wheaton,
  ``Statistical model predictions for particle ratios at s(NN)**(1/2) = 5.5-TeV,''
  Phys.\ Rev.\ C {\bf 74} (2006) 034903
  %doi:10.1103/PhysRevC.74.034903
  [hep-ph/0604237].

\bibitem{Andronic:2008gu}
  A.~Andronic, P.~Braun-Munzinger and J.~Stachel,
  ``Thermal hadron production in relativistic nuclear collisions: The Hadron mass spectrum, the horn, and the QCD phase transition,''
  Phys.\ Lett.\ B {\bf 673} (2009) 142
   Erratum: [Phys.\ Lett.\ B {\bf 678} (2009) 516]
  %doi:10.1016/j.physletb.2009.02.014, 10.1016/j.physletb.2009.06.021
  [arXiv:0812.1186 [nucl-th]].




\bibitem{Dumitru:1994vc}
  A.~Dumitru, U.~Katscher, J.~A.~Maruhn, H.~Stoecker, W.~Greiner and D.~H.~Rischke,
  ``Pion and thermal photon spectra as a possible signal for a phase transition,''
  Phys.\ Rev.\ C {\bf 51} (1995) 2166
  %doi:10.1103/PhysRevC.51.2166
  [hep-ph/9411358].
  %%CITATION = doi:10.1103/PhysRevC.51.2166;%%


\bibitem{Abelev:2013pqa}
  B.~B.~Abelev {\it et al.} [ALICE Collaboration],
  ``Two- and three-pion quantum statistics correlations in Pb-Pb collisions at $\sqrt{{s}_{NN}} =$ 2.76 TeV at the CERN Large Hadron Collider,''
  Phys.\ Rev.\ C {\bf 89} (2014) no.2,  024911
  %doi:10.1103/PhysRevC.89.024911
  [arXiv:1310.7808 [nucl-ex]].
  %%CITATION = doi:10.1103/PhysRevC.89.024911;%%


\bibitem{Begun:2013nga}
  V.~Begun, W.~Florkowski and M.~Rybczynski,
  ``Explanation of hadron transverse-momentum spectra in heavy-ion collisions at $\sqrt s_{NN} =$ 2.76 TeV within chemical non-equilibrium statistical hadronization model,''
  Phys.\ Rev.\ C {\bf 90} (2014) no.1,  014906
  %doi:10.1103/PhysRevC.90.014906
  [arXiv:1312.1487 [nucl-th]].


\bibitem{Begun:2015ifa}
  V.~Begun and W.~Florkowski,
  ``Bose-Einstein condensation of pions in heavy-ion collisions at the CERN Large Hadron Collider (LHC) energies,''
  Phys.\ Rev.\ C {\bf 91} (2015) 054909
  %doi:10.1103/PhysRevC.91.054909
  [arXiv:1503.04040 [nucl-th]].
  %%CITATION = doi:10.1103/PhysRevC.91.054909;%%


\bibitem{Sako:2014usa}
  M.~Sako {\it et al.},
  ``Beam energy dependence of charged pion ratio in $^{28}$Si + In reactions,''
  arXiv:1409.3322 [nucl-ex].
  %%CITATION = ARXIV:1409.3322;%%


\bibitem{Ishii:2016dln}
  M.~Ishii, H.~Kouno and M.~Yahiro,
  ``Model prediction for temperature dependence of meson pole masses from lattice QCD results on meson screening masses,''
  Phys.\ Rev.\ D {\bf 95} (2017) no.11,  114022
  %doi:10.1103/PhysRevD.95.114022
  [arXiv:1609.04575 [hep-ph]].
  %%CITATION = doi:10.1103/PhysRevD.95.114022;%%


\bibitem{Brandt:2014qqa}
  B.~B.~Brandt, A.~Francis, H.~B.~Meyer and D.~Robaina,
  ``Chiral dynamics in the low-temperature phase of QCD,''
  Phys.\ Rev.\ D {\bf 90} (2014) no.5,  054509
  %doi:10.1103/PhysRevD.90.054509
  [arXiv:1406.5602 [hep-lat]].
  %%CITATION = doi:10.1103/PhysRevD.90.054509;%%


\bibitem{Brandt:2015sxa}
  B.~B.~Brandt, A.~Francis, H.~B.~Meyer and D.~Robaina,
  ``Pion quasiparticle in the low-temperature phase of QCD,''
  Phys.\ Rev.\ D {\bf 92} (2015) no.9,  094510
  %doi:10.1103/PhysRevD.92.094510
  [arXiv:1506.05732 [hep-lat]].
  %%CITATION = doi:10.1103/PhysRevD.92.094510;%%

\bibitem{Fischer:2018sdj}
  C.~S.~Fischer,
  ``QCD at finite temperature and chemical potential from Dyson–Schwinger equations,''
  Prog.\ Part.\ Nucl.\ Phys.\  {\bf 105} (2019) 1
  %doi:10.1016/j.ppnp.2019.01.002
  [arXiv:1810.12938 [hep-ph]].


\bibitem{Gao:2020hwo}
  F.~Gao and M.~Ding,
  ``Thermal properties of $\pi$ and $\rho$ meson,''
  arXiv:2006.05909 [hep-ph].


\bibitem{Tripolt:2013jra}
  R.~A.~Tripolt, N.~Strodthoff, L.~von Smekal and J.~Wambach,
  ``Spectral Functions for the Quark-Meson Model Phase Diagram from the Functional Renormalization Group,''
  Phys.\ Rev.\ D {\bf 89} (2014) no.3,  034010
  %doi:10.1103/PhysRevD.89.034010
  [arXiv:1311.0630 [hep-ph]].

\bibitem{Wang:2017vis}
  Z.~Wang and P.~Zhuang,
  ``Meson spectral functions at finite temperature and isospin density with the functional renormalization group,''
  Phys.\ Rev.\ D {\bf 96} (2017) no.1,  014006
  %doi:10.1103/PhysRevD.96.014006
  [arXiv:1703.01035 [hep-ph]].


%\cite{Ebert:1992jx}
\bibitem{Ebert:1992jx}
  D.~Ebert, Y.~L.~Kalinovsky and M.~K.~Volkov,
  ``Mesons at finite temperature in the NJL model with gluon condensate,''
  Phys.\ Lett.\ B {\bf 301} (1993) 231.
  %doi:10.1016/0370-2693(93)90694-D
  %%CITATION = doi:10.1016/0370-2693(93)90694-D;%%


\bibitem{Xia:2013caa}
  T.~Xia, L.~He and P.~Zhuang,
  ``Three-flavor Nambu–Jona-Lasinio model at finite isospin chemical potential,''
  Phys.\ Rev.\ D {\bf 88} (2013) no.5,  056013
  %doi:10.1103/PhysRevD.88.056013
  [arXiv:1307.4622 [hep-ph]].


\bibitem{Xia:2014bla}
  T.~Xia, J.~Hu and S.~Mao,
  ``Quark-antiquark Scattering Phase Shift and Meson Spectral Function in Pion Superfluid,''
  Chin.\ Phys.\ C {\bf 43} (2019) no.5,  054103
  %doi:10.1088/1674-1137/43/5/054103
  [arXiv:1411.6713 [hep-ph]].


\bibitem{Liu:2018zag}
  H.~Liu, X.~Wang, L.~Yu and M.~Huang,
  ``Neutral and charged scalar mesons, pseudoscalar mesons, and diquarks in magnetic fields,''
  Phys.\ Rev.\ D {\bf 97} (2018) no.7,  076008
  %doi:10.1103/PhysRevD.97.076008
  [arXiv:1801.02174 [hep-ph]].
  %%CITATION = doi:10.1103/PhysRevD.97.076008;%%



\bibitem{Chao:2018ejd}
  J.~Chao, M.~Huang and A.~Radzhabov,
  ``Charged pion condensation in anti-parallel electromagnetic fields and nonzero isospin density,''
  Chin.\ Phys.\ C {\bf 44} (2020) no.3,  034105
  %doi:10.1088/1674-1137/44/3/034105
  [arXiv:1805.00614 [hep-ph]].


\bibitem{Xu:2020yag}
  K.~Xu, J.~Chao and M.~Huang,
  ``Spin polarization inducing diamagnetism, inverse magnetic catalysis and saturation behavior of charged pion spectra,''
  arXiv:2007.13122 [hep-ph].
  %%CITATION = ARXIV:2007.13122;%%





\bibitem{Shuryak:1990ie}
  E.~V.~Shuryak,
  ``Physics of the pion liquid,''
  Phys.\ Rev.\ D {\bf 42} (1990) 1764.
  %doi:10.1103/PhysRevD.42.1764
  %%CITATION = doi:10.1103/PhysRevD.42.1764;%%



%\cite{Son:2001ff}
\bibitem{Son:2001ff}
  D.~T.~Son and M.~A.~Stephanov,
  ``Pion propagation near the QCD chiral phase transition,''
  Phys.\ Rev.\ Lett.\  {\bf 88}, 202302 (2002)
  %doi:10.1103/PhysRevLett.88.202302
  [hep-ph/0111100].
  %%CITATION = doi:10.1103/PhysRevLett.88.202302;%%

\bibitem{Son:2002ci}
  D.~T.~Son and M.~A.~Stephanov,
  ``Real time pion propagation in finite temperature QCD,''
  Phys.\ Rev.\ D {\bf 66} (2002) 076011
  %doi:10.1103/PhysRevD.66.076011
  [hep-ph/0204226].







\bibitem{Maldacena:1997re}
  J.~M.~Maldacena,
  ``The large N limit of superconformal field theories and supergravity,''
  Adv.\ Theor.\ Math.\ Phys.\  {\bf 2}, 231 (1998)
  [Int.\ J.\ Theor.\ Phys.\  {\bf 38}, 1113 (1999)]  [arXiv:hep-th/9711200].
  %%CITATION = IJTPB,38,1113;%%


%\cite{Gubser:1998bc}
\bibitem{Gubser:1998bc}
  S.~S.~Gubser, I.~R.~Klebanov and A.~M.~Polyakov,
  ``Gauge theory correlators from non-critical string theory,''
  Phys.\ Lett.\  B {\bf 428}, 105 (1998)
  [arXiv:hep-th/9802109].
  %%CITATION = PHLTA,B428,105;%%


%\cite{Witten:1998qj}
\bibitem{Witten:1998qj}
  E.~Witten,
  ``Anti-de Sitter space and holography,''
  Adv.\ Theor.\ Math.\ Phys.\  {\bf 2}, 253 (1998)
  [arXiv:hep-th/9802150].
  %%CITATION = 00203,2,253;%%


\bibitem{Kovtun:2004de}
  P.~Kovtun, D.~T.~Son and A.~O.~Starinets,
  ``Viscosity in strongly interacting quantum field theories from black hole physics,''
  Phys.\ Rev.\ Lett.\  {\bf 94} (2005) 111601
  %doi:10.1103/PhysRevLett.94.111601
  [hep-th/0405231].

\bibitem{Teaney:2000cw}
  D.~Teaney, J.~Lauret and E.~V.~Shuryak,
  ``Flow at the SPS and RHIC as a quark gluon plasma signature,''
  Phys.\ Rev.\ Lett.\  {\bf 86} (2001) 4783
  %doi:10.1103/PhysRevLett.86.4783
  [nucl-th/0011058].

\bibitem{Huovinen:2001cy}
  P.~Huovinen, P.~F.~Kolb, U.~W.~Heinz, P.~V.~Ruuskanen and S.~A.~Voloshin,
  ``Radial and elliptic flow at RHIC: Further predictions,''
  Phys.\ Lett.\ B {\bf 503} (2001) 58
  %doi:10.1016/S0370-2693(01)00219-2
  [hep-ph/0101136].

\bibitem{Hirano:2005xf}
  T.~Hirano, U.~W.~Heinz, D.~Kharzeev, R.~Lacey and Y.~Nara,
  ``Hadronic dissipative effects on elliptic flow in ultrarelativistic heavy-ion collisions,''
  Phys.\ Lett.\ B {\bf 636} (2006) 299
  %doi:10.1016/j.physletb.2006.03.060
  [nucl-th/0511046].

\bibitem{Romatschke:2007mq}
  P.~Romatschke and U.~Romatschke,
  ``Viscosity Information from Relativistic Nuclear Collisions: How Perfect is the Fluid Observed at RHIC?,''
  Phys.\ Rev.\ Lett.\  {\bf 99} (2007) 172301
  %doi:10.1103/PhysRevLett.99.172301
  [arXiv:0706.1522 [nucl-th]].



\bibitem{Karch:2002sh}
  A.~Karch and E.~Katz,
  ``Adding flavor to AdS / CFT,''
  JHEP {\bf 0206} (2002) 043
  %doi:10.1088/1126-6708/2002/06/043
  [hep-th/0205236].

\bibitem{Babington:2003vm}
  J.~Babington, J.~Erdmenger, N.~J.~Evans, Z.~Guralnik and I.~Kirsch,
  ``Chiral symmetry breaking and pions in nonsupersymmetric gauge / gravity duals,''
  Phys.\ Rev.\ D {\bf 69} (2004) 066007
  %doi:10.1103/PhysRevD.69.066007
  [hep-th/0306018].

\bibitem{Kruczenski:2003uq}
  M.~Kruczenski, D.~Mateos, R.~C.~Myers and D.~J.~Winters,
  ``Towards a holographic dual of large N(c) QCD,''
  JHEP {\bf 0405} (2004) 041
  %doi:10.1088/1126-6708/2004/05/041
  [hep-th/0311270].

\bibitem{Sakai:2004cn}
  T.~Sakai and S.~Sugimoto,
  ``Low energy hadron physics in holographic QCD,''
  Prog.\ Theor.\ Phys.\  {\bf 113} (2005) 843
  %doi:10.1143/PTP.113.843
  [hep-th/0412141].

\bibitem{Sakai:2005yt}
  T.~Sakai and S.~Sugimoto,
  ``More on a holographic dual of QCD,''
  Prog.\ Theor.\ Phys.\  {\bf 114} (2005) 1083
  %doi:10.1143/PTP.114.1083
  [hep-th/0507073].

\bibitem{Erlich:2005qh}
  J.~Erlich, E.~Katz, D.~T.~Son and M.~A.~Stephanov,
  ``QCD and a holographic model of hadrons,''
  Phys.\ Rev.\ Lett.\  {\bf 95} (2005) 261602
  %doi:10.1103/PhysRevLett.95.261602
  [hep-ph/0501128].

\bibitem{Karch:2006pv}
  A.~Karch, E.~Katz, D.~T.~Son and M.~A.~Stephanov,
  ``Linear confinement and AdS/QCD,''
  Phys.\ Rev.\ D {\bf 74} (2006) 015005
  %doi:10.1103/PhysRevD.74.015005
  [hep-ph/0602229].

\bibitem{deTeramond:2005su}
  G.~F.~de Teramond and S.~J.~Brodsky,
  ``Hadronic spectrum of a holographic dual of QCD,''
  Phys.\ Rev.\ Lett.\  {\bf 94} (2005) 201601
  %doi:10.1103/PhysRevLett.94.201601
  [hep-th/0501022].

\bibitem{Gubser:2008ny}
  S.~S.~Gubser and A.~Nellore,
  ``Mimicking the QCD equation of state with a dual black hole,''
  Phys.\ Rev.\ D {\bf 78} (2008) 086007
  %doi:10.1103/PhysRevD.78.086007
  [arXiv:0804.0434 [hep-th]].

\bibitem{Gubser:2008yx}
  S.~S.~Gubser, A.~Nellore, S.~S.~Pufu and F.~D.~Rocha,
  ``Thermodynamics and bulk viscosity of approximate black hole duals to finite temperature quantum chromodynamics,''
  Phys.\ Rev.\ Lett.\  {\bf 101} (2008) 131601
  %doi:10.1103/PhysRevLett.101.131601
  [arXiv:0804.1950 [hep-th]].

\bibitem{DeWolfe:2010he}
  O.~DeWolfe, S.~S.~Gubser and C.~Rosen,
  ``A holographic critical point,''
  Phys.\ Rev.\ D {\bf 83} (2011) 086005
  %doi:10.1103/PhysRevD.83.086005
  [arXiv:1012.1864 [hep-th]].

\bibitem{Gursoy:2007cb}
  U.~Gursoy and E.~Kiritsis,
  ``Exploring improved holographic theories for QCD: Part I,''
  JHEP {\bf 0802} (2008) 032
  %doi:10.1088/1126-6708/2008/02/032
  [arXiv:0707.1324 [hep-th]].

\bibitem{Gursoy:2007er}
  U.~Gursoy, E.~Kiritsis and F.~Nitti,
  ``Exploring improved holographic theories for QCD: Part II,''
  JHEP {\bf 0802} (2008) 019
  %doi:10.1088/1126-6708/2008/02/019
  [arXiv:0707.1349 [hep-th]].



 %\cite{Aharony:1999ti}
\bibitem{Aharony:1999ti}
  O.~Aharony, S.~S.~Gubser, J.~M.~Maldacena, H.~Ooguri and Y.~Oz,
  ``Large N field theories, string theory and gravity,''
  Phys.\ Rept.\  {\bf 323}, 183 (2000)
  [arXiv:hep-th/9905111].
  %%CITATION = PRPLC,323,183;%%


%\cite{Erdmenger:2007cm}
\bibitem{Erdmenger:2007cm}
  J.~Erdmenger, N.~Evans, I.~Kirsch and E.~Threlfall,
  ``Mesons in Gauge/Gravity Duals - A Review,''
  Eur.\ Phys.\ J.\ A {\bf 35} (2008) 81
  [arXiv:0711.4467 [hep-th]].
  %%CITATION = ARXIV:0711.4467;%%
  %308 citations counted in INSPIRE as of 18 Jul 2015

%\cite{deTeramond:2012rt}
\bibitem{deTeramond:2012rt}
  G.~F.~de Teramond and S.~J.~Brodsky,
  ``Hadronic Form Factor Models and Spectroscopy Within the Gauge/Gravity Correspondence,''
  arXiv:1203.4025 [hep-ph].
  %%CITATION = ARXIV:1203.4025;%%
  %60 citations counted in INSPIRE as of 18 Jul 2015


%\cite{Adams:2012th}
\bibitem{Adams:2012th}
  A.~Adams, L.~D.~Carr, T.~Schafer, P.~Steinberg and J.~E.~Thomas,
  ``Strongly Correlated Quantum Fluids: Ultracold Quantum Gases, Quantum Chromodynamic Plasmas, and Holographic Duality,''
  New J.\ Phys.\  {\bf 14} (2012) 115009
  [arXiv:1205.5180 [hep-th]].

\bibitem{Brodsky:2014yha}
  S.~J.~Brodsky, G.~F.~de Teramond, H.~G.~Dosch and J.~Erlich,
  ``Light-Front Holographic QCD and Emerging Confinement,''
  Phys.\ Rept.\  {\bf 584} (2015) 1
  %doi:10.1016/j.physrep.2015.05.001
  [arXiv:1407.8131 [hep-ph]].








\bibitem{Gherghetta-Kapusta-Kelley}
T.~Gherghetta, J.~I.~Kapusta and T.~M.~Kelley,
  ``Chiral symmetry breaking in the soft-wall AdS/QCD model,''
Phys.\ Rev.\ D {\bf 79} (2009) 076003; % [arXiv:0902.1998 [hep-ph]].  %%CITATION = ARXIV:0902.1998;%%

\bibitem{Gherghetta-Kapusta-Kelley-2}
T.~M.~Kelley, S.~P.~Bartz and J.~I.~Kapusta,
  ``Pseudoscalar Mass Spectrum in a Soft-Wall Model of AdS/QCD,''
  Phys.\ Rev.\ D {\bf 83} (2011) 016002;
%[arXiv:1009.3009 [hep-ph]].  %%CITATION = ARXIV:1009.3009;%%

%\cite{Li:2012ay}
\bibitem{Li:2012ay}
  D.~Li, M.~Huang and Q.~S.~Yan,
  ``A dynamical soft-wall holographic QCD model for chiral symmetry breaking and linear confinement,''
  Eur.\ Phys.\ J.\ C {\bf 73} (2013) 2615
  [arXiv:1206.2824 [hep-th]].
  %%CITATION = ARXIV:1206.2824;%%
  %13 citations counted in INSPIRE as of 28 juil. 2015


%\cite{Li:2013oda}
\bibitem{Li:2013oda}
  D.~Li and M.~Huang,
  ``Dynamical holographic QCD model for glueball and light meson spectra,''
  JHEP {\bf 1311} (2013) 088
  [arXiv:1303.6929 [hep-ph]].
  %%CITATION = ARXIV:1303.6929;%%
  %13 citations counted in INSPIRE as of 28 juil. 2015



\bibitem{YLWu}
  Y.~-Q.~Sui, Y.~-L.~Wu, Z.~-F.~Xie and Y.~-B.~Yang,
  ``Prediction for the Mass Spectra of Resonance Mesons in the Soft-Wall AdS/QCD with a Modified 5D Metric,''
  Phys.\ Rev.\ D {\bf 81} (2010) 014024;  % [arXiv:0909.3887 [hep-ph]].  %%CITATION = ARXIV:0909.3887;%%

\bibitem{YLWu-1}
Y.~-Q.~Sui, Y.~-L.~Wu and Y.~-B.~Yang,
  ``Predictive AdS/QCD Model for Mass Spectra of Mesons with Three Flavors,''
  Phys.\ Rev.\ D {\bf 83} (2011) 065030.  %[arXiv:1012.3518 [hep-ph]].  %%CITATION = ARXIV:1012.3518;%%





\bibitem{Colangelo:2008us}
  P.~Colangelo, F.~De Fazio, F.~Giannuzzi, F.~Jugeau and S.~Nicotri,
  ``Light scalar mesons in the soft-wall model of AdS/QCD,''
   Phys.\ Rev.\ D {\bf 78}, 055009 (2008)  [arXiv:0807.1054 [hep-ph]].  %%CITATION = ARXIV:0807.1054;%%  %86 citations counted in INSPIRE as of 14 Mar 2013

\bibitem{Ballon-Bayona:2020qpq}
  A.~Ballon-Bayona and L.~A.~H.~Mamani,
  ``Nonlinear realization of chiral symmetry breaking in holographic soft wall models,''
  Phys.\ Rev.\ D {\bf 102} (2020) no.2,  026013
  %doi:10.1103/PhysRevD.102.026013
  [arXiv:2002.00075 [hep-ph]].

\bibitem{FolcoCapossoli:2019imm}
  E.~Folco Capossoli, M.~A.~Martín Contreras, D.~Li, A.~Vega and H.~Boschi-Filho,
  ``Hadronic Spectra from Deformed AdS Backgrounds,''
  Chin.\ Phys.\ C {\bf 44} (2020) no.6,  064104
  %doi:10.1088/1674-1137/44/6/064104
  [arXiv:1903.06269 [hep-ph]].

\bibitem{Contreras:2018hbi}
  M.~Á.~Martín Contreras, A.~Vega and S.~Cortés,
  ``Light Pseudoscalar and Axial Spectroscopy using AdS/QCD Modified Soft Wall Model,''
  Chin.\ J.\ Phys.\  {\bf 66} (2020) 715
  %doi:10.1016/j.cjph.2020.06.018
  [arXiv:1811.10731 [hep-ph]].



%\cite{Colangelo:2011sr}
\bibitem{Colangelo:2011sr}
  P.~Colangelo, F.~Giannuzzi, S.~Nicotri and V.~Tangorra,
  ``Temperature and quark density effects on the chiral condensate: An AdS/QCD study,''
  Eur.\ Phys.\ J.\ C {\bf 72} (2012) 2096  [arXiv:1112.4402 [hep-ph]].  %%CITATION = ARXIV:1112.4402;%%  %6 citations counted in INSPIRE as of 14 Jun 2013



\bibitem{Dudal:2015wfn}
  D.~Dudal, D.~R.~Granado and T.~G.~Mertens,
  ``No inverse magnetic catalysis in the QCD hard and soft wall models,''
  Phys.\ Rev.\ D {\bf 93}, no. 12, 125004 (2016)
  [arXiv:1511.04042 [hep-th]].

\bibitem{Chelabi:2015cwn}
  K.~Chelabi, Z.~Fang, M.~Huang, D.~Li and Y.~L.~Wu,
  ``Realization of chiral symmetry breaking and restoration in holographic QCD,''
  Phys.\ Rev.\ D {\bf 93}, no. 10, 101901 (2016) [arXiv:1511.02721 [hep-ph]].

\bibitem{Chelabi:2015gpc}
  K.~Chelabi, Z.~Fang, M.~Huang, D.~Li and Y.~L.~Wu,
  ``Chiral Phase Transition in the Soft-Wall Model of AdS/QCD,''
  JHEP {\bf 1604} (2016) 036
  [arXiv:1512.06493 [hep-ph]].

\bibitem{Fang:2015ytf}
  Z.~Fang, S.~He and D.~Li,
  ``Chiral and Deconfining Phase Transitions from Holographic QCD Study,''
  Nucl.\ Phys.\ B {\bf 907} (2016) 187
  [arXiv:1512.04062 [hep-ph]].

\bibitem{Li:2016gfn}
  D.~Li, M.~Huang, Y.~Yang and P.~H.~Yuan,
  ``Inverse Magnetic Catalysis in the Soft-Wall Model of AdS/QCD,''
  JHEP {\bf 1702} (2017) 030
  [arXiv:1610.04618 [hep-th]].

\bibitem{Li:2016smq}
  D.~Li and M.~Huang,
  ``Chiral phase transition of QCD with $N_f=2+1$ flavors from holography,''
  JHEP {\bf 1702} (2017) 042
  [arXiv:1610.09814 [hep-ph]].

\bibitem{Bartz:2016ufc}
  S.~P.~Bartz and T.~Jacobson,
  ``Chiral Phase Transition and Meson Melting from AdS/QCD,''
  Phys.\ Rev.\ D {\bf 94} (2016) 075022
  [arXiv:1607.05751 [hep-ph]].

\bibitem{Fang:2016nfj}
  Z.~Fang, Y.~L.~Wu and L.~Zhang,
  ``Chiral phase transition and meson spectrum in improved soft-wall AdS/QCD,''
  Phys.\ Lett.\ B {\bf 762} (2016) 86
  [arXiv:1604.02571 [hep-ph]].

\bibitem{Bartz:2017jku}
  S.~P.~Bartz and T.~Jacobson,
  ``Chiral phase transition at finite chemical potential in 2+1 -flavor soft-wall anti–de Sitter space QCD,''
  Phys.\ Rev.\ C {\bf 97} (2018) no.4,  044908
  [arXiv:1801.00358 [hep-ph]].

\bibitem{Fang:2018vkp}
  Z.~Fang, Y.~L.~Wu and L.~Zhang,
  ``Chiral Phase Transition with 2+1 quark flavors in an improved soft-wall AdS/QCD Model,''
  arXiv:1805.05019 [hep-ph].


\bibitem{Ghoroku:2005kg}
  K.~Ghoroku and M.~Yahiro,
  ``Holographic model for mesons at finite temperature,''
  Phys.\ Rev.\ D {\bf 73} (2006) 125010
  %doi:10.1103/PhysRevD.73.125010
  [hep-ph/0512289].



\bibitem{Cui:2013zha}
  L.~X.~Cui and Y.~L.~Wu,
  ``Thermal Mass Spectra of Scalar and Pseudo-Scalar Mesons in IR-improved Soft-Wall AdS/QCD Model with Finite Chemical Potential,''
  Mod.\ Phys.\ Lett.\ A {\bf 28} (2013) 1350132
  %doi:10.1142/S0217732313501320
  [arXiv:1302.4828 [hep-ph]].


\bibitem{Cui:2014oba}
  L.~X.~Cui, Z.~Fang and Y.~L.~Wu,
  ``Thermal Spectral Function and Deconfinement Temperature in Bulk Holographic AdS/QCD with Back Reaction of Bulk Vacuum,''
  Chin.\ Phys.\ C {\bf 40} (2016) no.6,  063101
  %doi:10.1088/1674-1137/40/6/063101
  [arXiv:1404.0761 [hep-ph]].


\bibitem{Nambu:1961tp}
  Y.~Nambu and G.~Jona-Lasinio,
  ``Dynamical Model of Elementary Particles Based on an Analogy with Superconductivity. I,''
  Phys.\ Rev.\  {\bf 122} (1961) 345.


\bibitem{Nambu:1961fr}
  Y.~Nambu and G.~Jona-Lasinio,
  ``Dynamical Model Of Elementary Particles Based On An Analogy With Superconductivity. II,''
  Phys.\ Rev.\  {\bf 124} (1961) 246.

\bibitem{Aggarwal:2010cw}
  M.~M.~Aggarwal {\it et al.} [STAR Collaboration],
  ``An Experimental Exploration of the QCD Phase Diagram: The Search for the Critical Point and the Onset of De-confinement,''
  arXiv:1007.2613 [nucl-ex].

\bibitem{Odyniec:2013aaa}
  G.~Odyniec,
  ``RHIC Beam Energy Scan Program: Phase I and II,''
  PoS CPOD {\bf 2013} (2013) 043.

\bibitem{Luo:2017faz}
  X.~Luo and N.~Xu,
  ``Search for the QCD Critical Point with Fluctuations of Conserved Quantities in Relativistic Heavy-Ion Collisions at RHIC : An Overview,''
  Nucl.\ Sci.\ Tech.\  {\bf 28} (2017) no.8,  112
  [arXiv:1701.02105 [nucl-ex]].



\bibitem{Albrecht:2010eg}
  D.~Albrecht and J.~Erlich,
  ``Pion condensation in holographic QCD,''
  Phys.\ Rev.\ D {\bf 82} (2010) 095002
  [arXiv:1007.3431 [hep-ph]].

\bibitem{Lee:2013oya}
  B.~H.~Lee, S.~Mamedov, S.~Nam and C.~Park,
  ``Holographic meson mass splitting in the Nuclear Matter,''
  JHEP {\bf 1308} (2013) 045
  [arXiv:1305.7281 [hep-th]].

\bibitem{Nishihara:2014nva}
  H.~Nishihara and M.~Harada,
  ``Enhancement of Chiral Symmetry Breaking from the Pion condensation at finite isospin chemical potential in a holographic QCD model,''
  Phys.\ Rev.\ D {\bf 89} (2014) no.7,  076001
  [arXiv:1401.2928 [hep-ph]].



\bibitem{Nishihara:2014nsa}
  H.~Nishihara and M.~Harada,
  ``Equation of state in the pion condensation phase in asymmetric nuclear matter using a holographic QCD model,''
  Phys.\ Rev.\ D {\bf 90} (2014) no.11,  115027
  [arXiv:1407.7344 [hep-ph]].


\bibitem{Mamedov:2015sha}
  S.~Mamedov,
  ``Meson effective mass in the isospin medium in hard-wall AdS/QCD model,''
  Eur.\ Phys.\ J.\ C {\bf 76} (2016) no.2,  83
  [arXiv:1504.05687 [hep-th]].

\bibitem{Lv:2018wfq}
  M.~Lv, D.~Li and S.~He,
  %``Pion condensation in a soft-wall AdS/QCD model,''
  JHEP {\bf 1911} (2019) 026
  doi:10.1007/JHEP11(2019)026
  [arXiv:1811.03828 [hep-ph]].

\bibitem{Cao:2020ske}
  X.~Cao, H.~Liu, D.~Li and G.~Ou,
  ``QCD phase diagram at finite isospin chemical potential and temperature in an IR-improved soft-wall AdS/QCD model,''
  Chin.\ Phys.\ C {\bf 44} (2020) no.8,  083106
  %doi:10.1088/1674-1137/44/8/083106
  [arXiv:2001.02888 [hep-ph]].

\bibitem{Rodrigues:2018chh}
  D.~M.~Rodrigues, D.~Li, E.~Folco Capossoli and H.~Boschi-Filho,
  ``Holographic Description of Chiral Symmetry Breaking in a Magnetic Field in 2+1 Dimensions with an Improved Dilaton,''
  EPL {\bf 128} (2019) no.6,  61001
  %doi:10.1209/0295-5075/128/61001
  [arXiv:1811.04117 [hep-ph]].

\bibitem{Rodrigues:2018pep}
  D.~M.~Rodrigues, D.~Li, E.~Folco Capossoli and H.~Boschi-Filho,
  ``Chiral symmetry breaking and restoration in 2+1 dimensions from holography: Magnetic and inverse magnetic catalysis,''
  Phys.\ Rev.\ D {\bf 98} (2018) no.10,  106007
  %doi:10.1103/PhysRevD.98.106007
  [arXiv:1807.11822 [hep-th]].

\bibitem{Rougemont:2017tlu}
  R.~Rougemont, R.~Critelli, J.~Noronha-Hostler, J.~Noronha and C.~Ratti,
  ``Dynamical versus equilibrium properties of the QCD phase transition: A holographic perspective,''
  Phys.\ Rev.\ D {\bf 96} (2017) no.1,  014032
  %doi:10.1103/PhysRevD.96.014032
  [arXiv:1704.05558 [hep-ph]].

\bibitem{Finazzo:2014cna}
  S.~I.~Finazzo, R.~Rougemont, H.~Marrochio and J.~Noronha,
  ``Hydrodynamic transport coefficients for the non-conformal quark-gluon plasma from holography,''
  JHEP {\bf 1502} (2015) 051
  %doi:10.1007/JHEP02(2015)051
  [arXiv:1412.2968 [hep-ph]].

\bibitem{Zollner:2018uep}
  R.~Zöllner and B.~Kämpfer,
  ``Phase structures emerging from holography with Einstein gravity -- dilaton models at finite temperature,''
  Eur.\ Phys.\ J.\ Plus {\bf 135} (2020) no.3,  304
  %doi:10.1140/epjp/s13360-020-00106-3
  [arXiv:1807.04260 [hep-th]].

\bibitem{Li:2011hp}
  D.~Li, S.~He, M.~Huang and Q.~S.~Yan,
  ``Thermodynamics of deformed AdS$_5$ model with a positive/negative quadratic correction in graviton-dilaton system,''
  JHEP {\bf 1109} (2011) 041
  %doi:10.1007/JHEP09(2011)041
  [arXiv:1103.5389 [hep-th]].

\bibitem{Cai:2012xh}
  R.~G.~Cai, S.~He and D.~Li,
  ``A hQCD model and its phase diagram in Einstein-Maxwell-Dilaton system,''
  JHEP {\bf 1203} (2012) 033
  %doi:10.1007/JHEP03(2012)033
  [arXiv:1201.0820 [hep-th]].

\bibitem{Chen:2019rez}
  X.~Chen, D.~Li, D.~Hou and M.~Huang,
  ``Quarkyonic phase from quenched dynamical holographic QCD model,''
  JHEP {\bf 2003} (2020) 073
  %doi:10.1007/JHEP03(2020)073
  [arXiv:1908.02000 [hep-ph]].


\bibitem{He:2020fdi}
  S.~He, Y.~Yang and P.~H.~Yuan,
  ``Analytic Study of Magnetic Catalysis in Holographic QCD,''
  arXiv:2004.01965 [hep-th].


\bibitem{Ballon-Bayona:2020xls}
  A.~Ballon-Bayona, H.~Boschi-Filho, E.~Folco Capossoli and D.~M.~Rodrigues,
  ``Criticality from EMD holography at finite temperature and density,''
  arXiv:2006.08810 [hep-th].

\bibitem{Mamani:2020pks}
  L.~A.~H.~Mamani, C.~V.~Flores and V.~T.~Zanchin,
  ``Phase diagram and compact stars in a holographic QCD model,''
  arXiv:2006.09401 [hep-th].


 %\cite{Cherman:2008eh}
\bibitem{Cherman:2008eh}
  A.~Cherman, T.~D.~Cohen and E.~S.~Werbos,
  ``The Chiral condensate in holographic models of QCD,''  Phys.\ Rev.\ C {\bf 79} (2009) 045203  [arXiv:0804.1096 [hep-ph]].






\bibitem{Chen:2018msc}
  J.~Chen, S.~He, M.~Huang and D.~Li,
  ``Critical exponents of finite temperature chiral phase transition in soft-wall AdS/QCD models,''
  JHEP {\bf 1901} (2019) 165
  %doi:10.1007/JHEP01(2019)165
  [arXiv:1810.07019 [hep-ph]].

\bibitem{Son:2002sd}
  D.~T.~Son and A.~O.~Starinets,
  ``Minkowski space correlators in AdS / CFT correspondence: Recipe and applications,''
  JHEP {\bf 0209} (2002) 042
  %doi:10.1088/1126-6708/2002/09/042
  [hep-th/0205051].

\bibitem{Abidin:2008hn}
  Z.~Abidin and C.~E.~Carlson,
  ``Gravitational Form Factors in the Axial Sector from an AdS/QCD Model,''
  Phys.\ Rev.\ D {\bf 77} (2008) 115021
  %doi:10.1103/PhysRevD.77.115021
  [arXiv:0804.0214 [hep-ph]].

\bibitem{Abidin:2019xwu}
    Z.~Abidin and P.~T.~P.~Hutauruk,
    ``Kaon form factor in holographic QCD,''
    Phys. Rev. D \textbf{100} (2019) no.5, 054026
    %doi:10.1103/PhysRevD.100.054026
    [arXiv:1905.08953 [hep-ph]].
%1 citations counted in INSPIRE as of 27 Aug 2020

\bibitem{Li:1997px}
  B.~A.~Li, C.~M.~Ko and W.~Bauer,
  ``Isospin physics in heavy ion collisions at intermediate-energies,''
  Int.\ J.\ Mod.\ Phys.\ E {\bf 7} (1998) 147
  [nucl-th/9707014].


\bibitem{Son:2000xc}
  D.~T.~Son and M.~A.~Stephanov,
  ``QCD at finite isospin density,''
  Phys.\ Rev.\ Lett.\  {\bf 86} (2001) 592
  [hep-ph/0005225].

\bibitem{Brandt:2017oyy}
  B.~B.~Brandt, G.~Endrodi and S.~Schmalzbauer,
  ``QCD phase diagram for nonzero isospin-asymmetry,''
  Phys.\ Rev.\ D {\bf 97} (2018) no.5,  054514
  [arXiv:1712.08190 [hep-lat]].



\bibitem{Stachel:2013zma}
    J.~Stachel, A.~Andronic, P.~Braun-Munzinger and K.~Redlich,
    ``Confronting LHC data with the statistical hadronization model,''
    J. Phys. Conf. Ser. \textbf{509} (2014), 012019
    %doi:10.1088/1742-6596/509/1/012019
    [arXiv:1311.4662 [nucl-th]].
    %219 citations counted in INSPIRE as of 28 Aug 2020


\bibitem{Pratt:1999ku}
    S.~Pratt and K.~Haglin,
    ``Hadronic phase space density and chiral symmetry restoration in relativistic heavy ion collisions,''
    Phys. Rev. C \textbf{59} (1999), 3304-3308
    %doi:10.1103/PhysRevC.59.3304
    %21 citations counted in INSPIRE as of 28 Aug 2020


\bibitem{Melo:2015wpa}
    I.~Melo and B.~Tomasik,
    ``Reconstructing the final state of Pb+Pb collisions at $\sqrt{s_{NN}}=2.76$ TeV,''
    J. Phys. G \textbf{43} (2016) no.1, 015102
    %doi:10.1088/0954-3899/43/1/015102
    [arXiv:1502.01247 [nucl-th]].
    %18 citations counted in INSPIRE as of 28 Aug 2020

\bibitem{Prorok:2015vxa}
    D.~Prorok,
    ``Single Freeze-Out, Statistics and Pion, Kaon and Proton Production in Central Pb-Pb Collisions at $\sqrt{s_{NN}} = 2.76$ TeV,''
    J. Phys. G \textbf{43} (2016) no.5, 055101
    %doi:10.1088/0954-3899/43/5/055101
    [arXiv:1508.07922 [nucl-th]].
    %8 citations counted in INSPIRE as of 28 Aug 2020

\end{thebibliography}
 \end{document}